\documentclass[a4paper,11pt]{article}
\pdfoutput=1 

\usepackage{jcappub}
\usepackage{indentfirst}
\usepackage[table]{xcolor}

\usepackage{cancel}
\usepackage{braket}
\usepackage{mathrsfs}
\numberwithin{equation}{section}
\newcolumntype{L}{>{$}l<{$}} 

\usepackage{tikz}	
\usetikzlibrary{quantikz}
\usetikzlibrary{backgrounds,fit,decorations.pathreplacing}

\allowdisplaybreaks

\makeatletter
\gdef\@fpheader{}
\g@addto@macro\bfseries{\boldmath}
\makeatother

\newcommand{\ie}{\textsl{i.e.~}}

\newcommand{\eg}{\textsl{e.g.~}}

\newcommand{\etc}{\textsl{etc.}}

\newcommand{\tr}{\mathrm{Tr}}

\newcommand{\dd}{\mathrm{d}}
\newcommand{\ee}{e}

\newcommand{\boldmathsymbol}[1]{{\ensuremath{\boldsymbol{#1}}}}

\newcommand{\uin}{\mathrm{in}}

\newcommand{\umax}{\mathrm{max}}

\newcommand{\Rea}{\Re \mathrm{e}\,}

\newcommand{\beq}{\begin{equation}}
\newcommand{\eeq}{\end{equation}}
\newcommand{\bea}{\begin{eqnarray}}
\newcommand{\eea}{\end{eqnarray}}

\newlength{\wsingfig}
\newlength{\wdblefig}
\newlength{\wquadfig}
\newlength{\wtriplefig}
\newcommand{\Eq}[1]{Eq.~(\ref{#1})}
\newcommand{\Eqs}[1]{Eqs.~(\ref{#1})}

\newcommand{\Refa}[1]{Ref.~{\cite{#1}}}
\newcommand{\Refs}[1]{Refs.~{\cite{#1}}}
\newcommand{\Sec}[1]{Sec.~\ref{#1}}
\newcommand{\Secs}[1]{Secs.~\ref{#1}}
\newcommand{\App}[1]{Appendix~\ref{#1}}

\newcommand{\ds}{\displaystyle}

\newcommand{\bs}{\boldsymbol}

\newcommand{\Spnew}{\mathscr{S}\kern-0.3em p}

\subheader{}

\title{Four-mode squeezed states: two-field quantum systems and the symplectic group $\mathrm{Sp}(4,\mathbb{R})$}

\author[a,b]{Thomas Colas,}
\affiliation[a]{Universit\'e Paris-Saclay, CNRS, Institut d'Astrophysique Spatiale, 91405, Orsay, France}
\emailAdd{thomas.colas@universite-paris-saclay.fr}

\author[a]{Julien Grain,}
\emailAdd{julien.grain@universite-paris-saclay.fr}

\author[b]{Vincent Vennin}
\affiliation[b]{Laboratoire Astroparticule et Cosmologie, Universit\'e Denis Diderot Paris 7, 10 rue
	Alice Domon et L\'eonie Duquet, 75013 Paris, France}
\emailAdd{vincent.vennin@apc.in2p3.fr}

\date{today}

\begin{document}
	\sloppy
	
	\abstract{We construct the four-mode squeezed states and study their physical properties. These states describe two linearly-coupled quantum scalar fields, which makes them physically relevant in various contexts such as cosmology. They are shown to generalise the usual two-mode squeezed states of single-field systems, with additional transfers of quanta between the fields. To build them in the Fock space, we use the symplectic structure of the phase space. For this reason, we first present a pedagogical analysis of the symplectic group $\mathrm{Sp}(4,\mathbb{R})$ and its Lie algebra, from which we construct the four-mode squeezed states and discuss their structure. We also study the reduced single-field system obtained by tracing out one of the two fields. This procedure being easier in the phase space, it motivates the use of the Wigner function which we introduce as an alternative description of the state. It allows us to discuss environmental effects in the case of linear interactions. In particular, we find that there is always a range of interaction coupling for which decoherence occurs without substantially affecting the power spectra (hence the observables) of the system.}
	
	
	
	\maketitle
	

\section{Introduction}
\label{sec:intro}
Two-mode squeezed states~\cite{BarnettRadmore,Caves:1985zz,Schumaker:1985zz} have been widely studied in the past for the important role they play in quantum optics (see \eg \Refs{Dodonov:2002rx, Schnabel:2016gdi} for reviews), but also in the cosmological context where they describe primordial density perturbations, amplified by gravitational instability from the vacuum quantum fluctuations~\cite{Grishchuk:1990bj,Grishchuk:1992tw,Albrecht:1992kf, Polarski:1995jg, Lesgourgues:1996jc, Kiefer:1998qe, Martin:2015qta}. In general, they characterise the quantum state of linear single-field systems, where each pair of Fourier modes is placed in a two-mode squeezed state~\cite{Grain:2019vnq}.

When more degrees of freedom are present however, two-mode squeezed state are insufficient and the squeezing formalism needs to be generalised to higher numbers of modes. This is why, in this work, we construct the four-mode squeezed states, which describe two linearly-coupled quantum scalar fields. The motivation behind this analysis is twofold. First, there are a number of situations where two-field systems are directly relevant, for instance during the inflationary phase our primordial universe underwent. Even if current cosmological data is consistent with single-field setups~\cite{Akrami:2018vks}, from a theoretical point of view, inflation takes place in a regime that is far beyond the reach of accelerators, and most physical setups that have been proposed to embed inflation contain extra scalar fields, for instance in the string-theoretic context~\cite{Turok:1987pg, Damour:1995pd, Kachru:2003sx, Krause:2007jk, Baumann:2014nda}. Those additional degrees of freedom are usually associated to entropic perturbations. Four-mode squeezed states would then naturally appear in two-field inflation models, and provide insight about multi-field cosmology in general. Second, this setup provides a way to investigate environmental effects in the case of linear interactions, by tracing over one of the two fields. More precisely, when the system of observational relevance couples to unobserved degrees of freedom (referred to as the ``environment''), quantum entanglement builds up between the system and the environment. This affects observational predictions and also leads to the quantum decoherence of the observable sector \cite{PhysRevD.24.1516, PhysRevD.26.1862, Joos:1984uk}. This phenomenon is usually investigated by the means of effective methods that only provide results that are perturbative in the interaction strength and that rely on additional assumptions, see \eg \Refa{Martin:2018zbe}. By considering that one of the two fields represents the observed system and the other field stands for the environment, the formalism we develop will allow us to go beyond those methods and present exact results. 

Let us stress that the explicit construction of the squeezed quantum states, especially in the Fock's space, is not only of formal interest. As we will explain, it provides important insight into the physical mechanisms at play in the dynamics of those states and in the emergence of peculiar properties such as quantum entanglement. Furthermore, it is required in a number of concrete computations (see for instance \Refs{2015JOSAB..32.2190O, Martin:2016nrr, Choudhury:2017qyl, Ando:2020kdz, Kanno:2021gpt}).

Although we are inspired by problems formulated in the context of cosmology, it is worth mentioning that the formalism we develop here is generic and broad in applicability. It does not require prior knowledge of the concepts and tools relevant in cosmology (which will only be mentioned in our concluding remarks for illustrative purpose), to which we plan to apply our results in separate publications. 

Let us now describe how this article is organised, and highlight its main results. In \Sec{sec:QuantumPhaseSpace}, we introduce the physical setup describing two free scalar fields, both at the classical and quantum levels, and we highlight the symplectic structure that underlies its phase space. This leads us to introducing the symplectic group in four dimensions, $\mathrm{Sp}(4,\mathbb{R})$, which we formally describe in \Sec{sec:Sp4R:toolkit}. This section reviews material that may also be found in other references on the same topic (apart from the fully factorised form of group elements, \Eq{eq:fullyfact}, which, up to our knowledge, is a new result), see \eg \Refs{Kim:2012av, Hasebe:2019ibg, Garcia-Chung:2020cag, Garcia-Chung:2020gxy, Chacon-Acosta:2021urw}. As a consequence, it may be skipped by those readers already familiar with the use of symplectic groups in quantum mechanics. Otherwise, it provides a self-contained presentation of the techniques employed in the rest of the paper. The Hamiltonians leading to the four-mode squeezed states are then built in \Sec{sec:QuantumDynamics}, where we also construct the evolution operator using theses results. In particular, we comment on the  physical interpretation of the generators of the Lie algebra, when acting in the Hamiltonian. As an example, we also briefly apply our formalism to describe two massless fields in a cosmological background. Finally, we use the previous results to write a tractable expression for the evolution operator from which four-mode squeezed states can be obtained. \Sec{sec:QuantumState} is devoted to the explicit construction of the four-mode squeezed states in the Fock basis, see \Eqs{eq:vacevolfinal} and~\eqref{eq:ampl}, which constitute one of the main results of this paper. An expansion around the limit where the two fields are uncoupled further allows us to discuss the physical interpretation of these formulas, to interpret the four-mode squeezed states in terms of particle transfer and to link their structure to the relevant microphysical parameters. We also derive the Wigner function of the system, which provides an alternative description of the state in the phase space. Although equivalent to the Fock-space description, its simple Gaussian form, built out of the power spectra of the configuration fields, makes some calculations simpler. Finally, in \Sec{sec:Decoherence}, we investigate environmental effects by tracing out one of the two fields and studying how the reduced quantum state of the first field is affected. This is done both at the level of the density matrix and of the Wigner function. Two independent calculations of the purity are thus performed, and then expanded in the small-coupling limit where they are shown to lead to the same result. \Sec{sec:conclu} presents our conclusions, and  the paper ends by four appendices to which various technical aspects of the calculations presented in the main text are deferred. 
\section{Quantum phase space of two free fields}
\label{sec:QuantumPhaseSpace}
\subsection{Two scalar fields in a homogeneous and isotropic background}
\label{sec:TwoFreeFields}
Let us consider two real-valued scalar fields $\phi_1(t,\vec{x})$ and $\phi_2(t,\vec{x})$, with conjugate momenta $\pi_1(t,\vec{x})$ and $\pi_2(t,\vec{x})$. These phase-space coordinates can be arranged into the four-dimensional vector $\boldsymbol{z}(\vec{x}) = (\phi_1(\vec{x}), \phi_2(\vec{x}), \pi_1(\vec{x}), \pi_2(\vec{x}))^{\mathrm{T}}$ where ``$\mathrm{T}$'' stands for the transpose and explicit time dependence is dropped for notational convenience. In this work, for simplicity, we focus on the case of free fields, for which the Hamiltonian $H$ is a local quadratic form, 
	\begin{align}
	\label{eq:Hamiltonian:Real:Space}
		H = \frac{1}{2}\int \dd^3 \vec{x} \boldsymbol{z}^{\mathrm{T}}(\vec{x}) \boldsymbol{H}(t)\boldsymbol{z}(\vec{x})\, .
	\end{align}
In the cosmological context, one may view $\phi_1$ and $\phi_2$ as two test fields (\ie they do not backreact on the background geometry), or as the perturbations of some cosmological fields where cosmological perturbation theory is carried out at leading order. In \Eq{eq:Hamiltonian:Real:Space}, $\boldsymbol{H}(t)$ is a four-by-four real symmetric matrix, which we assume does not depend on the spatial coordinate $\vec{x}$, so the background on which the fields evolve is homogeneous. Note that $\boldsymbol{H}(t)$ may however involve the gradient operator $\partial/\partial \vec{x}$ (though to positive powers only, to be compatible with the  locality assumption).

Phase space is equipped with the Poisson bracket
	\begin{align}
	\label{eq:Poisson:generic}
		\{F,G\} = \int \dd^3\vec{x} \left[\frac{\delta F}{\delta \phi_1(\vec{x})}\frac{\delta G}{\delta \pi_1(\vec{x})}  - \frac{\delta F}{\delta \pi_1(\vec{x})}\frac{\delta G}{\delta \phi_1(\vec{x})}  +
		\frac{\delta F}{\delta \phi_2(\vec{x})}\frac{\delta G}{\delta \pi_2(\vec{x})}  - \frac{\delta F}{\delta \pi_2(\vec{x})}\frac{\delta G}{\delta \phi_2(\vec{x})}  \right],
	\end{align}
which can be written in matricial form for the phase-space coordinates as	
	\begin{eqnarray}
	\label{eq:Poisson}
		\{\boldsymbol{z}(\vec{x}),\boldsymbol{z}^T(\vec{y})\} = \bs{\Omega} \delta^3(\vec{x} - \vec{y})\, .
	\end{eqnarray}
In this expression, $\delta$ is the Dirac distribution, the matrix $\boldsymbol{\Omega}$ is given by
\begin{eqnarray}
\label{eq:def:Omega}
		\boldsymbol{\Omega} = \begin{pmatrix}
			0 & \boldsymbol{I}_2 \\
			- \boldsymbol{I}_2 & 0
		\end{pmatrix},
\end{eqnarray}
where $\boldsymbol{I}_n$ is the $n \times n$ identity matrix, and the notation in \Eq{eq:Poisson} has to be understood as $\{\bs{z}_\mu (\vec{x}), \bs{z}_\nu (\vec{y})\} = \bs{\Omega}_{\mu \nu} \delta^3(\vec{x} - \vec{y})$ with $\mu,\nu=1\cdots 4$. The time evolution of any function $F$ of the phase-space field variables is given by 
	\begin{eqnarray}
	\label{eq:eom:generic}
		\Dot{F}(\boldsymbol{z}) = \{F(\boldsymbol{z}), H\}\, ,
	\end{eqnarray}
where a dot means differentiating with respect to the time variable. 
\subsubsection*{Fourier space}
Since the background is homogeneous, it is useful to work in Fourier space and to introduce
	\begin{eqnarray}
	\label{eq:Fourier}
		\boldsymbol{z}_{\vec{k}} = \begin{pmatrix} \phi_{1,\vec{k}}\\ \phi_{2, \vec{k}}\\\pi_{1, \vec{k}} \\ \pi_{2, \vec{k}}\end{pmatrix} = \int \frac{\mathrm{d}^3 \vec{x}}{(2\pi)^{3/2}} \boldsymbol{z}(\vec{x}) e^{-i\vec{k}.\vec{x}}.
	\end{eqnarray}
The condition for the fields to be real-valued, $\bs{z}(\vec{x})=\bs{z}^*(\vec{x})$, translates into $\boldsymbol{z}_{\vec{k}}^* = \boldsymbol{z}_{-\vec{k}}$. This means that half of the Fourier modes are enough to parametrise the entire (now complex) phase space. In practice, any integral over $k\in\mathbb{R}^{3}$ can be split into an integral over $\mathbb{R}^{3+} \equiv \mathbb{R}^2 \times \mathbb{R}^+$ and $\mathbb{R}^{3-} \equiv \mathbb{R}^2 \times \mathbb{R}^-$, where the latter can be related to the former by a simple change of integration variable $\vec{k}\to -\vec{k}$ and using the relation $\boldsymbol{z}_{\vec{k}}^* = \boldsymbol{z}_{-\vec{k}}$. For the Hamiltonian~\eqref{eq:Hamiltonian:Real:Space}, this leads to
\begin{eqnarray}
	\label{eq:quadhamilt}
		H = \int_{\mathbb{R}^{3+}} \mathrm{d}^3 \vec{k} \boldsymbol{z}_{\vec{k}}^\dag \boldsymbol{H}_k(t)\boldsymbol{z}_{\vec{k}}\, ,
\end{eqnarray}
which avoids double counting the degrees of freedom of the theory. In \Eq{eq:quadhamilt}, $\boldsymbol{H}_k(t)$ corresponds to $\boldsymbol{H}(t)$ where the spatial gradient $\partial/\partial\vec{x}$ is replaced with $i \vec{k}$. If we further assume the background to be isotropic, $\boldsymbol{H}_k(t)$ only depends on the norm $k=\vert \vec{k} \vert$ of $\vec{k}$, hence the notation.

Plugging \Eq{eq:Fourier} into \Eq{eq:Poisson:generic}, one can compute
	\begin{eqnarray}\label{eq:poissonk}
		\{\boldsymbol{z}_{\vec{k}},\boldsymbol{z}_{\vec{q}}^\dag\} = \bs{\Omega} \delta^3(\vec{k} - \vec{q})\, ,
	\end{eqnarray}
which is of the same form as \Eq{eq:Poisson}. This shows that the Poisson brackets are preserved when going to Fourier space (in the language that will be introduced in \Sec{sec:Symplectic:Structure}, the Fourier transform is a ``symplectic transformation''), so phase space can be equivalently parametrised with the Fourier coordinates, and the Poisson bracket~\eqref{eq:Poisson:generic} can also be written as
	\begin{eqnarray}
	\label{eq:Poisson:Fourier}
		\{F,G\} = \int_{\mathbb{R}^3} \dd^3 \vec{k} \left(\frac{\delta F}{\delta \phi_{1,\vec{k}}}\frac{\delta G}{\delta \pi^*_{1,\vec{k}}}  - \frac{\delta F}{\delta \pi_{1,\vec{k}}}\frac{\delta G}{\delta \phi^*_{1,\vec{k}}} + 
		\frac{\delta F}{\delta \phi_{2,\vec{k}}}\frac{\delta G}{\delta \pi^*_{2,\vec{k}}}  - \frac{\delta F}{\delta \pi_{2,\vec{k}}}\frac{\delta G}{\delta \phi^*_{2,\vec{k}}}  \right).
	\end{eqnarray}
Finally, applying the equation of motion~\eqref{eq:eom:generic} to the Fourier phase-space coordinates yields	
	\begin{eqnarray}\label{eq:zdyn}
		\dot{\boldsymbol{z}}_{\vec{k}} = \left(\bs{\Omega} \boldsymbol{H}_k\right)\boldsymbol{z}_{\vec{k}}\, ,
	\end{eqnarray}
where the explicit time dependence of $\boldsymbol{H}_k$ has been dropped for notational convenience.
This equation shows that, for free fields in a homogeneous and isotropic background, Fourier modes decouple and evolve independently. This implies that each Fourier sector can be studied separately, which greatly simplifies the analysis. From now on, we therefore focus on a single Fourier sector, \ie $\vec{k}$ in $\mathbb{R}^{3+}$ is fixed hereafter. 
\subsubsection*{Quantisation}
So far, linear Hamiltonian systems have been described at the classical level. We now follow the canonical quantisation prescriptions and promote the phase-space coordinates to quantum operators acting on a Hilbert space, $\widehat{\boldsymbol{z}}_{\vec{k}} := (\widehat{\phi}_{1,\vec{k}}, \widehat{\phi}_{2,\vec{k}}, \widehat{\pi}_{1,\vec{k}}, \widehat{\pi}_{2,\vec{k}})^{\mathrm{T}}$ (where, from now on, hats denote quantum operators). These operators satisfy the canonical commutation relations 
	\begin{eqnarray}
	\label{eq:Commutators:zk}
		\left[\widehat{\boldsymbol{z}}_{\vec{k}},\widehat{\boldsymbol{z}}_{\vec{q}}^\dag\right] = i \bs{\Omega} \delta^3(\vec{k} - \vec{q})\, ,
	\end{eqnarray}
which is the quantum analogue of \Eq{eq:poissonk} and where, hereafter, we work with $\hbar=1$. The Hamiltonian operator reads $\widehat{H} = \int_{\mathbb{R}^{3+}} \dd^3\vec{k} \widehat{\boldsymbol{z}}_{\vec{k}}^\dag \boldsymbol{H}_k\widehat{\boldsymbol{z}}_{\vec{k}}$, and the dynamics of any function of the quantum phase-space variables is given by the Heisenberg equation $\Dot{F}(\widehat{\boldsymbol{z}}_{\vec{k}}) = - i  [F(\widehat{\boldsymbol{z}}_{\vec{k}}), \widehat{H}]$. In particular, for the field variables themselves, this gives rise to $
\dot{\bs{\widehat{z}}}_{\vec{k}} = (\bs{\Omega} \bs{H}_k)\bs{\widehat{z}}_{\vec{k}}$, which directly transposes \Eq{eq:zdyn}. 
	
Creation and annihilation operators are defined in the usual way, \ie
\bea
\label{eq:a1:a2}
 \widehat{a}_{j,\vec{k}} &=& \frac{1}{\sqrt{2}} \left(\sqrt{k} \widehat{\phi}_{j,\vec{k}} + \frac{i}{\sqrt{k}} \widehat{\pi}_{j,\vec{k}}\right)
 \quad\text{for}\quad j=1,2\, ,
\eea
where the prefactors $\sqrt{k}$ and $1/\sqrt{k}$ are introduced for dimensional reasons. This can be written in matricial form as
\begin{eqnarray}
\label{eq:helpassquant}
		\boldsymbol{\widehat{a}}_{\vec{k}} = \begin{pmatrix} \widehat{a}_{1,\vec{k}}\\ \widehat{a}_{2, \vec{k}}\\\widehat{a}^\dag_{1, -\vec{k}} \\ \widehat{a}^\dag_{2, -\vec{k}}\end{pmatrix} = \bs{U} \boldsymbol{D}_k\boldsymbol{\widehat{z}}_{\vec{k}}\, ,
\end{eqnarray}
where $``\dag"$ stands for the conjugate transpose, and the matrices $\bs{U}$ and $\boldsymbol{D}_k$ are defined as
\bea
	\label{eq:helicity}
		\bs{U} = \frac{1}{\sqrt{2}}
		\begin{pmatrix}
		\bs{I}_2 &i \bs{I}_2\\
		\bs{I}_2 & -i \bs{I}_2
		\end{pmatrix}
		\quad\text{and}\quad
		\boldsymbol{D}_k = 
		\begin{pmatrix}
			\sqrt{k}  \bs{I}_2& 0 \\
			0 & & \bs{I}_2/\sqrt{k}
		\end{pmatrix}
\, .
\eea
One can check that $\bs{U}$ is a unitary matrix, \ie $\bs{U}\bs{U}^\dag = \bs{U}^\dag \bs{U} = \bs{I}_4$. In principle, $\boldsymbol{D}_k$ may be replaced with $\boldsymbol{M}_k\boldsymbol{D}_k$, where $\boldsymbol{M}_k$ is any (dimensionless) symplectic matrix (formally defined below in \Sec{sec:Symplectic:Structure}). It only leads to a different definition of the vacuum state (\ie the state that is annihilated by the annihilation operators)~\cite{Grain:2019vnq}. 
Let us note that the ordering in $\widehat{\bs{a}}_{\vec{k}}$ is different than in $\widehat{\bs{z}}_{\vec{k}}$, since in $\widehat{\bs{a}}_{\vec{k}}$ the first two entries concern the $\vec{k}$ sector and the last two entries the $-\vec{k}$ sector.\footnote{This is because the classical version of \Eq{eq:helpassquant} gives $\boldsymbol{z}_{-\vec{k}} = \boldsymbol{D}^{-1}_k \bs{U}^\dag \boldsymbol{a}_{-\vec{k}}$ and $\boldsymbol{z}^*_{\vec{k}} = \boldsymbol{D}^{-1}_k \bs{U}^{\mathrm{T}} \boldsymbol{a}^*_{\vec{k}}$. The reality condition $\boldsymbol{z}_{\vec{k}}^* = \boldsymbol{z}_{-\vec{k}}$ thus entails $\boldsymbol{a}_{-\vec{k}} = \bs{U}\bs{U}^{\mathrm{T}} \boldsymbol{a}^*_{\vec{k}}$, where $\bs{U}\bs{U}^{\mathrm{T}}=\begin{pmatrix} 0 & \bs{I}_2 \\ \bs{I}_2 & 0 \end{pmatrix}$, explaining the structure of the $\widehat{\bs{a}}_{\vec{k}}$ vector.}

The dynamics of the creation and annihilation operators is generated by the quadratic Hamiltonian $	\widehat{\mathcal{H}} = \int_{\mathbb{R}^{3+}} \mathrm{d}^3 \vec{k} \boldsymbol{\widehat{a}}_{\vec{k}}^\dag \bs{\mathcal{H}}_k\boldsymbol{\widehat{a}}_{\vec{k}}$, where $\bs{\mathcal{H}}_k$ reads \cite{Grain:2019vnq}
	\begin{eqnarray} \label{eq:helH}
		\bs{\mathcal{H}}_k = \bs{U} \left[(\boldsymbol{D}^{-1}_k)^{\mathrm{T}}\boldsymbol{H}_k\boldsymbol{D}^{-1}_k + (\boldsymbol{D}^{-1}_k)^{\mathrm{T}}\bs{\Omega}\dot{\boldsymbol{D}}^{-1}_k\right]\bs{U}^\dag\, .
	\end{eqnarray}
	The canonical commutation relations are given by
	\begin{eqnarray}\label{eq:acommut}
		\left[\widehat{\boldsymbol{a}}_{\vec{k}},\widehat{\boldsymbol{a}}_{\vec{q}}^\dag\right] = i \bs{\mathcal{J}} \delta^3(\vec{k} - \vec{q})
		\quad\text{where}\quad
		\bs{\mathcal{J}} = \bs{U} \bs{\Omega} \bs{U}^\dag= -i\begin{pmatrix}
			\bs{I}_2 & 0  \\
			0 & - \bs{I}_2
		\end{pmatrix}\, ,
	\end{eqnarray}
	and any function of the creation and annihilation operators evolves according to $\Dot{F}(\boldsymbol{\widehat{a}}_{\vec{k}}) = \left[F(\boldsymbol{\widehat{a}}_{\vec{k}}), \widehat{\mathcal{H}}\right]$. In particular, for the creation and annihilation operators themselves, one obtains
	\begin{eqnarray}\label{eq:adynJHk}
		\dot{\boldsymbol{\widehat{a}}}_{\vec{k}} = \left(\bs{\mathcal{J}} \bs{\mathcal{H}}_k\right)\boldsymbol{\widehat{a}}_{\vec{k}}.
	\end{eqnarray}
\subsection{Symplectic structure of the phase space}
\label{sec:Symplectic:Structure}
In \Sec{sec:TwoFreeFields}, we saw that the Fourier transform preserves the Poisson brackets of the phase-space variables. Another example of a transformation that  preserves the Poisson brackets is provided by the Hamiltonian evolution itself. Indeed, the equation of motion~\eqref{eq:zdyn} can be solved as
	\begin{eqnarray}\label{eq:greenformalism}
		\boldsymbol{z}_{\vec{k}}(t) = \boldsymbol{G}_k(t,t_{\uin}) \boldsymbol{z}_{\vec{k}}(t_{\uin}),
	\end{eqnarray}
where $\boldsymbol{G}_k$ is a $(4 \times 4)$-real matrix called the Green's matrix and that satisfies $\dot{\boldsymbol{G}_k} = \boldsymbol{\Omega}\boldsymbol{H}_k\boldsymbol{G}_k + \boldsymbol{I}_{4} \delta(t-t_{\uin})$, with initial condition $\boldsymbol{G}_k(t_{\uin},t_{\uin}) = \boldsymbol{I}_{4}$. Note that, as $\boldsymbol{H}_k$,  $\boldsymbol{G}_k$ only depends on the wavenumber $k$. One can also check that  $\boldsymbol{G}_k$ satisfies
	\begin{eqnarray}
	\label{eq:sympdef}
		\boldsymbol{G}_k^{\mathrm{T}} \boldsymbol{\Omega} \boldsymbol{G}_k = \boldsymbol{\Omega}.
	\end{eqnarray}
This is indeed obviously the case at initial time, and by plugging the equation of motion for  $\boldsymbol{G}_k$ in the time derivative of the left-hand-side of \Eq{eq:sympdef}, one obtains a vanishing result after using that $\boldsymbol{H}_k$ is symmetric and that $\boldsymbol{\Omega}^{\mathrm{T}} \boldsymbol{\Omega}=  - \bs{\Omega}^2 = \bs{I}_{4}$. 

In general, real matrices satisfying \Eq{eq:sympdef} are called symplectic, and they form the symplectic group $\mathrm{Sp}(4,\mathbb{R})$. They describe all possible reparametrisations of phase space through linear canonical transformations, \ie transformations that preserve the Poisson brackets. Indeed, consider  two phase-space coordinates $\boldsymbol{z}_{\vec{k}}$ and $\widetilde{\boldsymbol{z}}_{\vec{k}}$, related through a linear transformation
	\begin{eqnarray}\label{eq:cantransfo}
		\widetilde{\boldsymbol{z}}_{\vec{k}} = \boldsymbol{M}_k\boldsymbol{z}_{\vec{k}}.
	\end{eqnarray}
One can check that the Poisson brackets are preserved, \ie $ \{\boldsymbol{\widetilde{z}}_{\vec{k}},\boldsymbol{\widetilde{z}}_{\vec{k}}^\dag\}= \{\boldsymbol{z}_{\vec{k}},\boldsymbol{z}_{\vec{q}}^\dag\} = \bs{\Omega}$, if and only if  $\boldsymbol{M}_k \in \mathrm{Sp}(4,\mathbb{R})$ [\ie $\boldsymbol{M}_k$ satisfies \Eq{eq:sympdef}]. This ensures that the Poisson bracket between two arbitrary phase-space functions is the same when calculated with the $\boldsymbol{z}_{\vec{k}}$-variables or with the $\boldsymbol{\widetilde{z}}_{\vec{k}}$-variables. One can check that the equations of motion for the new set of canonical variables $\widetilde{\boldsymbol{z}}_{\vec{k}}$ are then given by Hamilton equations with the new Hamiltonian kernel
	\begin{align}
	\label{eq:Htransfo}
\widetilde{\boldsymbol{H}}_k =  (\boldsymbol{M}^{-1}_k)^{\mathrm{T}}\boldsymbol{H}_k\boldsymbol{M}^{-1}_k + (\boldsymbol{M}^{-1}_k)^{\mathrm{T}}\bs{\Omega}\dot{\boldsymbol{M}}^{-1}_k\, ,
	\end{align}
	and the dynamics is solved by $\widetilde{\boldsymbol{z}}_{\vec{k}}(t) = \widetilde{\boldsymbol{G}}_k(t,t_{\uin}) \widetilde{\boldsymbol{z}}_{\vec{k}}(t_{\uin})$ where $\widetilde{\boldsymbol{G}}_k(t,t_{\uin}) = \boldsymbol{M}_k(t) \boldsymbol{G}_k(t,t_{\uin}) \boldsymbol{M}^{-1}_k(t_{\uin})$. Symplectic transformations thus constitute a fundamental symmetry of the Hamiltonian phase space, and this is why the symplectic group for linear scalar-field systems is the main topic of the present work.
	
Let us note that in this framework, the dynamical evolution is nothing but a particular symplectic transformation since, as stressed above, the Green's matrix is symplectic. One can also check that for the transformation that goes from  $\boldsymbol{z}_{\vec{k}}(t)$ to $\boldsymbol{z}_{\vec{k}}(t_{\uin})$, generated by the matrix $\bs{G}_k^{-1}(t,t_{\uin})$, the new Hamiltonian vanishes, as can be shown by plugging the equation of motion for $\bs{G}_k$ into \Eq{eq:Htransfo}. This is consistent with the fact that the $\boldsymbol{z}_{\vec{k}}(t_{\uin})$ variables are indeed time independent, and in that case the dynamics is entirely contained in the canonical transformation that relates them with the primary variables $\boldsymbol{z}_{\vec{k}}(t)$. 

Finally, when one works with the creation and annihilation operators introduced in \Sec{sec:TwoFreeFields}, a similar description applies. The equation of motion given in \Eq{eq:adynJHk} can be solved in terms of the Green's matrix,
\begin{eqnarray}
\label{eq:adyn}
		\boldsymbol{\widehat{a}}_{\vec{k}}(t) &=& \bs{\mathcal{G}}_k(t,t_{\uin}) \boldsymbol{\widehat{a}}_{\vec{k}}(t_{\uin})\, ,
	\end{eqnarray}	
where $\bs{\mathcal{G}}_k(t,t_{\uin}) = \boldsymbol{U}\boldsymbol{{D}}_k\bs{G}_k(t,t_{\uin}) \boldsymbol{D}^{-1}_k\boldsymbol{U}^\dag$ such that $\bs{\mathcal{G}}_k^\dag\boldsymbol{\mathcal{J}}\bs{\mathcal{G}}_k=\boldsymbol{\mathcal{J}}$ and $\det(\bs{\mathcal{G}}_k)=1$. Similarly, one can check that any generic canonical transformation $\bs{\widehat{z}}_{\vec{k}} \to \boldsymbol{M}_k\bs{\widehat{z}}_{\vec{k}} $ gives rise to $\bs{\widehat{a}}_{\vec{k}}\to  \bs{\mathcal{M}}_{k} \bs{\widehat{a}}_{\vec{k}}$, where $\bs{\mathcal{M}}_{k}=\boldsymbol{U}\boldsymbol{{D}}_k\boldsymbol{M}_k\boldsymbol{D}^{-1}_k\boldsymbol{U}^\dag$ satisfies $\boldsymbol{\mathcal{M}}_k^\dag\boldsymbol{\mathcal{J}}\boldsymbol{\mathcal{M}}_k=\boldsymbol{\mathcal{J}}$ and $\det(\boldsymbol{\mathcal{M}}_k)=1$.
\section{$\mathrm{Sp}(4,\mathbb{R})$ toolkit}
\label{sec:Sp4R:toolkit}
In the previous section, we have seen how symplectic transformations naturally arise in the phase-space description of linear Hamiltonian systems, both as a fundamental reparametrisation symmetry and as a way to generate the dynamics. This is why in this section, we further study the mathematical structure of the symplectic group in four dimensions, which is relevant to discuss the physics of two scalar fields. Readers already familiar with the use of symplectic groups in quantum mechanics can easily skip this section, which mostly consists in a review of the mathematical tools employed in the rest of the paper. It may otherwise serve as a pedagogical introduction to the techniques employed in the subsequent calculations, and set out our main notations. 
\subsection{Generators and Lie algebra}
\label{subsec:liealg}
As explained around \Eq{eq:sympdef}, the group of symplectic $(4 \times 4)$-matrices, denoted $\mathrm{Sp}(4,\mathbb{R})$, is defined as
\begin{eqnarray}\label{eq:sympdef44}
\mathrm{Sp}(4,\mathbb{R}) = \{\bs{M} \in \mathcal{M}_{4} (\mathbb{R}) : \boldsymbol{M}^{\mathrm{T}} \boldsymbol{\Omega} \boldsymbol{M} = \boldsymbol{\Omega}\}\, ,
\end{eqnarray}
where $\mathcal{M}_{n}(\mathbb{R})$ is the set of $(n\times n)$-real matrices. Since $\boldsymbol{\Omega}^{\mathrm{T}} \boldsymbol{\Omega}=  - \bs{\Omega}^2 = \bs{I}_{4}$, one can show that, if $\bs{M}\in \mathrm{Sp}(4,\mathbb{R}) $, then $\bs{M}^{\mathrm{T}}\in \mathrm{Sp}(4,\mathbb{R}) $ (this is because \Eq{eq:sympdef44} leads to $\boldsymbol{M}^{\mathrm{T}} = - \boldsymbol{\Omega} \bs{M}^{-1} \boldsymbol{\Omega}$, which implies that $\bs{M} \bs{\Omega}\boldsymbol{M}^{\mathrm{T}} = \bs{\Omega}$).  One can also readily check that $\mathrm{Sp}(4,\mathbb{R})$ is indeed a group, and it follows from \Eq{eq:sympdef44} that all symplectic matrices have unit determinant~\cite{goldstein2002classical}.

$\mathrm{Sp}(4,\mathbb{R})$ is a Lie group, that is a continuous group whose multiplication and inversion operations are differentiable, so one can investigate global properties of the group by looking at its local or linearised version, given in terms of its so-called Lie algebra. The Lie algebra is a vector space under the bracket operation $\left[\bs{X},\bs{Y}\right] = \bs{X}\bs{Y} -\bs{Y}\bs{X}$ that completely captures the local structure of the group. When analysing Lie groups, finding a basis of this vector space, \ie a set of so-called ``generators'', and deriving their commutators, is of paramount importance. The number of generators specifies both the dimension of the Lie group and its associated Lie algebra. In what follows, the Lie algebra of $\mathrm{Sp}(4,\mathbb{R})$ is denoted $\mathfrak{sp}(4,\mathbb{R})$.

The exponential map allows us to connect the Lie algebra to its corresponding Lie group. For any $\bs{X} \in \mathfrak{sp}(4,\mathbb{R})$, $\boldsymbol{M} = \exp(\bs{X})$ is an element of $\mathrm{Sp}(4,\mathbb{R})$ (see \Refa{Chacon-Acosta:2021urw} for a recent complete characterisation of the exponential map of $\mathrm{Sp}(4,\mathbb{R})$). Note that the exponential map is not surjective, meaning that there exist elements of $\mathrm{Sp}(4,\mathbb{R})$ that cannot be written as $\exp(\bs{X})$. By plugging $\boldsymbol{M} = \exp(\bs{X})$ into \Eq{eq:sympdef44}, and upon writing the obtained formula as an expansion in $\bs{X}$, one finds that the Lie algebra is defined according to 
\begin{eqnarray}\label{eq:defalg}
\mathfrak{sp}(4,\mathbb{R}) = \{\bs{X} \in \mathcal{M}_{4} (\mathbb{R}) : \bs{\Omega} \bs{X} + \bs{X}^{\mathrm{T}} \bs{\Omega} = 0\}.
\end{eqnarray}
This allows one to write a generic element of the Lie algebra as 
\begin{eqnarray}\label{eq:alg}
\bs{X} = 
\begin{pmatrix}
\bs{A} & \bs{B} \\
\bs{C} & -\bs{A}^{\mathrm{T}}
\end{pmatrix}
\end{eqnarray}
where $\bs{A}$, $\bs{B}$ and $\bs{C}$ are three $(2\times2)$ real matrices,  $\bs{B}$ and $\bs{C}$ being symmetric. This implies that $\mathfrak{sp}(4,\mathbb{R})$ is a ten-dimensional vector space, since $\bs{A}$ contains four degrees of freedom while $\bs{B}$ and $\bs{C}$ being symmetric, they contain three degrees of freedom each. Denoting by $\boldsymbol{K}_i$, $i=1\cdots 10$, a basis of $10$ generators, a generic element of the Lie algebra can be decomposed as 
\begin{eqnarray}\label{eq:liealg}
	\bs{X} = \sum_{i=1}^{10} \alpha_i \boldsymbol{K}_i,~~~~\alpha_i\in \mathbb{R}.
\end{eqnarray}
Our next step is to exhibit such a basis.

To that end, we first introduce the Kronecker product. For two $(2\times 2)$-matrices
\begin{eqnarray}
\bs{A} = \begin{pmatrix}
a_{11} & a_{12} \\
a_{21} & a_{22}
\end{pmatrix}
\quad\text{and}\quad
\bs{B} = \begin{pmatrix}
b_{11} & b_{12} \\
b_{21} & b_{22}
\end{pmatrix}\, ,
\end{eqnarray}
the Kronecker product $\bs{A} \otimes \bs{B}$ is a $(4\times4)$-matrix defined as
\begin{eqnarray}
\bs{A} \otimes \bs{B} = \begin{pmatrix}
a_{11} \bs{B} & a_{12} \bs{B} \\
a_{21} \bs{B} & a_{22} \bs{B}
\end{pmatrix}  = \left(
\begin{array}{c c|c c}
a_{11} b_{11} & a_{11} b_{12} & a_{12} b_{11} & a_{12} b_{12} \\ 
a_{11} b_{21} & a_{11} b_{22} & a_{12} b_{21} & a_{12} b_{22} \\
\hline
a_{21} b_{11} & a_{21} b_{12} & a_{22} b_{11} & a_{22} b_{12} \\
a_{21} b_{21} & a_{21} b_{22} & a_{22} b_{21} & a_{22} b_{22}
\end{array}
\right).
\end{eqnarray}
The reason why this construction is useful is because, within the two-field system at hand, each field is individually described by an $\mathfrak{sp}(2,\mathbb{R})$ algebra. It is therefore natural for expect that $\mathfrak{sp}(4,\mathbb{R})$ contains products of elements of $\mathfrak{sp}(2,\mathbb{R})$ with themselves.  The generators of $\mathfrak{sp}(2,\mathbb{R})$ can be simply written in terms of the Pauli matrices~\cite{Grain:2019vnq},
\begin{eqnarray}
\mathfrak{sp}(2,\mathbb{R}) = \mathrm{Span}\{\bs{\sigma}_x, i \bs{\sigma}_y, \bs{\sigma}_z\}\, ,
\end{eqnarray}
with 
\begin{eqnarray}
\label{eq:Pauli}
\bs{\sigma}_x = 
\begin{pmatrix}
0 & 1 \\
1 & 0
\end{pmatrix}\, ,~~~~~~\bs{\sigma}_y = 
\begin{pmatrix}
0 & -i \\
i & 0
\end{pmatrix}\, ,~~~~~~\bs{\sigma}_z = 
\begin{pmatrix}
1 & 0 \\
0 & -1
\end{pmatrix}\, ,
\end{eqnarray}	
so  $\mathfrak{sp}(2,\mathbb{R})$ is of dimension $3$. This implies that over the $10$ generators of $\mathfrak{sp}(4,\mathbb{R})$, $6$ provide two copies of $\mathfrak{sp}(2,\mathbb{R})$ and describe the two sectors separately, and $4$ are related to the coupling between the two sectors. For $\bs{\sigma}_a,\bs{\sigma}_b \in \{\boldsymbol{I}_2, \bs{\sigma}_x, i\bs{\sigma}_y, \bs{\sigma}_z\}$, there are $16$ combinations $\bs{\sigma}_a\otimes \bs{\sigma}_b$ for $a,b\in\{0,\cdots,3\}$, namely $\boldsymbol{I}_4$ and the $15$ Dirac matrices \cite{Kim:2012av}. Among them, only $10$ are of the form~\eqref{eq:alg},\footnote{Formally, the $15$ Dirac matrices form the $\mathfrak{o}(3,3)$ algebra. $\mathfrak{sp}(4,\mathbb{R})$ is isomorphic to $\mathfrak{o}(3,2)$, which is a subalgebra of $\mathfrak{o}(3,3)$~\cite{Kim:2012av}.} which are listed in Table \ref{tab:gen}, where they are organised in three subsets. The so-called squeezing generators are diagonal. Through the exponential map, they give rise to group elements of the form $\exp(d_1 \bs{K}_1) = \mathrm{diag}(e^{d_1}, e^{-d_1}, e^{-d_1}, e^{d_1})$ and $\exp(d_2 \bs{K}_2) = \mathrm{diag}(e^{d_2}, e^{d_2}, e^{-d_2}, e^{-d_2})$ with $d_1$ and $d_2$ two real parameters. Therefore, they elongate one phase-space direction while contracting the other, hence their name. The two other kinds of generators are called rotations and boosts. When squared, rotations give $-\boldsymbol{I}_4$ and boosts give $\boldsymbol{I}_4$. Therefore, once exponentiated, rotations generate group elements of the form $\exp(\theta_i \bs{K}_i) = \cos{\theta_i}\boldsymbol{I}_4 + \sin{\theta_i} \bs{K}_i$ for $i \in \{3,4,5,6\}$ with $\theta_i \in \mathbb{R}$, while boosts generate $\exp(\alpha_i \bs{K}_i) = \cosh{\alpha_i}\boldsymbol{I}_4 + \sinh{\alpha_i} \bs{K}_i$ for $i \in \{7,8,9,10\}$ with $\alpha_i \in \mathbb{R}$, hence their name. 
\begin{table}[h!]
	\centering
	\resizebox{\columnwidth}{!}{%
		\renewcommand{\arraystretch}{1.5}
		\begin{tabular}{| L | L | L |}
			\hline
			\text{Squeezing} & \text{Rotation} & \text{Boost} \\
			\hline
			\bs{K}_1 = \bs{\sigma}_z \otimes \bs{\sigma}_z = \begin{pmatrix}
				1 & 0 & 0 & 0\\
				0 & -1 & 0 & 0\\
				0 & 0 & -1 & 0\\
				0 & 0 & 0 & 1
			\end{pmatrix} & \bs{K}_3= i\bs{\sigma}_y \otimes \boldsymbol{I}_2 = \begin{pmatrix}
				0 & 0 & 1 & 0\\
				0 & 0 & 0 & 1\\
				-1 & 0 & 0 & 0\\
				0 & -1 & 0 & 0
			\end{pmatrix} & \bs{K}_7 = \bs{\sigma}_x \otimes \boldsymbol{I}_2 = \begin{pmatrix}
				0 & 0 & 1 & 0\\
				0 & 0 & 0 & 1\\
				1 & 0 & 0 & 0\\
				0 & 1 & 0 & 0
			\end{pmatrix} \\
			\bs{K}_2 = \bs{\sigma}_z \otimes \boldsymbol{I}_2 = \begin{pmatrix}
				1 & 0 & 0 & 0\\
				0 & 1 & 0 & 0\\
				0 & 0 & -1 & 0\\
				0 & 0 & 0 & -1
			\end{pmatrix} & \bs{K}_4= i\bs{\sigma}_y \otimes \bs{\sigma}_z = \begin{pmatrix}
				0 & 0 & 1 & 0\\
				0 & 0 & 0 & -1\\
				-1 & 0 & 0 & 0\\
				0 & 1 & 0 & 0
			\end{pmatrix} & \bs{K}_8 = \bs{\sigma}_x \otimes \bs{\sigma}_z = \begin{pmatrix}
				0 & 0 & 1 & 0\\
				0 & 0 & 0 & -1\\
				1 & 0 & 0 & 0\\
				0 & -1 & 0 & 0
			\end{pmatrix} \\
			& \bs{K}_5 = \boldsymbol{I}_2 \otimes i\bs{\sigma}_y = \begin{pmatrix}
				0 & 1 & 0 & 0\\
				-1 & 0 & 0 & 0\\
				0 & 0 & 0 & 1\\
				0 & 0 & -1 & 0
			\end{pmatrix} & \bs{K}_9 = \bs{\sigma}_z \otimes \bs{\sigma}_x = \begin{pmatrix}
				0 & 1 & 0 & 0\\
				1 & 0 & 0 & 0\\
				0 & 0 & 0 & -1\\
				0 & 0 & -1 & 0
			\end{pmatrix}\\
			& \bs{K}_6 = i\bs{\sigma}_y \otimes \bs{\sigma}_x = \begin{pmatrix}
				0 & 0 & 0 & 1\\
				0 & 0 & 1 & 0\\
				0 & -1 & 0 & 0\\
				-1 & 0 & 0 & 0
			\end{pmatrix} & \bs{K}_{10} = \bs{\sigma}_x \otimes \bs{\sigma}_x = \begin{pmatrix}
				0 & 0 & 0 & 1\\
				0 & 0 & 1 & 0\\
				0 & 1 & 0 & 0\\
				1 & 0 & 0 & 0
			\end{pmatrix}\\
			\hline
	\end{tabular}}
	\caption{Generators of $\mathfrak{sp}(4,\mathbb{R})$ in the fundamental representation.}
	\label{tab:gen}
\end{table}

Now that we have explicitly obtained the generators of the Lie algebra, let us identify their respective role. The two $\mathfrak{sp}(2,\mathbb{R})$ algebras are given by $\{(\bs{K}_1 +\bs{K}_2)/2, (\bs{K}_3 + \bs{K}_4)/2, (\bs{K}_7 + \bs{K}_8)/2 \}$ and $\{(\bs{K}_1 -\bs{K}_2)/2, (\bs{K}_3 - \bs{K}_4)/2, (\bs{K}_7 - \bs{K}_8)/2 \}$, each of them being composed of a squeezing, a rotation and a boost. They act on each sector separately and would be enough to describe two non-interacting degrees of freedom. Formally, they generate the $\mathrm{Sp}(2,\mathbb{R}) \times\mathrm{Sp}(2,\mathbb{R})$ subgroup of $\mathrm{Sp}(4,\mathbb{R})$. The 4 remaining generators are associated with the coupling between the two sectors. They correspond to the rotations $\bs{K}_5$ and $\bs{K}_6$ and the boosts $\bs{K}_9$ and $\bs{K}_{10}$, which entangle the two sectors, as will be made clear in \Sec{subsec:quantumliealg}. One can indeed check that these 4 generators (and only them) have non-vanishing off-diagonal elements within the $(2\times 2)$ blocks, which, from the ordering of the phase-space variables in \Eq{eq:Fourier}, implies that they mix the two fields. This is why, hereafter, they will be referred to as the coupling generators.
 
To complete our description of the Lie algebra, let us finally provide the commutators between its generators. They can be obtained from the multiplication rule of the Kronecker product for square matrices,\footnote{Other properties of the Kronecker product that will be used in the following are that it is bilinear and associative, and that for square matrices, $\left(\bs{A}\otimes \bs{B}\right)^{-1} =  \bs{A}^{-1}\otimes \bs{B}^{-1}$, $\left(\bs{A}\otimes \bs{B}\right)^{*} =  \bs{A}^{*}\otimes \bs{B}^{*}$ and $\left(\bs{A}\otimes \bs{B}\right)^{T} =  \bs{A}^{T}\otimes \bs{B}^{T}$.\label{footnote:Kronecker}} 
\bea
\label{eq:Kronecker:Multiplication:Rule}
\left(\bs{A}\otimes \bs{B}\right) \left(\bs{C}\otimes \bs{D}\right) =  \left(\bs{A}\bs{C}\right)\otimes \left(\bs{B}\bs{D}\right)\, ,
\eea
together with the formula
\begin{eqnarray}
\label{eq:Commutators:Pauli}
\bs{\sigma}_i \bs{\sigma}_j = \delta_{ij}\bs{I}_2 + i \varepsilon_{ijk} \bs{\sigma}_k\quad\text{for}\quad\{i,j,k\} \in \{x,y,z\} \, ,
\end{eqnarray}
where $\delta_{ab}$ is the Kronecker symbol and $\varepsilon_{abc}$ is the Levi-Civita symbol.
For instance, one has $\left[\bs{K}_6, \bs{K}_4\right] = \left[ i \bs{\sigma}_y \otimes \bs{\sigma}_x, i \bs{\sigma}_y \otimes \bs{\sigma}_z\right]
= (i \bs{\sigma}_y)^2 \otimes \bs{\sigma}_x \bs{\sigma}_z - (i \bs{\sigma}_y)^2 \otimes \bs{\sigma}_z \bs{\sigma}_x
= - \bs{I}_2 \otimes (- i \bs{\sigma}_y) + \bs{I}_2 \otimes ( i \bs{\sigma}_y)
= 2 \bs{K}_5$. All other commutators are presented in \App{sec:commutrel}, where we also derive the various subalgebras. We do not reproduce these formulas here for display convenience, and will simply refer to \App{sec:commutrel} when needed.
\subsection{Bloch-Messiah decomposition}
\label{subsec:decomp}
The explicit derivation of the 10 generators of the Lie algebra allows one to decompose any element of the group (within the exponential map) onto these generators, upon exponentiating \Eq{eq:liealg}. This parametrisation of the exponential map is performed in \Refa{Chacon-Acosta:2021urw} (see also \Refs{Garcia-Chung:2020gxy,Garcia-Chung:2020cag}). When all 10 parameters are non vanishing, it however leads to expressions that may be cumbersome to manipulate, and which, as mentioned above, do not reach all the elements of the group. This is why, in this section, we turn our attention to an alternative decomposition, the so-called Bloch-Messiah decomposition (also sometimes called Euler decomposition)~\cite{Bloch:1962zj}, which allows one to write any symplectic matrix as
\begin{eqnarray}\label{eq:bm}
\bs{M} (\boldsymbol{\theta}, \boldsymbol{d}, \boldsymbol{\varphi}) = \bs{R}(\boldsymbol{\theta})\bs{Z}(\boldsymbol{d})\bs{R}(\boldsymbol{\varphi})\, .
\end{eqnarray}
Here, $\bs{R}(\boldsymbol{\theta}), \bs{R}(\boldsymbol{\varphi}) \in \mathrm{Sp}(4,\mathbb{R}) \cap \mathrm{SO}(4)$ are constructed from the four rotation generators and $\bs{Z}(\boldsymbol{d})$ from the two squeezing generators, \ie
\begin{eqnarray}
\label{eq:rot1} \bs{R}(\boldsymbol{\theta}) = \exp(\theta_3 \bs{K}_3 + \theta_4 \bs{K}_4 + \theta_5 \bs{K}_5 + \theta_6 \bs{K}_6)
\end{eqnarray} 
and a similar expression for $\bs{R}(\boldsymbol{\varphi})$ with $\bs{\varphi}=(\varphi_1,\varphi_2,\varphi_3,\varphi_4)$, and
\begin{eqnarray}
\bs{Z}(\boldsymbol{d}) = \exp(d_1 \bs{K}_1 + d_2 \bs{K}_2).
\end{eqnarray} 
The 8 parameters contained in $\bs{\theta}$ and $\bs{\varphi}$ are called the rotation parameters, while $\bs{d}$ contains the so-called squeezing parameters. Note that the parameters associated with the coupling generators are $\theta_5$, $\varphi_5$, $\theta_6$ and $\varphi_6$, which thus control the mixing between the two sectors.
 
One may note that only 6 out of the 10 generators of $\mathfrak{sp}(4,\mathbb{R})$ are involved in the Bloch-Messiah decomposition. It therefore provides a factorised expression of the group elements that is slightly more convenient to manipulate. Moreover, the three blocks of the decomposition can be further factorised down. Indeed, in  \App{sec:commutrel}, it is shown that $\bs{K}_1$ and $\bs{K}_2$ commute, so 
\begin{eqnarray}
\label{eq:fact1}
\bs{Z}(\boldsymbol{d}) = \exp(d_1 \bs{K}_1)\cdot\exp(d_2 \bs{K}_2).
\end{eqnarray} 
Regarding the rotation operators, still in \App{sec:commutrel}, it is shown that $\bs{K}_3$ commutes with the other three rotation generators, namely $\bs{K}_4, \bs{K}_5$ and $\bs{K}_6$ [see \Eq{eq:u1rot}], so $\bs{K}_3$ generates a separate $\mathrm{U}(1)$ Lie group; while $\bs{K}_4$, $\bs{K}_5$ and $\bs{K}_6$ generate a $\mathrm{SU}(2)$ Lie group [see \Eq{eq:su2}]. As a consequence, the four rotation generators form\footnote{The role played by the 4 rotation generators can be further understood as follows. As explained in \Sec{subsec:liealg}, $(\bs{K}_3 + \bs{K}_4)/2$ and  $(\bs{K}_3 - \bs{K}_4)/2$ generate separate rotations in the first and second sectors respectively. Thus $\bs{K}_3$ induces the same rotation in both sectors and $\theta_3$ can be thought of as a ``coherent phase'' (which is why $\bs{K}_3$ decouples from the other generators), while $\bs{K}_4$ generates opposite rotations in the two sectors and $\theta_4$ can be thought of as a ``phase shift''.  For the two coupling generators, $\bs{K}_5$ operates the same rotation in the position plane and in the momentum plane, so it can be understood as a field redefinition; while $\bs{K}_6$ operates a rotation that mixes positions and momenta of the two sectors.}
\begin{eqnarray}\label{eq:decomprot}
\mathrm{Sp}(4,\mathbb{R}) \cap \mathrm{SO}(4) \cong \mathrm{U}(2) \cong \mathrm{SU}(2) \times \mathrm{U}(1)\, ,
\end{eqnarray}
where $``\cong"$ indicates group isomorphisms. This leads to factorising
\begin{eqnarray}\label{eq:fact2}
\bs{R}(\bs{\theta}) = \exp(\theta_3 \bs{K}_3)\cdot\exp(\theta_4 \bs{K}_4 + \theta_5 \bs{K}_5 + \theta_6 \bs{K}_6)\, ,
\end{eqnarray}
and a similar expression for $\bs{R}(\bs{\varphi})$.
Finally, one can use the Baker-Campbell-Haussdorf formula 
\cite{puri2001mathematical, perelomov1986generalized, barnett2002methods, Truax:1985vk} to further factorise the remaining $\mathrm{SU}(2)$ part. This can be done by first introducing the complexified Lie algebra
\begin{eqnarray}
\bs{S}_z = \frac{1}{2i} \bs{K}_4, ~~~~~~ \bs{S}_+ = \frac{1}{2i} \left(\bs{K}_5 - i \bs{K}_6\right), ~~~~~~ \bs{S}_- = \frac{1}{2i} \left(\bs{K}_5 + i \bs{K}_6\right)\, ,
\end{eqnarray}  
where the notation is purposely reminiscent of spin physics. One can check that $\bs{S}_z^\dag = \bs{S}_z$ and $\bs{S}_+^\dag = \bs{S}_-$, and from the $\mathrm{SU}(2)$ commutation relations derived in \Eq{eq:su2}, one has
\begin{eqnarray}
\left[\bs{S}_z,\bs{S}_+\right] = \bs{S}_+,~~~~\left[\bs{S}_z,\bs{S}_-\right] = -\bs{S}_-, ~~~~\left[\bs{S}_+,\bs{S}_-\right] = 2\bs{S}_z\, .
\end{eqnarray} 
Expanding $\theta_4 \bs{K}_4 + \theta_5 \bs{K}_5 + \theta_6 \bs{K}_6$ onto $\bs{S}_z$, $\bs{S}_+$ and $\bs{S}_-$, the Baker-Campbell-Haussdorf formula thus leads to~\cite{Truax:1985vk}
\bea
\exp(\theta_4 \bs{K}_4 + \theta_5 \bs{K}_5 + \theta_6 \bs{K}_6) =\exp(p_+ \bs{S}_+) \exp(p_z \bs{S}_z) \exp(p_- \bs{S}_-)\, ,
\eea
where
\begin{align}
\label{eq:pz:p-:p+} 
p_z = -2\ln \left(\cos \theta -i \frac{\theta_4}{\theta} \sin \theta\right)\, , \qquad
p_- = \frac{-\tau^* \sin \theta}{\theta \cos \theta - i \theta_4\sin \theta}\, , \qquad
p_+ = \frac{\tau \sin \theta}{\theta \cos \theta - i \theta_4\sin \theta}\, ,
\end{align}
and where we have introduced
\bea \label{eq:thetavarphi}
\tau = - \theta_6 + i \theta_5\qquad\text{and}\qquad\theta = \sqrt{\theta_4^2 +\theta_5^2+ \theta_6^2}\, .
\eea
A similar expression can be found for $\bs{R}(\boldsymbol{\varphi})$, where $q_z, q_-, q_+$ denote the parameters analogue to $p_z, p_-, p_+$. Combining the above results, any element of $\mathrm{Sp}(4,\mathbb{R}) $ can be decomposed according to
\begin{equation}
\begin{aligned}
\label{eq:fullyfact}
\bs{M} = &\overbrace{\left[\exp(p_+ \bs{S}_+) \cdot \exp(p_z \bs{S}_z) \cdot \exp(p_- \bs{S}_-) \cdot \exp(\theta_3 \bs{K}_3)\right] }^{\bs{R}(\boldsymbol{\theta})} 
\cdot \overbrace{\left[\exp(d_1 \bs{K}_1)\cdot\exp(d_2 \bs{K}_2)\right] }^{\bs{Z}(\boldsymbol{d})} \\
&~~~~\cdot\overbrace{\left[\exp(q_+ \bs{S}_+) \cdot \exp(q_z \bs{S}_z) \cdot \exp(q_- \bs{S}_-) \cdot \exp(\varphi_3 \bs{K}_3)\right] }^{\bs{R}(\boldsymbol{\varphi})}\, .
\end{aligned}
\end{equation}
This fully factorised form will be of particular convenience when it comes to characterising the quantum states of the system in \Sec{sec:QuantumState}.
\subsection{Helicity basis}
\label{subsec:su22}
In \Sec{sec:TwoFreeFields}, the creation and annihilation operators have been introduced as an equivalent parametrisation of phase space, called the helicity basis. In the quantum mechanical context, it leads to the convenient occupation-number representation, which is why we now translate the above considerations into that basis.

We recall that when applying a canonical transformation, the helicity variables transform via matrices of the form $\bs{\mathcal{M}}_k = \bs{U} \bs{M}_k \bs{U}^\dag$ for $\bs{M}_k \in \mathrm{Sp}(4,\mathbb{R})$, which satisfy $\boldsymbol{\mathcal{M}}_k^\dag\boldsymbol{\mathcal{J}}\boldsymbol{\mathcal{M}}_k=\boldsymbol{\mathcal{J}}$ and $\det(\boldsymbol{\mathcal{M}}_k)=1$, see the discussion below \Eq{eq:adyn}. In particular, this is the case for the Green matrix $\bs{\mathcal{G}}_k(t,t_{\uin})$. Those two conditions define the $\mathrm{SU}(2,2)$ group, but given that $\mathrm{SU}(2,2)$ is a fifteen dimensional Lie group, $\mathrm{Sp}(4,\mathbb{R})$ cannot be isomorphic to the whole group and instead constitutes a ten dimensional subgroup, which we denote $\Spnew(4,\mathbb{R})$. More precisely, decomposing a generic matrix $\bs{M}_k \in \mathcal{M}_{4}(\mathbb{R})$ into blocks according to
\begin{eqnarray}
\bs{M}_k = \begin{pmatrix}
\bs{A} & \bs{B} \\
\bs{C} & \bs{D}
\end{pmatrix} 
\end{eqnarray}
with $\bs{A}, \bs{B}, \bs{C}, \bs{D} \in  \mathcal{M}_{2}(\mathbb{R})$, the condition $\bs{\mathcal{M}}_k = \bs{U} \bs{M}_k(t) \bs{U}^\dag$ can be written as 
\begin{eqnarray}\label{eq:helbloc}
	\bs{\mathcal{M}}_k= \begin{pmatrix}
		\bs{\mathcal{A}}  &   \bs{\mathcal{B}}   \\
		\bs{\mathcal{B}}^*    &  \bs{\mathcal{A}}^*
	\end{pmatrix},
\end{eqnarray}
where $\bs{\mathcal{A}} = (1/2)[(\bs{A} + \bs{D}) + i(\bs{C} -\bs{B})]$ and $\bs{\mathcal{B}} = (1/2)[(\bs{A} - \bs{D}) + i(\bs{C} +\bs{B})]$. As explained below \Eq{eq:sympdef44}, if $\bs{M}_k\in \mathrm{Sp}(4,\mathbb{R}) $ then $\bs{M}_k^{\mathrm{T}}\in \mathrm{Sp}(4,\mathbb{R}) $, which implies that $\bs{\mathcal{M}}_k^\dagger$ is also symplectic, hence $\bs{\mathcal{M}}_k \bs{\mathcal{J}} \bs{\mathcal{M}}_k^\dagger = \bs{\mathcal{J}}$. This leads to the two conditions
\begin{eqnarray}
\label{eq:mat1}   \bs{\mathcal{A}} \bs{\mathcal{A}}^\dag - \bs{\mathcal{B}} \bs{\mathcal{B}}^\dag = \boldsymbol{I}_2\quad\text{and}\quad
 \bs{\mathcal{A}}\bs{\mathcal{B}}^{\mathrm{T}} - \bs{\mathcal{B}}\bs{\mathcal{A}}^{\mathrm{T}} = 0\, .
\end{eqnarray}
Upon expanding the matrices $ \bs{\mathcal{A}} $ and $ \bs{\mathcal{B}} $ in terms of the so-called Bogolyubov coefficients,
\begin{eqnarray}\label{eq:Gdef2}
\bs{\mathcal{A}} = \begin{pmatrix}
\alpha_{11}   &   \alpha_{12}   \\
\alpha_{21}  &  \alpha_{22}
\end{pmatrix}~~~~~~~~~\text{and}~~~~~~~~\bs{\mathcal{B}} = \begin{pmatrix}
\beta_{11}   &   \beta_{12}   \\
\beta_{21}  &  \beta_{22}
\end{pmatrix},
\end{eqnarray}
\Eq{eq:mat1} leads to the four conditions  
\begin{eqnarray}
\label{eq:ct1}  \left|\alpha_{11}\right|^2 +\left|\alpha_{12}\right|^2 - \left|\beta_{11}\right|^2 - \left|\beta_{12}\right|^2   &=& 1\, , \\
\label{eq:ct2}    \left|\alpha_{21}\right|^2 +\left|\alpha_{22}\right|^2 - \left|\beta_{21}\right|^2 - \left|\beta_{22}\right|^2  &=& 1\, , \\
\label{eq:ct3}    \alpha_{11}\alpha_{21}^* + \alpha_{12}\alpha_{22}^* - \beta_{11} \beta_{21}^* - \beta_{12} \beta_{22}^* &=& 0\, , \\
\label{eq:ct4}   \alpha_{11}\beta_{21} + \alpha_{12}\beta_{22} - \alpha_{21}\beta_{11} - \alpha_{22}\beta_{12} &=& 0\, .
\end{eqnarray}
This fixes 2 real and 2 complex combinations out of the 8 complex Bogolyubov coefficients, that is 6 out of the 16 real parameters, and one recovers the 10 degrees of freedom of $\mathrm{Sp}(4,\mathbb{R})$.

One may note that the Green matrix $\bs{\mathcal{G}}_k(t,t_{\uin})$ being an element of $\Spnew(4,\mathbb{R})$, it can also be written in terms of  Bogolyubov coefficients according to \Eqs{eq:helbloc} and~\eqref{eq:Gdef2}. In that case, the relations~\eqref{eq:ct1}-\eqref{eq:ct4} translate the fact that the Poisson brackets (or their quantum analogue, the commutation relations) are preserved by the dynamical evolution. Moreover, when applying $\boldsymbol{\widehat{a}}_{\vec{k}}(t) = \bs{\mathcal{G}}_k(t,t_{\uin}) \boldsymbol{\widehat{a}}_{\vec{k}}(t_{\uin})$, one notices that $\bs{\mathcal{A}}$ and $\bs{\mathcal{A}}^*$ transform annihilation into annihilation operators, and creation into creation operators, respectively. Therefore, they maintain the overall excitation number while reshuffling the excitations between the different sectors. On the contrary, $\bs{\mathcal{B}}$ and $\bs{\mathcal{B}}^*$ convert annihilation into creation operators and conversely, so they create or annihilate new excitations. We finally note that $\alpha_{11}, \alpha_{22}, \beta_{11}$ and $\beta_{22}$ act on each sector separately while $\alpha_{12}, \alpha_{21}, \beta_{12}$ and $\beta_{21}$ mix the two sectors. As a consistency check, one can verify that when these mixing Bogolyubov coefficients vanish, \Eqs{eq:ct1}-\eqref{eq:ct4} reduce to the $\mathrm{Sp}(2,\mathbb{R})$-constraint on the Bogolyubov coefficients~\cite{Grain:2019vnq}, namely $\vert\alpha_{11}\vert^2-\vert\beta_{11}\vert^2=\vert\alpha_{22}\vert^2-\vert\beta_{22}\vert^2=1$.

Let us now study the infinitesimal properties of $\Spnew(4,\mathbb{R})$, as we did for $\mathrm{Sp}(4,\mathbb{R})$ in \Sec{subsec:liealg}. We observe that the ten generators of $\Spnew(4,\mathbb{R})$ can be found by simple correspondence with the ten generators of $\mathrm{Sp}(4,\mathbb{R})$, upon introducing  $\exp(\alpha_i \bs{L}_i)\equiv \bs{U}\exp(\alpha_i \bs{K}_i)\bs{U}^\dag = \exp(\alpha_i \bs{U} \bs{K}_i \bs{U}^\dag)$, where $i=1\cdots 10$ and where we have used the fact that $\bs{U}$ is unitary. Writing $K_i=\bs{\sigma}_{a_i}\otimes \bs{\sigma}_{b_i}$ and  $\bs{U} = u \otimes \boldsymbol{I}_2$ with 
\begin{eqnarray}
	u = \frac{1}{\sqrt{2}}\begin{pmatrix}
		1 & i\\
		1 & -i
	\end{pmatrix},
\end{eqnarray}
the generators of $\Spnew(4,\mathbb{R})$ can be calculated by means of \Eq{eq:Kronecker:Multiplication:Rule} and one has $ \bs{L}_i = \bs{U} \bs{K}_i \bs{U}^\dag = \left(\bs{u} \otimes \boldsymbol{I}_2\right)\left(\bs{\sigma}_{a_i}\otimes \bs{\sigma}_{b_i} \right) (\bs{u}^\dag \otimes \boldsymbol{I}_2) = \bs{u}\bs{\sigma}_{a_i}\bs{u}^\dag \otimes \bs{\sigma}_{b_i}$. Given that $\bs{u} \bs{\sigma}_x \bs{u}^\dag = -\bs{\sigma}_y$, $\bs{u} \bs{\sigma}_y \bs{u}^\dag = -\bs{\sigma}_z$ and $\bs{u} \bs{\sigma}_z \bs{u}^\dag = \bs{\sigma}_x$, one concludes that the generators of $\Spnew(4,\mathbb{R})$ are merely a reshuffling of those of $\mathrm{Sp}(4,\mathbb{R})$, where the detailed correspondence is given in Table~\ref{tab:gensu22}. In particular, one observes that rotations in the helicity basis are block diagonal (which generalises the  $\mathrm{Sp}(2,\mathbb{R})$ result where rotations in the helicity basis are $(2\times 2)$ diagonal matrices~\cite{Grain:2019vnq}). Note that the commutation relations between the $\bs{L}_i$ operators directly follow from those between the $\bs{K}_i$ operators given in \App{sec:commutrel}.
\begin{table}[h!]
	\centering
	\renewcommand{\arraystretch}{1.5}
	$\begin{tabular}{| L | L | L |}
	\hline
	\text{Squeezing} & \text{Rotation} & \text{Boost} \\
	\hline
	\bs{L}_1 = \bs{K}_8 & \bs{L}_3 = -i\bs{K}_2 & \bs{L}_7 = i\bs{K}_3 \\
	\bs{L}_2 = \bs{K}_7 & \bs{L}_4 = -i \bs{K}_1 & \bs{L}_8 = i\bs{K}_4 \\
	& \bs{L}_5 = \bs{K}_5 & \bs{L}_9 = \bs{K}_{10}\\
	& \bs{L}_6 = -i\bs{K}_9 & \bs{L}_{10} = i\bs{K}_6\\
	\hline
	\end{tabular}$
	\caption{Generators of the helicity basis $\Spnew(4,\mathbb{R})$ in the fundamental representation, $\bs{L}_i = \bs{U} \bs{K}_i \bs{U}^\dag$, where the $\bs{K}_i$ generators are given in Table~\ref{tab:gen}.}
	\label{tab:gensu22}
\end{table}

The Bloch-Messiah decomposition~\eqref{eq:bm} can also be performed in the helicity basis,
\begin{eqnarray}\label{eq:bmhel}
\bs{\mathcal{M}} (\boldsymbol{\theta}, \boldsymbol{d}, \boldsymbol{\varphi})    = \bs{\mathcal{R}}(\boldsymbol{\theta})\bs{\mathcal{Z}}(\boldsymbol{d})\bs{\mathcal{R}}(\boldsymbol{\varphi})\, ,
\end{eqnarray}
where $\bs{\mathcal{Z}} = \bs{U} \bs{Z} \bs{U}^\dag$ and $\bs{\mathcal{R}} = \bs{U} \bs{R} \bs{U}^\dag$ are the squeezing matrix and the rotation matrix in the helicity basis. This expression allows us to connect the Bogolyubov coefficients with the squeezing and rotation parameters. Using the above results, an explicit calculation yields
\begin{align}
\label{eq:bogolsqueez1}
		\alpha_{11} =& e^{-i(\theta_3 + \varphi_3)} \left(\cos \theta -i \frac{\theta_4}{\theta} \sin \theta\right) \left(\cos \varphi -i \frac{\varphi_4}{\varphi} \sin \varphi\right) \cosh r_1\nonumber \\
		& -e^{-i(\theta_3 + \varphi_3)} \left[ \left(\theta_5 -i \theta_6\right) \frac{\sin \theta}{\theta} \right]  \left[ \left(\varphi_5 +i \varphi_6\right) \frac{\sin \varphi}{\varphi} \right] \cosh r_2\, ,
\\
		\alpha_{12} =& e^{-i(\theta_3 + \varphi_3)} \left(\cos \theta -i \frac{\theta_4}{\theta} \sin \theta\right) \left[ \left(\varphi_5 -i \varphi_6\right) \frac{\sin \varphi}{\varphi} \right]   \cosh r_1 \nonumber\\
		& +e^{-i(\theta_3 + \varphi_3)}  \left[ \left(\theta_5 -i \theta_6\right) \frac{\sin \theta}{\theta} \right]  \left(\cos \varphi +i \frac{\varphi_4}{\varphi} \sin \varphi\right)\cosh r_2\, ,
\\
		\alpha_{21} =& -  e^{-i(\theta_3 + \varphi_3)} \left[ \left(\theta_5 + i \theta_6\right) \frac{\sin \theta}{\theta} \right]  \left(\cos \varphi -i \frac{\varphi_4}{\varphi} \sin \varphi\right) \cosh r_1 \nonumber\\
		& -e^{-i(\theta_3 + \varphi_3)} \left(\cos \theta +i \frac{\theta_4}{\theta} \sin \theta\right)  \left[ \left(\varphi_5 +i \varphi_6\right) \frac{\sin \varphi}{\varphi} \right] \cosh r_2\, ,
\\
		\alpha_{22} =& -e^{-i(\theta_3 + \varphi_3)} \left[ \left(\theta_5 +i \theta_6\right) \frac{\sin \theta}{\theta} \right]  \left[ \left(\varphi_5 -i \varphi_6\right) \frac{\sin \varphi}{\varphi} \right]  \cosh r_1\nonumber \\
		& +e^{-i(\theta_3 + \varphi_3)} \left(\cos \theta +i \frac{\theta_4}{\theta} \sin \theta\right) \left(\cos \varphi +i \frac{\varphi_4}{\varphi} \sin \varphi\right)  \cosh r_2\, ,
\end{align}
and
\begin{align}
		\beta_{11} =& e^{-i(\theta_3 - \varphi_3)} \left(\cos \theta -i \frac{\theta_4}{\theta} \sin \theta\right) \left(\cos \varphi +i \frac{\varphi_4}{\varphi} \sin \varphi\right) \sinh r_1 \nonumber\\
		&-e^{-i(\theta_3 - \varphi_3)} \left[ \left(\theta_5 -i \theta_6\right) \frac{\sin \theta}{\theta} \right]  \left[ \left(\varphi_5 -i \varphi_6\right) \frac{\sin \varphi}{\varphi} \right] \sinh r_2\, ,
\\
		\beta_{12} =& e^{-i(\theta_3 - \varphi_3)} \left(\cos \theta -i \frac{\theta_4}{\theta} \sin \theta\right) \left[ \left(\varphi_5 +i \varphi_6\right) \frac{\sin \varphi}{\varphi} \right]   \sinh r_1 \nonumber\\
		&+e^{-i(\theta_3 - \varphi_3)}  \left[ \left(\theta_5 -i \theta_6\right) \frac{\sin \theta}{\theta} \right]  \left(\cos \varphi -i \frac{\varphi_4}{\varphi} \sin \varphi\right)\sinh r_2\, ,
\\
		\beta_{21} =& -  e^{-i(\theta_3 - \varphi_3)} \left[ \left(\theta_5 + i \theta_6\right) \frac{\sin \theta}{\theta} \right]  \left(\cos \varphi +i \frac{\varphi_4}{\varphi} \sin \varphi\right) \sinh r_1\nonumber \\
		&-e^{-i(\theta_3 - \varphi_3)} \left(\cos \theta +i \frac{\theta_4}{\theta} \sin \theta\right)  \left[ \left(\varphi_5 -i \varphi_6\right) \frac{\sin \varphi}{\varphi} \right] \sinh r_2\, ,
\\
	\label{eq:bogolsqueez8}	
		\beta_{22} =& -e^{-i(\theta_3 - \varphi_3)} \left[ \left(\theta_5 +i \theta_6\right) \frac{\sin \theta}{\theta} \right]  \left[ \left(\varphi_5 +i \varphi_6\right) \frac{\sin \varphi}{\varphi} \right]  \sinh r_1\nonumber \\
		&+e^{-i(\theta_3 - \varphi_3)} \left(\cos \theta +i \frac{\theta_4}{\theta} \sin \theta\right) \left(\cos \varphi -i \frac{\varphi_4}{\varphi} \sin \varphi\right)  \sinh r_2\, ,
\end{align}
where $\theta = \sqrt{\theta^2_4 +\theta^2_5 + \theta^2_6}$ and $\varphi = \sqrt{\varphi^2_4 +\varphi^2_5 + \varphi^2_6}$ have already been defined in \Eq{eq:thetavarphi}, and where we have introduced 
\bea
r_1 = d_1 + d_2\quad \text{and} \quad r_2 = d_2 - d_1\, .
\eea 
One can check that if the rotation parameters associated with the coupling generators vanish, \ie if $\theta_5=\theta_6=\varphi_5=\varphi_6=0$, then the mixing Bogolyubov coefficients vanish too, \ie $\alpha_{12}=\alpha_{21}=\beta_{12}=\beta_{21}=0$, and the link between the Bogolyubov coefficients and the squeezing and rotation parameters reduces to the one obtained in $\mathrm{Sp}(2,\mathbb{R})$~\cite{Grain:2019vnq}. 

Finally, a fully factorised form for the elements of $\Spnew(4,\mathbb{R})$ can be obtained from transposing \Eq{eq:fullyfact}, which gives rise to
\begin{equation}\label{eq:fullyfactsu22}
\begin{aligned}
\bs{\mathcal{M}} = &\overbrace{\left[\exp(p_+ \bs{L}_+) \cdot \exp(p_z \bs{L}_z) \cdot \exp(p_- \bs{L}_-) \cdot \exp(\theta_3 \bs{L}_3)\right] }^{\bs{\mathcal{R}}(\boldsymbol{\theta})} 
\cdot\overbrace{\left[\exp(d_1 \bs{L}_1)\cdot\exp(d_2 \bs{L}_2)\right] }^{\bs{\mathcal{Z}}(\boldsymbol{d})} \\
&~~~~\cdot
\overbrace{\left[\exp(q_+ \bs{L}_+) \cdot \exp(q_z \bs{L}_z) \cdot \exp(q_- \bs{L}_-) \cdot \exp(\varphi_3 \bs{L}_3)\right] }^{\bs{\mathcal{R}}(\boldsymbol{\varphi})}\, ,
\end{aligned}
\end{equation}
where $\bs{L}_z = \bs{U}\bs{S}_z\bs{U}^\dag = \bs{L}_4/(2 i)$, $\bs{L}_+ = \bs{U}\bs{S}_+\bs{U}^\dag=\left(\bs{L}_5 - i \bs{L}_6\right)/(2i)$ and $\bs{L}_- = \bs{U}\bs{S}_-\bs{U}^\dag=\left(\bs{L}_5 + i \bs{L}_6\right)/(2 i)$.
\subsection{Quantum representation}
\label{subsec:quantumliealg}
Since the Green's matrix is an element of the symplectic group, in order to describe the dynamics in the occupation-number representation, one first needs to derive the quantum representation of the elements of $\Spnew(4,\mathbb{R})$. This can be done by following the procedure outlined in \Refs{PhysRevA.36.3868, SIMON1987223, Arvind:1995ab} and presented in details in Appendix B of \Refa{Grain:2019vnq}. It consists in first linearising the elements of the $\Spnew(4,\mathbb{R})$ Lie group,
\begin{eqnarray}
\bs{\mathcal{M}}_k \simeq \boldsymbol{I}_4 + \sum^{10}_{a = 1} \varepsilon^a_k \bs{L}_a,
\end{eqnarray}
where $\varepsilon^a_k$ are (small) real parameters. This generates a canonical transformation on the helicity variables, given by 
\bea
\label{eq:Canonical:Transf:Infinitesimal:Dir}
\widehat{\boldsymbol{a}}'_{\vec{k}} \simeq \widehat{\boldsymbol{a}}_{\vec{k}} + \sum^{10}_{a = 1} \varepsilon^a_k \bs{L}_a \widehat{\boldsymbol{a}}_{\vec{k}}\, .
\eea
Our goal is to find a set of operators $ \widehat{L}^{\vec{k}}_a$  such that this can be written as a unitary transformation  $\widehat{\boldsymbol{a}}'_{\vec{k}} = \widehat{{\mathcal{M}}}^\dag \widehat{\boldsymbol{a}}_{\vec{k}} \widehat{{\mathcal{M}}}$, with
\bea
\widehat{{\mathcal{M}}} \simeq \widehat{\mathbb{I}} + \int_{\mathbb{R}^3} \dd^3 \vec{q} \sum^{10}_{a = 1} \varepsilon^a_{\vec{q}} \widehat{L}^{\vec{q}}_a\, .
\eea
Note that $\widehat{L}^{\vec{q}}_a$ has to be anti-hermitian for $\widehat{{\mathcal{M}}}$ to be unitary. By expanding this transformation in $\epsilon_{\vec{q}}^a$, one obtains $
\widehat{\boldsymbol{a}}'_{\vec{k}} \simeq \widehat{\boldsymbol{a}}_{\vec{k}} + \left[\widehat{\boldsymbol{a}}_{\vec{k}}, \int_{\mathbb{R}^3} \mathrm{d}^3q \sum^{10}_{a = 1} \varepsilon^a_q \widehat{L}^{\vec{q}}_a \right]$ (where the commutator is performed on each entry of $\widehat{\boldsymbol{a}}_{\vec{k}}$). Since this formula should match \Eq{eq:Canonical:Transf:Infinitesimal:Dir}, and given the commutation relations~\eqref{eq:acommut}, this suggests to look for  $\widehat{L}^{\vec{q}}_a$ operators that are quadratic in the creation and annihilation operators,
\begin{eqnarray}\label{eq:La}
\widehat{L}^{\vec{q}}_a = \widehat{\boldsymbol{a}}^\dag_{\vec{q}} \bs{\mathcal{Q}}^a \widehat{\boldsymbol{a}}_{\vec{q}},
\end{eqnarray}
where $\bs{\mathcal{Q}}^a$ is anti-hermitian since $\widehat{L}^{\vec{q}}_a$ is. Making use of \Eq{eq:acommut}, this leads to
$ \left[\widehat{\boldsymbol{a}}_{\vec{k}}, \int_{\mathbb{R}^3} \mathrm{d}^3q \sum^{10}_{a = 1} \varepsilon^a_q \widehat{L}^{\vec{q}}_a \right] = i \bs{\mathcal{J}} \sum^{10}_{a = 1} \varepsilon^a_k \bs{\mathcal{Q}}^a \widehat{\boldsymbol{a}}_{\vec{k}}$. By identification with \Eq{eq:Canonical:Transf:Infinitesimal:Dir}, this gives $\bs{L}_a = i \bs{\mathcal{J}} \bs{\mathcal{Q}}^a$, which can be inverted as
\bea
\bs{\mathcal{Q}}^a = i \bs{\mathcal{J}}\bs{L}_a\, .
\eea
This yields the generators $\widehat{L}^{\vec{q}}_a $ of the quantum representation listed in Table~\ref{tab:qgen}. 
\begin{table}[h!]
	\centering
	\renewcommand{\arraystretch}{1.5}
	$\begin{tabular}{| L | L |}
		\hline
		\text{Squeezing} &  \widehat{L}^{\vec{k}}_1 = \left(\widehat{a}^\dag_{1, \vec{k}}\widehat{a}^\dag_{1, -\vec{k}} - \widehat{a}_{1, \vec{k}} \widehat{a}_{1, -\vec{k}} \right) - \left(\widehat{a}^\dag_{2, \vec{k}}\widehat{a}^\dag_{2, -\vec{k}} - \widehat{a}_{2, \vec{k}} \widehat{a}_{2, -\vec{k}} \right)\\
		& \widehat{L}^{\vec{k}}_2 = \left(\widehat{a}^\dag_{1, \vec{k}}\widehat{a}^\dag_{1, -\vec{k}} - \widehat{a}_{1, \vec{k}} \widehat{a}_{1, -\vec{k}} \right) + \left(\widehat{a}^\dag_{2, \vec{k}}\widehat{a}^\dag_{2, -\vec{k}} - \widehat{a}_{2, \vec{k}} \widehat{a}_{2, -\vec{k}} \right) \\
		\hline
		\text{Rotation} & \widehat{L}^{\vec{k}}_3 = -i\left(\widehat{a}^\dag_{1, \vec{k}}\widehat{a}_{1, \vec{k}} + \widehat{a}^\dag_{1, -\vec{k}}\widehat{a}_{1, -\vec{k}} + 1 \right) -i\left(\widehat{a}^\dag_{2, \vec{k}}\widehat{a}_{2, \vec{k}} + \widehat{a}^\dag_{2, -\vec{k}}\widehat{a}_{2, -\vec{k}} + 1  \right) \\
		&  \widehat{L}^{\vec{k}}_4 = -i\left(\widehat{a}^\dag_{1, \vec{k}}\widehat{a}_{1, \vec{k}} + \widehat{a}^\dag_{1, -\vec{k}}\widehat{a}_{1, -\vec{k}} + 1 \right) +i\left(\widehat{a}^\dag_{2, \vec{k}}\widehat{a}_{2, \vec{k}} + \widehat{a}^\dag_{2, -\vec{k}}\widehat{a}_{2, -\vec{k}} + 1  \right)\\
		& \widehat{L}^{\vec{k}}_5 = \left(\widehat{a}^\dag_{1, \vec{k}}\widehat{a}_{2, \vec{k}} + \widehat{a}^\dag_{1, -\vec{k}} \widehat{a}_{2, -\vec{k}} \right) - \left(\widehat{a}^\dag_{2, \vec{k}}\widehat{a}_{1, \vec{k}} + \widehat{a}^\dag_{2, -\vec{k}} \widehat{a}_{1, -\vec{k}} \right)\\
		& \widehat{L}^{\vec{k}}_6 = -i\left(\widehat{a}^\dag_{1, \vec{k}}\widehat{a}_{2, \vec{k}} + \widehat{a}^\dag_{1, -\vec{k}} \widehat{a}_{2, -\vec{k}} \right) -i\left(\widehat{a}^\dag_{2, \vec{k}}\widehat{a}_{1, \vec{k}} + \widehat{a}^\dag_{2, -\vec{k}} \widehat{a}_{1, -\vec{k}} \right)\\
		\hline
		\text{Boost} & \widehat{L}^{\vec{k}}_7 = i\left(\widehat{a}^\dag_{1, \vec{k}}\widehat{a}^\dag_{1, -\vec{k}} + \widehat{a}_{1, \vec{k}} \widehat{a}_{1, -\vec{k}} \right) + i\left(\widehat{a}^\dag_{2, \vec{k}}\widehat{a}^\dag_{2, -\vec{k}} + \widehat{a}_{2, \vec{k}} \widehat{a}_{2, -\vec{k}} \right) \\
		& \widehat{L}^{\vec{k}}_8 = i\left(\widehat{a}^\dag_{1, \vec{k}}\widehat{a}^\dag_{1, -\vec{k}} + \widehat{a}_{1, \vec{k}} \widehat{a}_{1, -\vec{k}} \right) - i\left(\widehat{a}^\dag_{2, \vec{k}}\widehat{a}^\dag_{2, -\vec{k}} + \widehat{a}_{2, \vec{k}} \widehat{a}_{2, -\vec{k}} \right)\\
		& \widehat{L}^{\vec{k}}_9 = \left(\widehat{a}^\dag_{1, \vec{k}}\widehat{a}^\dag_{2, -\vec{k}} + \widehat{a}^\dag_{2, \vec{k}}\widehat{a}^\dag_{1, -\vec{k}} \right) - \left(\widehat{a}_{1, \vec{k}}\widehat{a}_{2, -\vec{k}} + \widehat{a}_{2, \vec{k}}\widehat{a}_{1, -\vec{k}} \right)\\
		& \widehat{L}^{\vec{k}}_{10} = i\left(\widehat{a}^\dag_{1, \vec{k}}\widehat{a}^\dag_{2, -\vec{k}} + \widehat{a}^\dag_{2, \vec{k}}\widehat{a}^\dag_{1, -\vec{k}} \right) +i\left(\widehat{a}_{1, \vec{k}}\widehat{a}_{2, -\vec{k}} + \widehat{a}_{2, \vec{k}}\widehat{a}_{1, -\vec{k}} \right)\\
		\hline
	\end{tabular}$
	\caption{Quantum representation of the generators of $\Spnew(4,\mathbb{R})$.}
	\label{tab:qgen}
\end{table}

Another set of generators that is obtained from straightforward linear combinations of the $\widehat{L}^{\vec{k}}_a $ operators is given by
\begin{eqnarray}
\label{eq:qgensp2R_1}    
\widehat{a}^\dag_{i, \vec{k}}\widehat{a}^\dag_{i, -\vec{k}}\, ,\qquad\widehat{a}_{i, \vec{k}}\widehat{a}_{i, -\vec{k}}\, ,\qquad\widehat{a}^\dag_{i, \vec{k}}\widehat{a}_{i, \vec{k}} + \widehat{a}^\dag_{i, -\vec{k}}\widehat{a}_{i, -\vec{k}} + 1 \qquad \text{with}\quad i=1,2\, ,
\end{eqnarray}
which correspond to the two $\mathfrak{sp}(2,\mathbb{R})$ subalgebra identified below \Eq{eq:Pauli}, and 
\begin{align}
\label{eq:qgenH22_1}     
\widehat{a}^\dag_{i, \vec{k}}\widehat{a}^\dag_{j, -\vec{k}} + \widehat{a}^\dag_{j, \vec{k}}\widehat{a}^\dag_{i, -\vec{k}} \, ,\quad
\widehat{a}_{i, \vec{k}}\widehat{a}_{j, -\vec{k}} + \widehat{a}_{j, \vec{k}}\widehat{a}_{i, -\vec{k}}   \, ,\quad  \widehat{a}^\dag_{i, \vec{k}}\widehat{a}_{j, \vec{k}} + \widehat{a}^\dag_{i, -\vec{k}} \widehat{a}_{j, -\vec{k}}
 \quad \text{with}\quad i,j=1,2\, ,
 \end{align}
which correspond to the coupling generators. 
  The operators in \Eqs{eq:qgensp2R_1}  generate two-mode creation, two-mode annihilation and a number counting operation respectively, acting on each sector separately.
  The first and second mixing generators in \Eq{eq:qgenH22_1} originate from the two mixing boosts ($\widehat{L}^{\vec{k}}_9$ and $\widehat{L}^{\vec{k}}_{10}$). They correspond to the creation and annihilation of entangled pairs of particles with opposite momenta in the two sectors.
  The third mixing generator in \Eq{eq:qgenH22_1} originates from the two mixing rotations ($\widehat{L}^{\vec{k}}_5$ and $\widehat{L}^{\vec{k}}_6$). It does not lead to net particle creation, but rather transfer excitations from one sector to the other.  One can check that all these operations preserve momentum conservation.
\section{Quantum dynamics}
\label{sec:QuantumDynamics}
In \Sec{sec:QuantumPhaseSpace}, we explained how the symplectic group $\mathrm{Sp}(4,\mathbb{R})$ naturally appears in the description of physical systems made of two free fields. Having then studied its mathematical structure in \Sec{sec:Sp4R:toolkit}, we now want to apply these tools to describe the dynamics of two-field systems. To this end, we explore the structure of the system's Hamiltonian in \Sec{subsec:quantumdyn}, before applying our findings to an explicit example. Then, we provide in \Sec{subsec:evolop} a tractable expression of the evolution operator.
\subsection{Hamiltonian structure} 
\label{subsec:quantumdyn}
In this section, we express the Hamiltonian in the occupation-number representation, to see which interactions are allowed in the theory on generic grounds. The only constraint we have on the Hamiltonian is that it stems from a real symmetric kernel, \ie $\bs{H}_k$ is a real symmetric matrix. From \Eq{eq:helH}, $\bs{\mathcal{H}}_k$ can thus be written as $\bs{\mathcal{H}}_k = \bs{U}\bs{N}_k \bs{U}^\dag$, 
where $\bs{N}_k=(\boldsymbol{D}^{-1}_k)^{\mathrm{T}}\boldsymbol{H}_k\boldsymbol{D}^{-1}_k + (\boldsymbol{D}^{-1}_k)^{\mathrm{T}}\bs{\Omega}\dot{\boldsymbol{D}}^{-1}_k$ is a real symmetric matrix (the first term is obviously symmetric, while for the second term, this can be shown by differentiating with respect to time the symplectic relation~\eqref{eq:sympdef} satisfied by $\bs{D}^{-1}_k$, and by using that $\bs{\Omega}^{\mathrm{T}} = - \bs{\Omega}$), which we parametrise as
\begin{eqnarray}\label{eq:decomp}
\bs{N}_k = \begin{pmatrix}
\bs{A} & \bs{C} \\
\bs{C}^{\mathrm{T}} & \bs{B}
\end{pmatrix},
\end{eqnarray}
with $\bs{A}, \bs{B}, \bs{C} \in \mathcal{M}_{2} (\mathbb{R})$ and $\bs{A}^{\mathrm{T}} = \bs{A}$, $\bs{B}^{\mathrm{T}} = \bs{B}$. This leads to
\begin{eqnarray}\label{eq:JHk}
\bs{\mathcal{H}}_k = \begin{pmatrix}
\mathfrak{a}   &   \mathfrak{b}   \\
\mathfrak{b}^*    &  \mathfrak{a}^*
\end{pmatrix}
\end{eqnarray}
with $ \mathfrak{a} = (\bs{A} + \bs{B}) + i(\bs{C}^{\mathrm{T}}-\bs{C})$ and $ \mathfrak{b} =  (\bs{A} - \bs{B}) + i(\bs{C}^{\mathrm{T}}+\bs{C})$. Inverting those formulas, one obtains that $\mathfrak{a}+\mathfrak{a}^* = 2(A+B)$ and $\mathfrak{a}-\mathfrak{a}^* = 2i(C^{\mathrm{T}}-C)$, so the real part of $\mathfrak{a}$ should be symmetric while its imaginary part should be antisymmetric; and  $\mathfrak{b}+\mathfrak{b}^* = 2(A-B)$ and $\mathfrak{b}-\mathfrak{b}^* = 2i(C^{\mathrm{T}}+C)$, so both the real and imaginary parts of $\mathfrak{b}$ should be symmetric.
This imposes that $\mathfrak{a}$ and $\mathfrak{b}$ are of the form
\begin{eqnarray}
 \label{eq:JHk2}
\mathfrak{a} = \begin{pmatrix}
F_{1,k}  &   F_{1\leftrightarrow2,k}e^{i\varphi_k}  \\
F_{1\leftrightarrow2, k} \ee^{-i\varphi_k}    & F_{2,k}
\end{pmatrix}\text{,}~~~~ \mathfrak{b} = \begin{pmatrix}
R_{1, k}e^{i\Theta_{1, k}}  &   R_{1 \leftrightarrow 2, k}e^{i\xi_k}  \\
R_{1 \leftrightarrow 2, k}e^{i\xi_k}   & R_{2, k}e^{i\Theta_{2, k}}
\end{pmatrix}\, ,
\end{eqnarray}
where $F_{1,k}$, $F_{1\leftrightarrow2,k}$, $F_{2,k}$, $\varphi_k$, $R_{1, k}$, $R_{1 \leftrightarrow 2, k}$, $R_{2, k}$, $\Theta_{1, k}$, $\Theta_{2, k}$ and $\xi_k$ are ten real parameters, that in general depend on time, and which indeed saturate the ten degrees of freedom contained in $\bs{N}_k$. In fact, since the Green matrix is generated from the (integrated) Hamiltonian, those ten parameters fully determine the squeezing and rotation parameters, or equivalently the Bogolyubov coefficients, that characterise the dynamics. 

The Hamiltonian $\widehat{\mathcal{H}} = \int_{\mathbb{R}^{3+}} \mathrm{d}^3\vec{k} \boldsymbol{\widehat{a}}_{\vec{k}}^\dag \bs{\mathcal{H}}_k\boldsymbol{\widehat{a}}_{\vec{k}}$ thus contain three terms, 
\begin{eqnarray}\label{eq:Htotal}
\widehat{\mathcal{H}} &=&  \int_{\mathbb{R}^{3+}} \mathrm{d}^3\vec{k} \left(\widehat{\mathcal{H}}_{1,\vec{k}} + \widehat{\mathcal{H}}_{2,\vec{k}} + \widehat{\mathcal{H}}_{1 \leftrightarrow 2,\vec{k}}\right)\, ,
\end{eqnarray}
with 
\begin{align}
\label{eq:Hi}
\widehat{\mathcal{H}}_{i,\vec{k}} =& F_{i, k}\left( \widehat{a}^\dag_{i, \vec{k}}\widehat{a}_{i, \vec{k}} + \widehat{a}^\dag_{i, -\vec{k}}\widehat{a}_{i, -\vec{k}} + 1 \right)  +  R_{i, k}\left(e^{i\Theta_{i, k}} \widehat{a}^\dag_{i, \vec{k}} \widehat{a}^\dag_{i, -\vec{k}} + \text{h.c.} \right)
\quad\text{for}\quad i=1,2\, ,\\
\label{eq:H12}
	\widehat{\mathcal{H}}_{1 \leftrightarrow 2,\vec{k}} =& F_{1\leftrightarrow2, k}e^{i\varphi_k} \left(\widehat{a}^\dag_{1, \vec{k}}\widehat{a}_{2, \vec{k}} + \widehat{a}^\dag_{1, -\vec{k}} \widehat{a}_{2, -\vec{k}}\right) 
+ R_{1 \leftrightarrow 2, k}e^{i\xi_k} \left(\widehat{a}^\dag_{1, \vec{k}}\widehat{a}^\dag_{2, -\vec{k}} + \widehat{a}^\dag_{2, \vec{k}}\widehat{a}^\dag_{1, -\vec{k}}\right) + \text{h.c.} \, .
\end{align}
The Hamiltonian can thus be decomposed onto the quantum generators listed in Table~\ref{tab:qgen} (this is because $\bs{\mathcal{JH}}_k$ is an element of the Lie algebra), hence we can benefit from their physical interpretation discussed in \Sec{subsec:quantumliealg}. The components $\widehat{\mathcal{H}}_{1}$ and $\widehat{\mathcal{H}}_{2}$ drive each sector separately and are made of two terms: a harmonic part controlled by $F_{i, k}$ that does not induce particle creation, and a parametric part controlled by $R_{i, k}$ that changes the particle content. Physically, this parametric amplification is related to the presence of an external field (the electric field for the Schwinger effect~\cite{Martin:2007bw}, the gravitational field for the Hawking effect, space-time curvature for the physics of cosmological perturbations, \etc). The component $\widehat{\mathcal{H}}_{1 \leftrightarrow 2}$ is an interaction term between the two sectors, and also contains two contributions: a ``transferring'' part, controlled by $F_{1\leftrightarrow2, k}(t)$ and built from the two mixing rotation operators, that transfers particles from one sector to the other; and an ``entangling'' part, controlled by $R_{1\leftrightarrow2, k}$ and built from the two mixing boost operators, that creates or annihilates joint pairs of particles in the two sectors.

Let us note that even when the entangling part is absent, \ie when $R_{1\leftrightarrow2, k}=0$, entanglement between the two sectors can still indirectly arise from first creating particles in the two sectors separately with the parametric terms, and then transferring those particles between sectors by means of the transferring term. 
\subsubsection*{Example of two massless fields in a cosmological background}
\label{subsec:example}
Let us illustrate the formalism introduced above with a simple example. We consider two massless scalar fields $\phi$ and $\chi$ on a Friedmann-Lema\^itre-Robertson-Walker spatially flat geometry,
\begin{eqnarray}\label{eq:metric}
\mathrm{d}s^2 = a^2(\eta) \left(- \mathrm{d}\eta^2 + \delta^{ij} \mathrm{d}x_i \mathrm{d}x_j \right)\, ,
\end{eqnarray}	
where $a$ is the scale factor and $\eta$ is the conformal time. The Ricci scalar $R$ of this metric is given by $R=6a''/a^3$, where a prime denotes derivation with respect to conformal time. The action is given by
\begin{eqnarray}
S = - \int \mathrm{d}^4x \sqrt{-\det g} \left(\frac{1}{2} g^{\mu \nu}\partial_\mu \phi \partial_\nu \phi +\frac{R}{2} \zeta \phi^2+ \frac{1}{2} g^{\mu \nu}\partial_\mu \chi \partial_\nu \chi +\frac{R}{2} \zeta \chi^2 + \lambda^2 \phi \chi \right)\, ,
\end{eqnarray}
where $\zeta$ is the conformal coupling constant and $\lambda$ is a coupling parameter that has dimension of a mass. Expanding the scalar fields into Fourier modes as in \Eq{eq:Fourier} and making use of the reality prescription $\phi_{\vec{k}}^* = \phi_{-\vec{k}}$ and $\chi_{\vec{k}}^* = \chi_{-\vec{k}}$, with the the metric~\eqref{eq:metric} the action reads
\begin{eqnarray}
\begin{aligned}
S \equiv \int \mathrm{d}\eta L  = & - \int \mathrm{d}\eta \int_{\mathbb{R}^{3+}} \dd^3\vec{k} \left[a^2  \phi'_{-\vec{k}} \phi'_{\vec{k}} + \left( k^2 a^2 + R \zeta a^4 \right) \phi_{-\vec{k}} \phi_{\vec{k}}
\right. \\ & \left.
+  a^2 \chi'_{-\vec{k}} \chi'_{\vec{k}} + \left(k^2 a^2 + R \zeta a^4 \right) \chi_{-\vec{k}} \chi_{\vec{k}}
+ \lambda^2 a^4 \left( \phi_{-\vec{k}} \chi_{\vec{k}} + \chi_{-\vec{k}} \phi_{\vec{k}}\right) \right]\, ,
\end{aligned}
\end{eqnarray}
which also defines the Lagrangian density $L$.
The conjugate momenta can be identified as $p^\phi_{\vec{k}} = a^2 \phi'_{-\vec{k}}$ and $p^\chi_{\vec{k}} = a^2 \chi'_{-\vec{k}}$, and upon performing a Legendre transform, one obtains a Hamiltonian of the form~\eqref{eq:quadhamilt}, with
\begin{eqnarray}
\bs{H}_k =   \begin{pmatrix}
k^2 a^2 + R \zeta a^4 & \lambda^2 a^4 & 0 & 0  \\
\lambda^2 a^4 & k^2 a^2 + R \zeta a^4 & 0 & 0  \\
0 & 0 & 1/a^2 & 0  \\
0 & 0 & 0 & 1/a^2
\end{pmatrix}
\qquad\text{and}\qquad
\boldsymbol{z}_{\vec{k}} = \begin{pmatrix}
\phi_{\vec{k}}\\
\chi_{\vec{k}}\\
p^\phi_{\vec{k}}\\
p^\chi_{\vec{k}}
\end{pmatrix} \, .
\end{eqnarray}
The momentum sector of the Hamiltonian (\ie the bottom-right block) can be simplified by performing a canonical transformation of the form~\eqref{eq:cantransfo}, $\widetilde{\boldsymbol{z}}_{\vec{k}} = \boldsymbol{M}_k\boldsymbol{z}_{\vec{k}}$, with
\bea
\bs{M}_k = \begin{pmatrix}
a & 0 & 0 & 0\\
0 & a & 0 & 0\\
a' & 0 & 1/a & 0 \\
0 & a' & 0 & 1/a
\end{pmatrix}\, .
\eea
One can check that $\bs{M}_k$ satisfies the symplectic relation~\eqref{eq:sympdef}, and that the new Hamiltonian kernel, given by \Eq{eq:Htransfo}, reads
\bea
\widetilde{\bs{H}}_k =  \begin{pmatrix}
k^2 + R \zeta a^2 - a''/a &  \lambda^2 a^2 & 0 & 0  \\
\lambda^2 a^2 & k^2 + R \zeta a^2 - a''/a& 0 & 0  \\
0 & 0 & 1 & 0  \\
0 & 0 & 0 & 1
\end{pmatrix}\, .
\eea
The $\tilde{\bs{z}}_{\vec{k}}$ variables are usually referred to as the Mukhanov-Sasaki variables, and in the helicity basis, \Eq{eq:helH} gives rise to a Hamiltonian kernel of the form~\eqref{eq:JHk}, with 
\bea
\begin{aligned}
\label{eq:H:param:deSitter}
k R_{i,k}  &= \frac{a^2}{2}R \zeta-\frac{a''}{2a}
\, ,\qquad 
k F_{i,k}  = k^2 + k R_{i,k}
\, ,\qquad 
F_{1 \leftrightarrow2, k} = R_{1 \leftrightarrow2, k} = \frac{\lambda^2 a^2}{2k}
\, ,
\end{aligned}
\eea
 and where all the phases vanish, \ie $\varphi_k=\Theta_{1, k}=\Theta_{2, k}=\xi_k=0$. This allows one to relate the four Hamiltonian contributions identified in \Sec{subsec:quantumdyn}, namely the harmonic, parametric, transferring and entangling terms, to the microphysical parameters of the problem. In particular, one can see that the parametric term is generated by the external ``field'' $a(t)$ (since when the scale factor is a constant, $R_{i,k}=0$), and that the transferring and entangling terms are controlled by the coupling parameter $\lambda$, in agreement with the discussion in \Sec{subsec:quantumdyn}.
 
Since $R=6a''/a^3$, the parametric term is given by $kR_{i,k}=(3\zeta-1/2)a''/a$. Therefore, if one sets $\zeta=1/6$, the parametric term vanishes, and the harmonic term simply becomes $F_{i,k}=k$, as in flat space time. This is because, in that case, the fields are conformally coupled to the metric, which removes the effect of space-time expansion that otherwise generates parametric amplification. This implies that, in the absence of direct coupling (\ie if $\lambda=0$), there is no particle creation, and the above system is made of uncoupled harmonic oscillators. However, in the presence of direct coupling, both the transferring and the entangling terms become non vanishing, which guarantees that entangled pairs of particles are created and exchanged between the two fields. This is because the interaction term breaks conformal invariance, as noticed in \Refa{Matacz:1992mk}. 
\subsection{Evolution operator}
\label{subsec:evolop} 
We now investigate the integrated dynamics. As explained in \Sec{subsec:su22}, the Green's matrix in the helicity basis, $\bs{\mathcal{G}}_k(t,t_\uin)$, belongs to $\Spnew(4,\mathbb{R})$. Its quantum analogue, the so-called evolution operator $\widehat{\mathcal{U}}_{\vec{k}}(t, t_\uin)$, thus lies in the quantum representation of $\Spnew(4,\mathbb{R})$, which was studied in \Sec{subsec:quantumliealg}. In this section, we make use of the formal results derived above to derive a tractable expression for the evolution operator. 

Since the Green matrix is separable in Fourier space for free fields, the evolution operator is also factorisable as
\bea
\label{eq:modedecomp}
\widehat{\mathcal{U}}(t, t_\uin) = \prod\limits_{\vec{k} \in \mathbb{R}^{3+}} \widehat{\mathcal{U}}_{\vec{k}}(t, t_\uin)\, ,
\eea
where $\widehat{\mathcal{U}}_{\vec{k}}(t, t_\uin) \in \Spnew(4,\mathbb{R})$. Making use of the Bloch-Messiah decomposition presented in \Secs{subsec:decomp} and~\ref{subsec:su22}, one can write
\bea
\label{eq:BMrappel}
\widehat{\mathcal{U}}_{\vec{k}}(t, t_\uin) = \widehat{\mathcal{R}}_{\vec{k}}(\boldsymbol{\theta}_k) \cdot \widehat{\mathcal{Z}}_{\vec{k}}(\boldsymbol{d}_k) \cdot \widehat{\mathcal{R}}_{\vec{k}}(\boldsymbol{\varphi}_k)\, ,
\eea
see \Eq{eq:bmhel}, where the squeezing and rotation parameters depend only on the norm of the wavevector, since this is also the case for the Green's matrix. Further factorisation can be obtained by the procedure outlined in \Sec{subsec:su22} and leading to \Eq{eq:fullyfactsu22}, and replacing the generators in \Eq{eq:fullyfactsu22} by their expression in the quantum representation given in Table~\ref{tab:qgen}, one finds that the evolution operator involves three types of operation only, namely 
\bea
\label{eq:phase}
\widehat{\mathcal{R}}_{i, \vec{k}}(\theta_{k}) &\equiv& \exp\left[-i \theta_{k} \left(\widehat{a}^\dagger_{i,\vec{k}}\widehat{a}_{i,\vec{k}}+\widehat{a}^\dagger_{i,-\vec{k}}\widehat{a}_{i,-\vec{k}}+ 1 \right) \right]\, ,\\
\label{eq:squeezing}
\widehat{\mathcal{Z}}_{i, \vec{k}}(r_{k}) &\equiv&  \exp \left[ r_{k} \left(\widehat{a}^\dag_{i,\vec{k}}\widehat{a}^\dag_{i,-\vec{k}} - \widehat{a}_{i,\vec{k}} \widehat{a}_{i,-\vec{k}} \right)\right]\, ,\\
\label{eq:transfer}
\widehat{\mathcal{R}}_{i \rightarrow j, \vec{k}}(p_{k}) &\equiv& \exp\left[ip_{k} \left( \widehat{a}^\dag_{j, \vec{k}}\widehat{a}_{i, \vec{k}} + \widehat{a}^\dag_{j_{\vec{-k}}} \widehat{a}_{i, -\vec{k}}\right)\right]\, .
\eea
The operators $\widehat{\mathcal{R}}_{i, \vec{k}}(\theta_{k})$ are constructed from the rotation generators and induce global phase shifts in sector $i$, without changing the particle content.
The squeezing operators $\widehat{\mathcal{Z}}_{i, \vec{k}}(r_{k})$ create pairs of entangled particles in each sector separately, and $\widehat{\mathcal{R}}_{i \rightarrow j, \vec{k}}(p_{k})$ transfers particles from one sector to the other without changing their overall number. 
In order to express the operators $\widehat{\mathcal{R}}_{\vec{k}}(\boldsymbol{\theta}_{k}), \widehat{\mathcal{R}}_{\vec{k}}(\boldsymbol{\varphi}_{k})$ and $\widehat{\mathcal{Z}}_{\vec{k}}(\boldsymbol{d}_{k})$ that appear in \Eq{eq:BMrappel} in terms of $\widehat{\mathcal{R}}_{i, \vec{k}}, \widehat{\mathcal{R}}_{i \rightarrow j, \vec{k}}$ and $\widehat{\mathcal{Z}}_{i, \vec{k}}$, we adopt a diagrammatic representation analogous to quantum circuits:
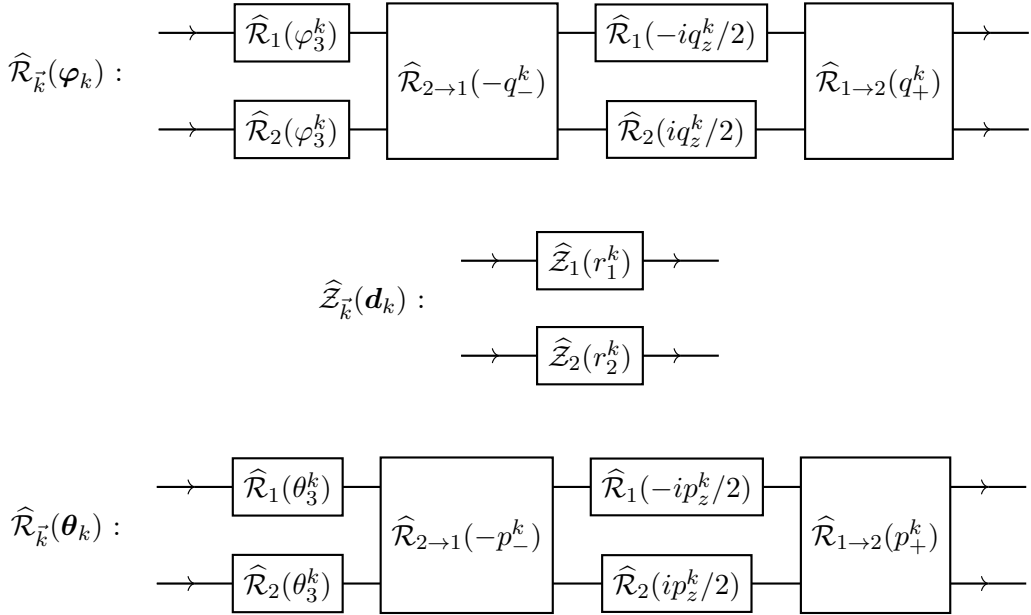
\begin{center}
		\begin{tikzpicture}
		\node[scale=1.0] {
		$\widehat{\mathcal{R}}_{\vec{k}}(\boldsymbol{\varphi}_{k})\ $:
			\begin{quantikz}
				\arrow[r] & \qw & \gate{\widehat{\mathcal{R}}_1(\varphi^{k}_3)} & \gate[wires=2]{\widehat{\mathcal{R}}_{2 \rightarrow 1}(-q^{k}_{-})} & \gate{\widehat{\mathcal{R}}_1(-i q^{k}_z /2)} & \gate[wires=2]{\widehat{\mathcal{R}}_{1 \rightarrow 2}(q^{k}_{+})} \arrow[r] & \qw & \qw \\
				\arrow[r] & \qw & \gate{\widehat{\mathcal{R}}_2(\varphi^{k}_3)} & & \gate{\widehat{\mathcal{R}}_2(i q^{k}_z /2)} & \arrow[r] & \qw & \qw
			\end{quantikz}
		};
	\end{tikzpicture}
\end{center}

\begin{center}
	\begin{tikzpicture}
		\node[scale=1.0] {
		$\widehat{\mathcal{Z}}_{\vec{k}}(\boldsymbol{d}_{k})\ $:
			\begin{quantikz}
				\arrow[r] & \qw & \gate{\widehat{\mathcal{Z}}_1(r^{k}_1)}  \arrow[r] & \qw & \qw \\
				\arrow[r] & \qw & \gate{\widehat{\mathcal{Z}}_2(r^{k}_2)}  \arrow[r] & \qw & \qw
			\end{quantikz}
		};
	\end{tikzpicture}
\end{center}

\begin{center}
\begin{tikzpicture}
	\node[scale=1.0] {
	$\widehat{\mathcal{R}}_{\vec{k}}(\boldsymbol{\theta}_{k})\ $:
		\begin{quantikz}
			\arrow[r] & \qw & \gate{\widehat{\mathcal{R}}_1(\theta^{k}_3)} & \gate[wires=2]{\widehat{\mathcal{R}}_{2 \rightarrow 1}(-p^{k}_{-})} & \gate{\widehat{\mathcal{R}}_1(-i p^{k}_z /2)} & \gate[wires=2]{\widehat{\mathcal{R}}_{1 \rightarrow 2}(p^{k}_{+})} \arrow[r] & \qw & \qw \\
			\arrow[r] & \qw & \gate{\widehat{\mathcal{R}}_2(\theta^{k}_3)} & & \gate{\widehat{\mathcal{R}}_2(i p^{k}_z /2)} & \arrow[r] & \qw & \qw
		\end{quantikz}
	};
\end{tikzpicture}
\end{center}
In those graphical representations, the top line stands for operations performed on the first sector, the bottom line on the second sector, and entangling operations are displayed with joint boxes. To run a ``circuit'', one successively applies the operations from the left to right (along the direction shown with the arrows). The parameters entering the circuits were introduced in \Sec{subsec:decomp}. The squeezing parameters $r^{k}_1 = d^{k}_1 + d^{k}_2$ and $r^{k}_2 = d^{k}_2 - d^{k}_1$ control the two-mode creation in each sector, while the mixing parameters $p^{k}_z, p^{k}_-, p^{k}_+$ and $q^{k}_z,q^{k}_-, q^{k}_+$ control the entanglement between the two sectors. This shows that dynamical evolution can be seen as successive applications of phase rotations within each sector, creations of particles with opposite momenta in each subspace, and particle transfers between the two sectors.
\section{Quantum state}
\label{sec:QuantumState}
We are now in a position to write down the quantum state of the system in the occupation-number basis. To that hand, we first need to equip our Hilbert space with a Fock-space structure. In practice, we impose that the creation and annihilation operators are ladder operators for the Hamiltonian at initial time $t_\uin$, \ie 
\begin{eqnarray}
\label{eq:ladder}
	\left[\widehat{\mathcal{H}},\widehat{a}^\dag_{i, \pm\vec{k}}\right] = c_{i,k} \widehat{a}^\dag_{i, \pm\vec{k}}~~~~~~~~~~ \text{and}~~~~~~~~~~ \left[\widehat{\mathcal{H}},\widehat{a}_{i, \pm\vec{k}}\right] = - c_{i,k} \widehat{a}_{i,\pm\vec{k}}
\end{eqnarray}
at $t_\uin$, for $i=1,2$ and where $c_{i,k}$ are real parameters. Among the various terms~\eqref{eq:Hi} and~\eqref{eq:H12} that the Hamiltonian may contain, only number counting operators, $\widehat{N}_{i,\pm\vec{k}} \equiv \widehat{a}^\dag_{i,\pm\vec{k}}\widehat{a}_{i,\pm\vec{k}}$, give commutators of the form~\eqref{eq:ladder}. This imposes that only the harmonic terms, \ie those controlled by $F_{i,k}$, can be present at initial time, leading to $c_{i,k}=F_{i,k}(t_\uin)$; while one must have $R_{i,k}(t_\uin)=R_{1 \leftrightarrow 2, k}(t_\uin)=F_{1\leftrightarrow2, k}(t_\uin) = 0$. This is for instance the case in the example discussed in \Sec{subsec:example} if the expansion is initially accelerating, since one can check that, then, in the asymptotic past (\ie in the limit $a \to 0$), only $F_{i,k}\simeq k$ survives in \Eq{eq:H:param:deSitter}. More generally, this is true in inflating backgrounds where Fourier modes get blue-shifted below the Hubble scale at early time, and hereafter we will assume that this condition is indeed satisfied. 

We then build the Fock basis of the initial Hilbert space,
\bea
\mathcal{E}(t_\uin) = \prod\limits_{\vec{k} \in \mathbb{R}^{3+}} \mathcal{E}^{(1)}_{\vec{k}}(t_\uin) \otimes \mathcal{E}^{(1)}_{-\vec{k}}(t_\uin) \otimes \mathcal{E}^{(2)}_{\vec{k}}(t_\uin) \otimes \mathcal{E}^{(2)}_{-\vec{k}}(t_\uin),
\eea
which is a quadripartite system. The four-mode vacuum state is the one annihilated by all four annihilation operators, denoted by 
\bea
\label{eq:inivac4}
\ket{\cancel{0}(t_\uin)} = \prod\limits_{\vec{k} \in \mathbb{R}^{3+}}  \ket{0^{(1)}_{\vec{k}},~0^{(1)}_{-\vec{k}},~ 0^{(2)}_{\vec{k}},~ 0^{(2)}_{-\vec{k}}}(t_\uin)\, ,
\eea
and the rest of the Fock space can be built by successive applications of creation operators, which leads to the Fock states
\begin{align}
\label{eq:Fock:definition}
	\ket{m^{(1)}_{\vec{k}},n^{(1)}_{-\vec{k}}, s^{(2)}_{\vec{k}}, t^{(2)}_{-\vec{k}}}(t_\uin) = \frac{\left(\widehat{a}^\dag_{1, \vec{k}}\right)^m}{m!} \frac{\left(\widehat{a}^\dag_{1, -\vec{k}}\right)^n}{n!}
	\frac{\left(\widehat{a}^\dag_{2, \vec{k}}\right)^{s}}{s!} \frac{\left(\widehat{a}^\dag_{2, -\vec{k}}\right)^{t}}{t!} \ket{0^{(1)}_{\vec{k}},0^{(1)}_{-\vec{k}}, 0^{(2)}_{\vec{k}}, 0^{(2)}_{-\vec{k}}}(t_\uin). 
\end{align}
From now on, we drop the argument $t_\uin$ to make the notation lighter. 
\subsection{Four-mode squeezed state}
\label{sec:vacevol}
The evolved vacuum is obtained by application of the evolution operator on the initial vacuum\footnote{Note that, in place of the initial vacuum, one may consider an initial state of the form $\ket{\Psi(t_\uin)} =  \prod_{\vec{k} } \widehat{\mathcal{A}}_{\vec{k}}(t_\uin)  \ket{\cancel{0}(t_\uin)}$, where $\widehat{\mathcal{A}}_{\vec{k}}(t_\uin) \in \Spnew(4,\mathbb{R})$. Such states are often referred to as ``alpha vacua''~\cite{Einhorn:2003xb} (where further restrictions on $\widehat{\mathcal{A}}_{\vec{k}}(t_\uin)$ sometimes apply). Since the product of two elements of $\Spnew(4,\mathbb{R})$ still lies in $\Spnew(4,\mathbb{R})$, the evolved alpha vacua are still of the form~\eqref{eq:vacevol}, so our discussion encompasses this possibility.\label{footnote:alpha:vacuua}}
\bea
\label{eq:vacevol}
\ket{\cancel{0}(t)} = \widehat{\mathcal{U}}(t, t_\uin)\ket{\cancel{0}(t_\uin)}.
\eea
In this section, we present its explicit expression in the Fock space, where we derive the most generic form of a four-mode squeezed state.

In \App{sec:vacappendix}, we apply the operators appearing in the circuits sketched in \Sec{subsec:evolop} onto the vacuum state~\eqref{eq:inivac4} one after the other, and derive the following expression for the evolved vacuum state $\ket{\cancel{0}(t)} = \prod_{\vec{k} \in \mathbb{R}^{3+}} \ket{\cancel{0}_{\vec{k}}(t)}$,
\bea
\label{eq:vacevolfinal}
\ket{\cancel{0}_{\vec{k}}(t)} =   \sum\limits_{n,m = 0}^\infty \sum\limits_{s,t = -n}^m c_{k}(n,m,s,t) \ket{(n+s)^{(1)}_{\vec{k}},(n+t)^{(1)}_{-\vec{k}}, (m-s)^{(2)}_{\vec{k}}, (m-t)^{(2)}_{-\vec{k}}},
\eea
with
\begin{equation}
\label{eq:ampl}
\begin{aligned}
		&\kern-1em c_{k}(n,m,s,t) =  \frac{e^{-2i\left[\theta^{k}_3(n+m+1) + \varphi^{k}_3\right]}}{\cosh{r^{k}_1} \cosh{r^{k}_2}}(-1)^{n+m}   e^{p^{k}_z(m-n)} \tanh^n (r^{k}_1) \tanh^m (r^{k}_2) (ip^{k}_+)^{-s-t} \frac{m!}{n!}\\
		& \quad \sqrt{\frac{(m-s)!(m-t)!}{(n+s)!(n+t)!}}
		\sum\limits_{i = \max(0, s)}^{m}\frac{\left(p^{k}_- p^{k}_+ e^{-p^{k}_z} \right)^i (n+i)!}{i!(i-s)!(m-i)!}
		\sum\limits_{j = \max(0, t)}^{m}\frac{\left(p^{k}_- p^{k}_+ e^{-p^{k}_z} \right)^j (n+j)!}{j!(j-t)!(m-j)!}
\end{aligned}
\end{equation} 
(and where the summation index $t$ should not be confused with the time parameter). 
This generic expression of four-mode squeezed states expanded in the Fock basis is one of the main results of this paper. It features an infinite tower of entangled states, characterised by non-trivial numbers of excitations. More precisely, there are 6 indices being summed over, which can be interpreted as follows. The indices $i$ and $j$ are internal and can be resumed within each Fock state separately, their physical interpretation will be made clearer below.\footnote{The internal indices $i$ and $j$ can be resumed in terms of the hypergeometric function $\prescript{}{2}{F_1}$, leading to
\bea
\begin{aligned}
\label{eq:resum}
			c_{k}(n,m,s,t) = & \frac{e^{-2i \left[ \varphi^{k}_3 + \theta^{k}_3(n+m+1) \right]}}{\cosh{r^{k}_1} \cosh{r^{k}_2}} (-1)^{n+m} e^{p^{k}_z(m-n)} \tanh^n\left( r^{k}_1\right) \tanh^m\left( r^{k}_2 \right) \left(p^{k}_- p^{k}_+ e^{-p^{k}_z} \right)^{\max{(0,s)} + \max{(0,t)}} \\ 
			&   (ip^{k}_+)^{-s-t} \frac{m!}{n!}\sqrt{\frac{(m-s)!(m-t)!}{(n+s)!(n+t)!}} \frac{1}{\vert s\vert!\vert t\vert!} \frac{[n+\max{(0,s)}]![n+\max{(0,t)}]!}{[m-\max{(0,s)}]![m-\max{(0,t)}]!} \\
			& \prescript{}{2}{F_1}\left[ -m+\max{(0,s)}, 1+n+\max{(0,s)}, 1-s+2\max{(0,s)}, -p^{k}_- p^{k}_+ e^{-p^{k}_z} \right] \\
			& \prescript{}{2}{F_1}\left[-m+\max{(0,t)}, 1+n+\max{(0,t)}, 1-t+2\max{(0,t)}, -p^{k}_- p^{k}_+ e^{-p^{k}_z}\right]\, .
\end{aligned}
\eea
} 
The indices $n$ and $m$ are related to the creation of entangled pairs of excitations with opposite wavevector inside each sector separately, and arise from the squeezing operation~\eqref{eq:squeezing}. The indices $s$ and $t$ are related to the transfer of excitations between the two sectors, with wavevector $\vec{k}$ and $-\vec{k}$ respectively. They arise from the transferring operation~\eqref{eq:transfer} and entangle the two sectors.
They are negative when excitations go from the first sector to the second sector, positive otherwise, and their bounds guarantee that the number of excitations remains non negative. One should also note that $c_{k}(m,n,s,t)$ is invariant when swapping $n,m$ and $s,t$, \ie $\vec{k}$ and $-\vec{k}$, which is a consequence of statistical isotropy. The form of the expansion~\eqref{eq:vacevolfinal} is therefore a direct consequence of the symmetries of the problem, although the precise expression of $c_{k}(m,n,s,t)$ given in \Eq{eq:ampl} required a non-trivial calculation.

Let us stress that only seven out of the ten $\mathrm{Sp}(4,\mathbb{R})$ squeezing and rotation parameters enter \Eq{eq:ampl}. This is because, as explained in \App{sec:vacappendix}, the rotation $\widehat{\mathcal{R}}_{\vec{k}}(\boldsymbol{\varphi}_{k})$ only adds a global phase to the initial vacuum state, since it is invariant under rotations (in practice, this global phase is irrelevant, which reduces the number of effective parameters down to six). As a consequence, $q^{k}_-$, $q^{k}_+$ and $q^{k}_z$ are ineffective (had we started from a different initial state, those parameters would have entered the final result, see footnote~\ref{footnote:alpha:vacuua}). In practice, once the Hamiltonian of the system is specified, the dynamics can be integrated, which yields the seven relevant squeezing and rotation parameters, $r^{k}_1$, $r^{k}_2$, $\varphi^{k}_3$, $\theta^{k}_3$, $p^{k}_-$, $p^{k}_+$ and $p^{k}_z$, and thus fully determines the quantum state of the system at any time.

As a consistency check, one may verify that in the decoupled limit, the product of two two-mode squeezed states is recovered. Setting the two mixing parameters $\theta_5^k$ and $\theta_6^k$ to zero, \Eq{eq:thetavarphi} leads to $\theta^k=\vert \theta_4^k\vert$ and $\tau_k=0$, so \Eq{eq:pz:p-:p+} gives $p^{k}_- = p^{k}_+ = 0$ and $p^{k}_z = 2i\theta^{k}_4$. In \Eq{eq:ampl}, the overall factor $(p^k_-)^{i+j}$ is therefore non vanishing only when $i+j=0$, \ie $i=j=0$ since $i$ and $j$ are non negative. Then, the overall factor $(p^{k}_+)^{i+j-s-t}$ is non vanishing only when $s+t=0$, but since $s\leq i = 0$ and $t\leq j = 0$, this implies that $s=t=0$, so only one term remains. This gives rise to
\bea
\label{eq:Prod:tmss}
\ket{\cancel{0}_{\vec{k}}(t)} &=& \ee^{-2i \left(\theta_3^k+\varphi_3^k\right)}
\sum_{n=0}^\infty  
\overbrace{\frac{(-1)^n\ee^{-2in(\theta_3^k+\theta_4^k)}}{\cosh r_1^k}\tanh^n(r_1^k)}^{c_{1,k}(n)} \left\vert n^{(1)}_{\vec{k}}  ,n^{(1)}_{-\vec{k}}\right\rangle
\nonumber \\  & & \times
\sum_{m=0}^\infty  \underbrace{\frac{(-1)^m\ee^{-2im(\theta_3^k-\theta_4^k)}}{\cosh r_2^k}\tanh^m (r_2^k) }_{c_{2,k}(m)}\left\vert   m^{(2)}_{\vec{k}}  ,m^{(2)}_{-\vec{k}}\right\rangle ,
\eea
which defines the two-mode squeezed states coefficients $c_{i,k}(n)$.
Up to a global irrelevant phase, this is indeed the product of two, uncoupled and disentangled, two-mode squeezed states~\cite{Grain:2019vnq}.
\subsubsection*{Expansion around the uncoupled limit}
In order to gain some physical insight into the structure of the four-mode squeezed states, and having in mind possible comparisons with perturbative techniques that expand in the amplitude of the interaction Hamiltonian (\ie in the terms in the Hamiltonian that mix the two sectors), let us now expand the evolved vacuum state~\eqref{eq:ampl} around the uncoupled limit~\eqref{eq:Prod:tmss}. From \Eq{eq:thetavarphi}, one can see that the coupling parameters, \ie $\theta_5^k$, $\varphi_5^k$, $\theta_6^k$ and $\varphi_6^k$, only appear through the combination $\tau_{k} = - (\theta^{k}_6 -i \theta^{k}_5) = \vert\tau_{k}\vert e^{i \,\mathrm{arg}(\tau_{k})}$, since $\theta^k=\sqrt{(\theta_4^k)^2+\vert\tau_{k}\vert^2}$, and given that $\varphi_5^k$ and $\varphi_6^k$ are irrelevant, as explained above. As a consequence, an expansion around the uncoupled limit is an expansion in $\vert\tau_{k}\vert $.
Upon expanding \Eq{eq:ampl} up to quadratic order in $\vert\tau_{k}\vert $, one obtains
\begin{align}
\label{eq:perturbstate}
\ket{\cancel{0}_{\vec{k}}(t)} =& \sum\limits_{n,m = 0}^\infty  c_{1,k}(n)c_{2,k}(m) \Bigg\{
\left| n^{(1)}_{\vec{k}},n^{(1)}_{-\vec{k}}, m^{(2)}_{\vec{k}}, m^{(2)}_{-\vec{k}}\right> 
\nonumber \\ & 
+\left|\tau_{k}\right| \bigg[	\mathcal{F}_k(n,m+1)\left|1_{1\rightarrow2}\right> 
		-\mathcal{F}_k^*(n+1,m)\left|1_{2\rightarrow1}\right> 
		\bigg]\nonumber \\
		 &+\frac{\left|\tau_{k}\right|^2}{2} \bigg[
		\mathcal{F}_k(n,m+2)\mathcal{F}_k(n-1,m+1)\left|2_{1\rightarrow2} \right> 
		+\mathcal{F}^*_k(n+2,m)\mathcal{F}^*_k(n+1,m-1) \left|2_{2\rightarrow1} \right>
		\nonumber \\ &
		+2\mathcal{F}_k^2(n,m+1) \left|1\text{-}1_{1\to2}\right>
		+2 (\mathcal{F}_k^*)^2(n+1,m) \left| 1\text{-}1 _{2\to1}\right>
		 \nonumber \\
		 &-2\mathcal{F}_k(n,m) \mathcal{F}_k^*(n+1,m+1) \left|2_{1\leftrightarrow 2} \right> 
		+2  \mathcal{G}_{k}(n,m) \left|n^{(1)}_{\vec{k}},n^{(1)}_{-\vec{k}}, m^{(2)}_{\vec{k}}, m^{(2)}_{-\vec{k}}\right> 
		\bigg]
		\Bigg\},
\end{align}
where the Fock states that appear in this expansion are labeled by the number of particles being transferred from one sector to the other (while respecting statistical isotropy), and are given by
\begin{align}
	 \left|1_{1\rightarrow2}\right> \simeq &	\left|(n-1)^{(1)}_{\vec{k}},n^{(1)}_{-\vec{k}}, (m+1)^{(2)}_{\vec{k}}, m^{(2)}_{-\vec{k}}\right> + \left|n^{(1)}_{\vec{k}},(n-1)^{(1)}_{-\vec{k}}, m^{(2)}_{\vec{k}}, (m+1)^{(2)}_{-\vec{k}}\right>\nonumber  \\
	 \left|1_{2\rightarrow1}\right> = &\left|(n+1)^{(1)}_{\vec{k}},n^{(1)}_{-\vec{k}}, (m-1)^{(2)}_{\vec{k}}, m^{(2)}_{-\vec{k}}\right> + \left|n^{(1)}_{\vec{k}},(n+1)^{(1)}_{-\vec{k}}, m^{(2)}_{\vec{k}}, (m-1)^{(2)}_{-\vec{k}}\right> \nonumber \\
	 \left|2_{1\rightarrow2} \right> = & \left|(n-2)^{(1)}_{\vec{k}},n^{(1)}_{-\vec{k}}, (m+2)^{(2)}_{\vec{k}}, m^{(2)}_{-\vec{k}}\right>+ \left|n^{(1)}_{\vec{k}},(n-2)^{(1)}_{-\vec{k}}, m^{(2)}_{\vec{k}}, (m+2)^{(2)}_{-\vec{k}}\right> \nonumber  \\
	 \left|2_{2\rightarrow1} \right> = & \left|(n+2)^{(1)}_{\vec{k}},n^{(1)}_{-\vec{k}}, (m-2)^{(2)}_{\vec{k}}, m^{(2)}_{-\vec{k}}\right> + \left|n^{(1)}_{\vec{k}},(n+2)^{(1)}_{-\vec{k}}, m^{(2)}_{\vec{k}}, (m-2)^{(2)}_{-\vec{k}}\right>\nonumber  \\
	 \left|2_{1\leftrightarrow 2}\right> = & \left|(n+1)^{(1)}_{\vec{k}},(n-1)^{(1)}_{-\vec{k}}, (m-1)^{(2)}_{\vec{k}}, (m+1)^{(2)}_{-\vec{k}}\right> 
	  \nonumber \\ &
	+ \left|(n-1)^{(1)}_{\vec{k}},(n+1)^{(1)}_{-\vec{k}}, (m+1)^{(2)}_{\vec{k}}, (m-1)^{(2)}_{-\vec{k}}\right> \nonumber \\
	 \left| 1\text{-}1 _{1\to2}\right> = &  \left|(n-1)^{(1)}_{\vec{k}},(n-1)^{(1)}_{-\vec{k}}, (m+1)^{(2)}_{\vec{k}}, (m+1)^{(2)}_{-\vec{k}}\right>\nonumber \\
	 \left| 1\text{-}1 _{2\to1}\right> = &\left|(n+1)^{(1)}_{\vec{k}},(n+1)^{(1)}_{-\vec{k}}, (m-1)^{(2)}_{\vec{k}}, (m-1)^{(2)}_{-\vec{k}}\right>
\end{align}
with the weighting functions
\begin{align}
\label{eq:calF:def}
\mathcal{F}_k(n,m) &=  i \frac{\sin \theta^{k}_4}{\theta^{k}_4} \ee^{i[\theta^{k}_4 +\mathrm{arg}(\tau_{k})]}\sqrt{n m}\, ,\\
\mathcal{G}_{k}(n,m) &= -2\mathcal{F}_k(n+1,m) \mathcal{F}^*_k(n+1,m) - i(n-m)\frac{\theta^{k}_4 - e^{i{\theta^{k}_4}}\sin \theta^{k}_4}{(\theta^{k}_4)^2}.
\label{eq:calG:def}
\end{align}
These expressions can be interpreted as follows. At linear order in $\tau$, the only effect of the interaction is to add contributions from states where one particle has been exchanged between the two sectors [those are displayed in the second line of \Eq{eq:perturbstate}]. The amplitude of these additional states is controlled by $\mathcal{F}_k$, which thus measures the rate at which particles transfer. It involves the phase $\theta_4^k$ through a function of order one, and the product of the numbers of particles in the two sectors, which mostly determines its amplitude. At quadratic order in $\tau$, the new states that appear in the expansion are obtained by exchanging two particles: either two particles with the same wavevector transfer from one sector to the other [third line of  \Eq{eq:perturbstate}], or two particles with opposite wavevector transfer from one sector to the other [fourth line of  \Eq{eq:perturbstate}], or two particles with opposite wavevector and from opposite sectors change sector [first term in the fifth line of \Eq{eq:perturbstate}], or, finally, one particle changes sector and then moves back to its original sector [last term in the fifth line of \Eq{eq:perturbstate}]. The amplitude of those states are now controlled by squared powers of $\mathcal{F}_k$, the only exception being the last state, controlled by $\mathcal{G}_k$, which also involves the second term in \Eq{eq:calG:def}. This can be interpreted as follows. When a particle transiently visits the opposite sector and is then reinstated, its journey to the other side modifies its state if the two sectors evolve differently. This is the case if $n\neq m$, which is why the second term in \Eq{eq:calG:def} is controlled by $n-m$.

One concludes that an expansion in the amplitude of the interaction, around the uncoupled limit, is essentially an expansion in the number of particles being exchanged: at order $\vert \tau_k\vert^p$, $p$ particles are transferred, which allows one to predict the form of the new states that appear in the expansion (\eg $\left|p_{1\rightarrow2} \right> = \left|(n- p )^{(1)}_{\vec{k}},n^{(1)}_{-\vec{k}}, (m+ p )^{(2)}_{\vec{k}}, m^{(2)}_{-\vec{k}}\right> + \left|n^{(1)}_{\vec{k}},(n- p )^{(1)}_{-\vec{k}}, m^{(2)}_{\vec{k}}, (m+ p )^{(2)}_{-\vec{k}}\right> $, \etc). In particular, one can see that the diagonal elements, \ie the amplitudes associated to the states of the form $\left| n^{(1)}_{\vec{k}},n^{(1)}_{-\vec{k}}, m^{(2)}_{\vec{k}}, m^{(2)}_{-\vec{k}}\right>$, where, within each sector, there are the same number of excitations in each wavevector, receive contributions from even powers of $\vert \tau_k\vert$ only. More generally, one can show that $c_k(n,m,s,t)=\mathcal{O}(\vert \tau\vert^{\vert s \vert + \vert t \vert})$, and that $c_k(n,m,s,t)$ only receives contributions of order $\vert\tau\vert^{\vert s \vert + \vert t \vert +2q}$, where $q$ is a positive integer number counting the number of particles being sent away and then sent back to their original sector. This is described by the sums over $i$ and $j$ in \Eq{eq:ampl}, where $i$ counts the number of quanta travelling from sector 1 to sector 2 and travelling back, and $j$ counts the number of quanta travelling along the opposite journey. This completes the physical interpretation of all six summation indices appearing in \Eq{eq:ampl}.
\subsection{Phase-space representation}
\label{subsec:wigner}
Although the quantum state derived in the Fock basis in \Sec{sec:vacevol} is enough to fully characterise the system, and in spite of the clear physical interpretation it yields, the seemingly complicated structure of \Eq{eq:ampl} may call for alternative, simpler representations of the state. In this section, we provide such an alternative description by means of the Wigner function~\cite{PhysRevA.36.3868, SIMON1987223, doi:10.1119/1.2957889}.
\subsubsection*{Gaussian state}
As we will see, the reason why the Wigner function is a convenient tool is that, the dynamics being linear, the evolved vacuum state is a Gaussian state, which is fully characterised by its two-point function
\begin{eqnarray}
\label{eq:fullsigma}
\bs{\Sigma}_{\vec{k}, \vec{q}}(t) = \bra{\cancel{0}} \widehat{\boldsymbol{z}}_{\vec{k}}(t)\widehat{\boldsymbol{z}}^\dag_{\vec{q}}(t) \ket{\cancel{0}}\, ,
\end{eqnarray}
which here is expressed in the Heisenberg picture. Using the statistical isotropy and homogeneity we have already invoked, different Fourier modes are uncoupled and one has
\begin{eqnarray}
\bs{\Sigma}_{\vec{k}, \vec{q}}(t) = \bs{\Sigma}_{k}(t) \delta^3(\vec{k} - \vec{q})\, .
\end{eqnarray}
Our first goal is to relate $ \bs{\Sigma}_{k}$ to the Bogolyubov coefficients introduced in \Sec{subsec:su22}, since those describe the dynamical evolution of the system. In the helicity basis~\eqref{eq:helpassquant}, \Eq{eq:fullsigma} gives rise to
\begin{eqnarray}
\label{eq:hel2pt}
\bs{\Sigma}_{k}(t) = \boldsymbol{D}^{-1}_k \bs{U}^\dag \bra{\cancel{0}_{\vec{k}}} \widehat{\boldsymbol{a}}_{\vec{k}}(t) \widehat{\boldsymbol{a}}^\dag_{\vec{k}}(t) \ket{\cancel{0}_{\vec{k}}}\bs{U} \boldsymbol{D}^{-1}_k\, ,
\end{eqnarray}
where the ladder operators evolve according to \Eq{eq:adyn},\footnote{In the quantum representation constructed in \Sec{subsec:quantumliealg}, the evolution of the ladder operators is rather given by $\widehat{\boldsymbol{a}}_{\vec{k}}(t) = \widehat{\mathcal{U}}^{\dag}_{\vec{k}}(t, t_\uin)\widehat{\boldsymbol{a}}_{\vec{k}}(t_\uin)  \widehat{\mathcal{U}}_{\vec{k}}(t, t_\uin)$, but one can show that this is strictly equivalent to \Eq{eq:adyn}. Indeed, upon writing the evolution operator $\widehat{\mathcal{U}}_{\vec{k}}$ in the fully factorised form derived in \Sec{subsec:evolop}, computing the commutators between the ten elementary operations and the $\widehat{\boldsymbol{a}}_{\vec{k}}(t_\uin)$ operators, and relating the Bogolyubov coefficients to the squeezing and rotation parameters using \Eqs{eq:bogolsqueez1}-\eqref{eq:bogolsqueez8}, one can check that the same result~\eqref{eq:P1}-\eqref{eq:P3} is obtained.\label{footnote:link:G:U}} \ie by means of the Green matrix $\bs{\mathcal{G}}_k(t,t_\uin)$. Since the Green matrix is an element of $\Spnew(4,\mathbb{R})$, it can be written down in terms of Bogolyubov coefficients as in \Eqs{eq:helbloc} and~\eqref{eq:Gdef2}. Moreover, the ordering of the ladder operators in \Eq{eq:helpassquant} is such that
\begin{eqnarray}
\bra{\cancel{0}} \widehat{\boldsymbol{a}}_{\vec{k}}(t_\uin) \widehat{\boldsymbol{a}}^\dag_{\vec{k}}(t_\uin) \ket{\cancel{0}} = \begin{pmatrix} 1 & 0 \\ 0 & 0\end{pmatrix} \otimes \boldsymbol{I}_2\, ,
\end{eqnarray} 
so one obtains
\begin{eqnarray}
\label{eq:twoptmatrix}
\bs{\Sigma}_{k}(t) = \underbrace{\begin{pmatrix}
		\bs{\mathrm{Cov}}^{(\phi \phi)}_{k} & \bs{\mathrm{Cov}}^{(\phi \pi)}_{k} \\
		\bs{\mathrm{Cov}}^{(\phi \pi)\mathrm{T}}_{k} & \bs{\mathrm{Cov}}^{(\pi \pi)}_{k}
	\end{pmatrix}
}_{\bs{\mathrm{Cov}}_{k}(t)} + \frac{i}{2} \bs{\Omega},
\end{eqnarray}
where the second term, $i\bs{\Omega}/2$, stems from the quantum commutator~\eqref{eq:Commutators:zk}, and 
which defines the covariance matrix $\bs{\mathrm{Cov}}_{k}(t)$. It is a real, symmetric, positive and semi-definite $4\times 4$ matrix, where $\bs{\mathrm{Cov}}^{(\phi \phi)}_{k}$, $\bs{\mathrm{Cov}}^{(\pi \pi)}_{k}$ and $\bs{\mathrm{Cov}}^{(\phi \pi)}_{k}$ are $2\times2$ matrices containing the position-position spectra, momentum-momentum spectra and position-momentum spectra, reading
\begin{align}
\label{eq:P1}
&  \bs{\mathrm{Cov}}^{(\phi \phi)}_{k}  = \frac{1}{k}\begin{pmatrix}
\frac{1}{2}\left(|\alpha_{11} + \beta^*_{11}|^2 + |\alpha_{12} + \beta^*_{12}|^2\right) & \textrm{Re}\left[\beta_{11}\left(\beta_{21}^* + \alpha_{21}\right) + \beta_{12}\left(\beta_{22}^* + \alpha_{22}\right)\right]  \\ 
\textrm{Re}\left[\beta_{11}\left(\beta_{21}^* + \alpha_{21}\right) + \beta_{12}\left(\beta_{22}^* + \alpha_{22}\right)\right] & \frac{1}{2}\left(|\alpha_{21} + \beta^*_{21}|^2 + |\alpha_{22} + \beta^*_{12}|^2\right)
\end{pmatrix}\\
\label{eq:P2} 
& \bs{\mathrm{Cov}}^{(\pi \pi)}_{k}  = k\begin{pmatrix}
\frac{1}{2}\left(|\alpha_{11} - \beta^*_{11}|^2 + |\alpha_{12} - \beta^*_{12}|^2\right) & \textrm{Re}\left[\beta_{11}\left(\beta_{21}^* - \alpha_{21}\right) + \beta_{12}\left(\beta_{22}^* - \alpha_{22}\right)\right] \\ 
\textrm{Re}\left[\beta_{11}\left(\beta_{21}^* - \alpha_{21}\right) + \beta_{12}\left(\beta_{22}^* - \alpha_{22}\right)\right] & \frac{1}{2}\left(|\alpha_{21} - \beta^*_{21}|^2 + |\alpha_{22} - \beta^*_{12}|^2\right)
\end{pmatrix}\\
\label{eq:P3}
&  \bs{\mathrm{Cov}}^{(\phi \pi)}_{k}  = \begin{pmatrix}
	\textrm{Im}\left(\alpha_{11}\beta_{11} + \alpha_{12}\beta_{12}\right)  &  - \textrm{Im}\left[\beta_{11}\left(\beta_{21}^* - \alpha_{21}\right) + \beta_{12}\left(\beta_{22}^* - \alpha_{22}\right)\right] \\ 
	\textrm{Im}\left[\beta_{11}\left(\beta_{21}^* + \alpha_{21}\right) + \beta_{12}\left(\beta_{22}^* + \alpha_{22}\right)\right] & \textrm{Im}\left(\alpha_{21}\beta_{21} + \alpha_{22}\beta_{22}\right) 
\end{pmatrix}.
\end{align}
An expression of $\bs{\mathrm{Cov}}_{k}(t)$ in terms of the squeezing and rotation parameters can also be obtained from the Bloch-Messiah decomposition of $\bs{\mathcal{G}}_k(t,t_\uin)$ given in \Eq{eq:bmhel}, where one notes that the squeezing and rotation parameters entering that decomposition are the same as those appearing in the evolution operator $\widehat{\mathcal{U}}_{\vec{k}}(t, t_\uin)$ given in \Eq{eq:BMrappel}, see footnote~\ref{footnote:link:G:U}. In order to write the result in a compact form, we introduce the complex parameter 
\begin{align}\label{eq:tau1}
    \widetilde{\tau}_k\equiv \ee^{-\frac{p_z}{2}}=\cos(\theta^k)+i\theta^k_4\mathrm{sinc}(\theta^k),
\end{align}
in terms of which
the position-position power spectra read
\begin{align}
\label{eq:P11} \mathrm{Cov}^{(\phi \phi)}_{11,k} =& \frac{1}{2k} \Bigg\{ \left|\widetilde{\tau}_k\right|^2\bigg[\cosh(2 r^{k}_1) + \cos(2 \theta^{k}_3+2 \mathrm{arg}\widetilde{\tau}_k) \sinh(2 r^{k}_1)\bigg]  \nonumber \\ &
	+ \mathrm{sinc}^2(\theta^k) \left|\tau_k\right|^2\bigg[\cosh(2 r^{k}_2) - \cos(2 \theta^{k}_3+2 \mathrm{arg}\tau_k) \sinh(2 r^{k}_2)\bigg]\Bigg\},\\
\mathrm{Cov}^{(\phi \phi)}_{22,k} =& \frac{1}{2k} \Bigg\{ \left|\widetilde{\tau}_k\right|^2\bigg[\cosh(2 r^{k}_2) + \cos(2 \theta^{k}_3-2 \mathrm{arg}\widetilde{\tau}_k) \sinh(2 r^{k}_2)\bigg]  \nonumber \\ &
	+ \mathrm{sinc}^2(\theta^k)\left|\tau_k\right|^2\bigg[\cosh(2 r^{k}_1) - \cos(2 \theta^{k}_3-2 \mathrm{arg}\tau_k) \sinh(2 r^{k}_1)\bigg]\Bigg\},\\
\mathrm{Cov}^{(\phi \phi)}_{12, k}  = & \frac{1}{2k} \mathrm{sinc}(\theta^k) \left|\widetilde{\tau}_k\right| \left|\tau_k\right| \Bigg\{  \sin( \mathrm{arg}\tau_k +\mathrm{arg}\widetilde{\tau}_k) \left[\cosh(2 r^{k}_2) - \cosh(2 r^{k}_1)\right]     \nonumber \\
		& + \sin( 2 \theta^{k}_3 - \mathrm{arg}\tau_k +\mathrm{arg}\widetilde{\tau}_k)  \sinh(2 r^{k}_1) 
		+\sin( 2 \theta^{k}_3 + \mathrm{arg}\tau_k -\mathrm{arg}\widetilde{\tau}_k)  \sinh(2 r^{k}_2) \Bigg\} .
\end{align}
The momentum-momentum power spectra are given by
\begin{align}
\label{eq:P33}		
\mathrm{Cov}^{(\pi \pi)}_{11,k} = & \frac{k}{2} \Bigg\{ \left|\widetilde{\tau}_k\right|^2\bigg[\cosh(2 r^{k}_1) - \cos(2 \theta^{k}_3+2 \mathrm{arg}\widetilde{\tau}_k) \sinh(2 r^{k}_1)\bigg]  \nonumber \\ &
	+ \mathrm{sinc}^2(\theta^k)\left|\tau_k\right|^2\bigg[\cosh(2 r^{k}_2) + \cos(2 \theta^{k}_3+2 \mathrm{arg}\tau_k) \sinh(2 r^{k}_2)\bigg]\Bigg\}, \\
		\mathrm{Cov}^{(\pi \pi)}_{22,k} = & \frac{k}{2} \Bigg\{ \left|\widetilde{\tau}_k\right|^2\bigg[\cosh(2 r^{k}_2) - \cos(2 \theta^{k}_3-2 \mathrm{arg}\widetilde{\tau}_k) \sinh(2 r^{k}_2)\bigg]  \nonumber \\ &
	+ \mathrm{sinc}^2(\theta^k)\left|\tau_k\right|^2\bigg[\cosh(2 r^{k}_1) + \cos(2 \theta^{k}_3-2 \mathrm{arg}\tau_k) \sinh(2 r^{k}_1)\bigg]\Bigg\},\\
\mathrm{Cov}^{(\pi \pi)}_{12, k}  = & \frac{k}{2} \mathrm{sinc}(\theta^k) \left|\widetilde{\tau}_k\right| \left|\tau_k\right| \Bigg\{  \sin( \mathrm{arg}\tau_k +\mathrm{arg}\widetilde{\tau}_k) \left[\cosh(2 r^{k}_2) - \cosh(2 r^{k}_1)\right]     \nonumber \\
		& - \sin( 2 \theta^{k}_3 - \mathrm{arg}\tau_k +\mathrm{arg}\widetilde{\tau}_k)  \sinh(2 r^{k}_1) 
		 - \sin( 2 \theta^{k}_3 + \mathrm{arg}\tau_k -\mathrm{arg}\widetilde{\tau}_k)  \sinh(2 r^{k}_2) \Bigg\}, 
\end{align}	
and the position-momentum power spectra by
\begin{align}
\label{eq:P13}		
\mathrm{Cov}^{(\phi \pi)}_{11,k} = & \frac{1}{2} \Bigg[ -\left|\widetilde{\tau}_k\right|^2\sin(2 \theta^{k}_3+2 \mathrm{arg}\widetilde{\tau}_k)\sinh(2 r^{k}_1) \nonumber  \\
	& +\mathrm{sinc}^2(\theta^k)\left|\tau_k\right|^2\sin(2 \theta^{k}_3+2   \mathrm{arg}\tau_k)\sinh(2 r^{k}_2)\Bigg],\\
	 \mathrm{Cov}^{(\phi \pi)}_{22,k} = & \frac{1}{2} \Bigg[ -\left|\widetilde{\tau}_k\right|^2\sin(2 \theta^{k}_3-2 \mathrm{arg}\widetilde{\tau}_k)\sinh(2 r^{k}_2) \nonumber  \\ & +\mathrm{sinc}^2(\theta^k)\left|\tau_k\right|^2\sin(2 \theta^{k}_3-2 \mathrm{arg}\tau_k)\sinh(2 r^{k}_1)\Bigg],\\
	\mathrm{Cov}^{(\phi \pi)}_{12, k}  = & \frac{1}{2} \mathrm{sinc}(\theta^k) \left|\widetilde{\tau}_k\right| \left|\tau_k\right| \Bigg\{  \cos( \mathrm{arg}\tau_k +\mathrm{arg}\widetilde{\tau}_k) \left[\cosh(2 r^{k}_1) - \cosh(2 r^{k}_2)\right]     \nonumber \\
		& + \cos( 2 \theta^{k}_3 - \mathrm{arg}\tau_k +\mathrm{arg}\widetilde{\tau}_k)  \sinh(2 r^{k}_1)  
		+ \cos( 2 \theta^{k}_3 + \mathrm{arg}\tau_k -\mathrm{arg}\widetilde{\tau}_k)  \sinh(2 r^{k}_2) \Bigg\}, \\
	\mathrm{Cov}^{(\phi \pi)}_{21, k} = & \frac{1}{2} \mathrm{sinc}(\theta^k) \left|\widetilde{\tau}_k\right| \left|\tau_k\right| \Bigg\{ \cos( \mathrm{arg}\tau_k +\mathrm{arg}\widetilde{\tau}_k) \left[\cosh(2 r^{k}_2) - \cosh(2 r^{k}_1)\right]     \nonumber \\
	&	+ \cos( 2 \theta^{k}_3 - \mathrm{arg}\tau_k +\mathrm{arg}\widetilde{\tau}_k)  \sinh(2 r^{k}_1)
		+ \cos( 2 \theta^{k}_3 + \mathrm{arg}\tau_k -\mathrm{arg}\widetilde{\tau}_k)  \sinh(2 r^{k}_2) \Bigg\}, 
\end{align}	
where the remaining spectra are obtained from the symmetry of $\bs{\mathrm{Cov}}_{k}(t)$. We stress that, contrary to what was done around \Eq{eq:perturbstate}, there is no expansion in $\tau_k$ in these expressions. In practice, solving the dynamics of the system yields the Bogolyubov coefficients, or equivalently the squeezing and rotation parameters, from which the covariance matrix can be obtained from the above expressions. Finally, let us note that in the absence of interaction, $\widetilde{\tau}_k=e^{i\theta^k_4}$ and $\tau_k=0$, and \Eqs{eq:P11}, \eqref{eq:P33} and \eqref{eq:P13} reduce to the power spectra describing a two-mode squeezed state (see \eg Eqs.~(6.18), (6.19) and (6.20) of \Refa{Grain:2019vnq}). This allows us to highlight the great similarity in the structure of the power spectra obtained from the two-mode and the four-mode squeezed states. For instance, in the expression of the diagonal elements of $\bs{\mathrm{Cov}}^{(\phi \phi)}_{k}$, $\bs{\mathrm{Cov}}^{(\pi \pi)}_{k}$ and $\bs{\mathrm{Cov}}^{(\phi \pi)}_{k}$, the first line is given by $|\tilde{\tau}_k|^2$ multiplied the power spectrum of a two-mode squeezed state in the sector of interest (position-position, momentum-momentum and position-momentum respectively). Since $|\tilde{\tau}_k|^2\leq1$, this parameter can be interpreted as describing a loss of power in a given sector induced by its couplings to the other sector. On the contrary, the second line in these expressions is given by $|\tau_k|^2$ multiplied by the power spectrum of a two-mode squeezed state in the other sector, and thus corresponds to an increase of power in a given sector coming from the opposite sector.
\subsubsection*{Wigner function}
Let us now introduce the Wigner function. By inverting \Eq{eq:a1:a2}, one can see that the field position and momentum operators, $\widehat{\phi}_{i,\vec{k}}$ and $\widehat{\pi}_{i,\vec{k}}$, involve creation and annihilation operators with opposite wavevectors. For convenience, it is useful to treat the two sectors $\vec{k}$ and $-\vec{k}$ separately, and to introduce the new position and momentum variables~\cite{Martin:2015qta}
\bea
\label{eq:posWigner}	
\widehat{\mathfrak{q}}_{i, \vec{k}} &=& \frac{1}{\sqrt{2k}}\left(\widehat{a}_{i,\vec{k}} + \widehat{a}^{\dag}_{i,\vec{k}}\right),\\
\label{eq:momWigner}	
\widehat{\mathfrak{p}}_{i, \vec{k}} &=& -i \sqrt{\frac{k}{2}}\left(\widehat{a}_{i,\vec{k}} - \widehat{a}^{\dag}_{i,\vec{k}}\right),
\eea
with $i=1,2$, which do not mix opposite Fourier modes. When going from $\widehat{\phi}_{i,\vec{k}}$ and $\widehat{\pi}_{i,\vec{k}}$ to $\widehat{\mathfrak{q}}_{i, \vec{k}}$ and $\widehat{\mathfrak{p}}_{i, \vec{k}} $, one simply decomposes the four complex field variables, $\boldsymbol{z}_{\vec{k}} = (\phi_{1, \vec{k}}, \phi_{2, \vec{k}}, \pi_{1, \vec{k}}, \pi_{2, \vec{k}})^{\mathrm{T}}$ into eight real field variables, $\boldsymbol{\mathfrak{q}}_{\vec{k}} \equiv (\mathfrak{q}_{1, \vec{k}}, \mathfrak{q}_{1, -\vec{k}}, \mathfrak{q}_{2, \vec{k}}, \mathfrak{q}_{2, -\vec{k}}, \mathfrak{p}_{1, \vec{k}}, \mathfrak{p}_{1, -\vec{k}}, \mathfrak{p}_{2, \vec{k}}, \mathfrak{p}_{2, -\vec{k}})^{\mathrm{T}}$, according to
\begin{align}
\phi_{i,\vec{k}} &= \frac{1}{2}\left[\left(\mathfrak{q}_{i, \vec{k}} + \mathfrak{q}_{i, -\vec{k}}\right) + \frac{i}{k}\left(\mathfrak{p}_{i, \vec{k}} - \mathfrak{p}_{i, -\vec{k}}\right)\right],\\
\pi_{i,\vec{k}} &= \frac{1}{2i}\left[k\left(\mathfrak{q}_{i, \vec{k}} - \mathfrak{q}_{i, -\vec{k}}\right) + i\left(\mathfrak{p}_{i, \vec{k}} + \mathfrak{p}_{i, -\vec{k}}\right)\right].
\end{align}
This transformation can be readily inverted, and is canonical.\footnote{This can be seen by introducing $\widehat{\phi}^{\mathrm{R}}_{i,\vec{k}} = (\widehat{\phi}_{i,\vec{k}}+\widehat{\phi}^{\dagger}_{i,\vec{k}})/\sqrt{2}$, $\widehat{\phi}^{\mathrm{I}}_{i,\vec{k}} = (\widehat{\phi}_{i,\vec{k}}-\widehat{\phi}^{\dagger}_{i,\vec{k}})/(\sqrt{2}i)$ and similarly for $\widehat{\pi}^{\mathrm{R}}_{i,\vec{k}}$ and $\widehat{\pi}^{\mathrm{I}}_{i,\vec{k}}$, and by showing that the $8\times 8$ matrix that relates those variables to $\boldsymbol{\mathfrak{q}}_{\vec{k}}$ satisfies the symplectic relation~\eqref{eq:sympdef} in 8 dimensions.} The helicity basis can be described by the eight-dimensional version of \Eq{eq:helpassquant}, \ie $\widehat{\boldsymbol{\mathfrak{a}}}_{\vec{k}} \equiv (\widehat{a}_{1,\vec{k}}$, $\widehat{a}_{1,-\vec{k}}$, $\widehat{a}_{2,\vec{k}}$, $\widehat{a}_{2,-\vec{k}}$, $\widehat{a}^{\dag}_{1,\vec{k}}$, $\widehat{a}^{\dag}_{1,-\vec{k}}$, $\widehat{a}^{\dag}_{2,\vec{k}}$, $\widehat{a}^{\dag}_{2,-\vec{k}})^{\mathrm{T}}$, which is related to the vector $\widehat{\boldsymbol{\mathfrak{q}}}_{\vec{k}}$ via
\bea
\label{eq:hel88}
\widehat{\boldsymbol{\mathfrak{a}}}_{\vec{k}} = \left[\left(\bs{U} \boldsymbol{D}_k\right) \otimes \boldsymbol{I}_2 \right]\widehat{\boldsymbol{\mathfrak{q}}}_{\vec{k}}\, ,
\eea
and whose commutators are given by
\bea
	\left[\widehat{\boldsymbol{\mathfrak{a}}}_{\vec{k}},\widehat{\boldsymbol{\mathfrak{a}}}_{\vec{q}}^\dag\right] = i (\bs{\mathcal{J}} \otimes \boldsymbol{I}_2 )\delta^3(\vec{k} - \vec{q}) .
\eea

The real variables $\boldsymbol{\mathfrak{q}}_{\vec{k}}$ can be used to introduce the Wigner-Weyl transform \cite{1927ZPhy...46....1W, 1946Phy....12..405G, 1949PCPS...45...99M}, which allows one to connect quantum operators $\widehat{O}_{\vec{k}}$ to classical functions in the phase space $\widetilde{\mathcal{O}}_{\vec{k}} (\boldsymbol{\mathfrak{q}}_{\vec{k}})$, according to
\begin{align}
\label{eq:Wigner-Weyl}
&\widetilde{\mathcal{O}}_{\vec{k}} (\boldsymbol{\mathfrak{q}}_{\vec{k}}) = \frac{1}{(2\pi)^4} \int_{\mathbb{R}^4} \dd x_1 \dd x_2 \dd y_1 \dd y_2 e^{- i \mathfrak{p}_{1, \vec{k}} x_1 - i \mathfrak{p}_{2, \vec{k}} x_2  - i \mathfrak{p}_{1, -\vec{k}} y_1 - i \mathfrak{p}_{2,- \vec{k}} y_2} \nonumber \\
&\left< \mathfrak{q}_{1, \vec{k}} + \frac{x_1}{2}, \mathfrak{q}_{2, \vec{k}} + \frac{x_2}{2}, \mathfrak{q}_{1,- \vec{k}} + \frac{y_1}{2}, \mathfrak{q}_{2, -\vec{k}} + \frac{y_2}{2} \right|\widehat{O}_{\vec{k}} \left| \mathfrak{q}_{1, \vec{k}} - \frac{x_1}{2}, \mathfrak{q}_{2, \vec{k}} - \frac{x_2}{2}, \mathfrak{q}_{1,- \vec{k}} - \frac{y_1}{2}, \mathfrak{q}_{2, -\vec{k}} - \frac{y_2}{2}\right> .
\end{align}

In general, the quantum state of a system can be equivalently described in terms of its density matrix $\widehat{\rho}(t) = \ket{\cancel{0}(t)} \bra{\cancel{0}(t)}=\prod_{\vec{k} \in \mathbb{R}^{3+}} \widehat{\rho}_{\vec{k}}(t)$, or in terms of its Wigner function $W( t) = \prod_{\vec{k} \in \mathbb{R}^{3+}}W_{\vec{k}}(\boldsymbol{\mathfrak{q}}_{\vec{k}}, t)$, which is the Wigner-Weyl transform of the density matrix~\cite{2008AmJPh..76..937C, 2019arXiv191111703S}.

The inverse Wigner-Weyl transform, which allows one to go from the Wigner function to the density matrix, is given in \App{sec:directproof}, see \Eq{eq:inverseWeyl}. The Wigner functions $W_{\vec{k}}(\bs{\mathfrak{q}}_{\vec{k}}, t)$ can be interpreted as \textit{quasi} distribution functions in the sense that expectation values of quantum operators can be expressed as
\bea
\label{eq:expectvalue}
\bra{\cancel{0}_{\vec{k}}(t)}\widehat{\mathcal{O}}_{\vec{k}}\ket{\cancel{0}_{\vec{k}}(t)} = \mathrm{Tr}\left[\widehat{\rho}_{\vec{k}}(t) \widehat{\mathcal{O}}_{\vec{k}}\right] = (2\pi)^4 \int \dd \boldsymbol{\mathfrak{q}}_{\vec{k}}~ \widetilde{\mathcal{O}}_{\vec{k}} (\boldsymbol{\mathfrak{q}}_{\vec{k}}) W_{\vec{k}}(\boldsymbol{\mathfrak{q}}_{\vec{k}},t),
\eea
where the $(2\pi)^4$ prefactor can be absorbed in the normalisation of the Wigner function if needed. For Gaussian states, as shown \eg in Appendix G of \Refa{Martin:2015qta}, the Wigner functions are Gaussian functions (hence the statement that the state is ``Gaussian'') and read
\bea
\label{eq:gaussianWigner}
	W_{\vec{k}}(\boldsymbol{\mathfrak{q}}_{\vec{k}}, t) = \frac{1}{(2\pi)^4 \sqrt{\det{\bs{\mathrm{Cov}}_{8\times8,k}}}}e^{-\frac{1}{2}\boldsymbol{\mathfrak{q}}_{\vec{k}}^{\mathrm{T}} \bs{\mathrm{Cov}}^{-1}_{8\times8,k} \boldsymbol{\mathfrak{q}}_{\vec{k}}}\, ,
\eea
where $\bs{\mathrm{Cov}}_{8\times8,k}$ is the $8\times8$ covariance matrix in the field basis $\boldsymbol{\mathfrak{q}}_{\vec{k}}$, defined through
\bea
	\bra{\cancel{0}} \widehat{\boldsymbol{\mathfrak{q}}}_{\vec{k}}(t)\widehat{\boldsymbol{\mathfrak{q}}}^\dag_{\vec{k}}(t) \ket{\cancel{0}} = \bs{\mathrm{Cov}}_{8\times8,k}(t) + \frac{i}{2} \bs{\Omega} \otimes \boldsymbol{I}_2\, .
\eea
It can be computed in a similar way as what was done around  \Eq{eq:hel2pt}, and \Eq{eq:hel88} gives rise to
\bea
\label{eq:covmat}
\bs{\mathrm{Cov}}_{8\times8,k}(t) = \bs{\mathrm{Cov}}_{k}(t) \otimes \boldsymbol{I}_2\, ,
\eea
where the inverse is simply given by $\bs{\mathrm{Cov}}^{-1}_{8\times8,k}= \bs{\mathrm{Cov}}_{k}^{-1} \otimes \boldsymbol{I}_2$, see footnote~\ref{footnote:Kronecker}.

The advantage of working with the Wigner functions $W_{\vec{k}}(\boldsymbol{\mathfrak{q}}_{\vec{k}}, t)$ rather than the quantum state or the density matrix in the Fock basis, lies in the simplicity of the Gaussian function~\eqref{eq:gaussianWigner}. It describes entirely the quantum state of the system from the knowledge of its two-point correlation functions, and can therefore greatly simplify some calculations, as will be made explicit in the next section. The drawback of this approach is that the entangled structure, otherwise easily interpretable from \Eq{eq:ampl}, is now hidden in the details of the power spectra of the system.
\section{Decoherence}
\label{sec:Decoherence}
Having determined in \Sec{sec:QuantumState} the quantum state of a linear two-field system, both in the Fock basis where one obtains direct products of four-mode squeezed states and at the level of the Wigner function, we now consider the case where one of the two fields is unobservable and can be traced over. This corresponds to situations where measurements can only be performed on the first field, dubbed the ``system'', while the second field, dubbed the ``environment'', cannot be directly accessed. When the two fields become entangled, this leads to the concept of quantum decoherence~\cite{Joos:1984uk}, and we choose to illustrate the usefulness of the various tools introduced above by studying this notion for the systems at hand. 
\subsection{Reduced density matrix}
\label{subsec:rhored}
Let $\widehat{O}=\widehat{O}_1\otimes \widehat{\mathbb{I}}_2$ be a quantum operator describing an observable of the first field only, where $ \widehat{\mathbb{I}}_2$ denotes the identity acting in the second-field sector (\ie the environment). Its expectation value is given by 
$\mathrm{Tr}(\widehat{\rho} \widehat{O}) = \mathrm{Tr}_1\mathrm{Tr}_2 [\widehat{\rho}(\widehat{O}_1\otimes \widehat{\mathbb{I}}_2)]=\mathrm{Tr}_1 [\mathrm{Tr}_2(\widehat{\rho}) \widehat{O}_1]$, see \Eq{eq:expectvalue}, where $\mathrm{Tr}_i$ denotes the trace over the degrees of freedom contained in the field $i$. The expectation value of $\widehat{O}$ can therefore be obtained from the reduced density matrices
\bea
\label{eq:traceout}
\widehat{\rho}_{\vec{k},\mathrm{red}} = \mathrm{Tr}_2 (\widehat{\rho}_{\vec{k}})
=\sum_{n,m = 0}^\infty \left[\boldsymbol{I}^{(1)}_2 \otimes \bra{n^{(2)}_{\vec{k}}, m^{(2)}_{-\vec{k}}}\right]\widehat{\rho}_{\vec{k}}\left[\boldsymbol{I}^{(1)}_2 \otimes\ket{n^{(2)}_{\vec{k}},m^{(2)}_{-\vec{k}}}\right]
\eea
by tracing over in the first-field sector (\ie the system). In \Eq{eq:traceout}, for explicitness, the trace over the environmental degrees of freedom has been expanded in the Fock basis. The reduced density matrix thus contains all accessible information about the system.

If the two fields are entangled, the reduced density matrix follows non-unitary evolution, and describes a mixed (as opposed to pure) state. This can be described by the so-called purity
\begin{eqnarray}
\label{eq:puritydef}
	\gamma_{\vec{k}} = \tr(\widehat{\rho}^2_{\vec{k},\mathrm{red}})\, .
\end{eqnarray}
For a pure state, $\rho^2=\rho$ and $\gamma=1$, while mixed states have $1/d\leq \gamma\leq 1$ in general, where $d$ is the dimension of the Fock space of the system (which is infinite in the present case) and the limit $\gamma\to 1/d$ corresponds to a maximally decohered state. Therefore, purity provides a measure of the information loss into the environment, hence of decoherence. Note that it is simply related to the linear entropy $S_{\vec{k},\mathrm{lin}} = 1 - \gamma_{\vec{k}}$, which itself provides a lower bound to the entanglement entropy $S_{\vec{k},\mathrm{ent}}=-\tr(\rho\ln\rho)$, and which, as the entanglement entropy, characterises one's ignorance about the state of a system.

For the system at hand, one can check that if the two fields are uncoupled, then the reduced density matrix is pure. Indeed, if the quantum state is given by \Eq{eq:Prod:tmss}, the reduced density matrix reads
\begin{align}
\widehat{\rho}_{\vec{k},\mathrm{red}} = & \tr_2 \left(
\sum_{n,m,\bar{n},\bar{m}=0}^\infty
c_{1,k}(n) c_{1,k}^*(\bar{n})  c_{2,k}(m) c_{2,k}^*(\bar{m}) 
\left\vert n^{(1)}_{\vec{k}}  ,n^{(1)}_{-\vec{k}}
m^{(2)}_{\vec{k}}  ,m^{(2)}_{-\vec{k}}\right\rangle
\left\langle \bar{n}^{(1)}_{\vec{k}}  ,\bar{n}^{(1)}_{-\vec{k}}
\bar{m}^{(2)}_{\vec{k}}  ,\bar{m}^{(2)}_{-\vec{k}}\right\vert \right)\nonumber\\
=& 
\sum_{m=0}^\infty \left\vert c_{2,k}(m) \right\vert^2
\sum_{n,\bar{n}=0}^\infty c_{1,k}(n) c_{1,k}^*(\bar{n}) 
\left\vert n^{(1)}_{\vec{k}}  ,n^{(1)}_{-\vec{k}}\right\rangle \left\langle \bar{n}^{(1)}_{\vec{k}}  ,\bar{n}^{(1)}_{-\vec{k}}\right\vert \, ,
\label{eq:trref}
\end{align}
where we have performed the trace in the Fock basis as in \Eq{eq:traceout}. From the expression of the $c_{i,k}(n)$ coefficients given in \Eq{eq:Prod:tmss}, one can easily check that $\sum_n |c_{i,k}(n)|^2=1$, hence the reduced density matrix is the one of a (pure) two-mode squeezed state, and using again the identity $\sum_n |c_{i,k}(n)|^2=1$, it has purity $\gamma_{\vec{k}}=1$.

In general, for the evolved vacuum state given in \Eq{eq:vacevolfinal}, the density matrix $\widehat{\rho}_{\vec{k}}(t)= \ket{\cancel{0}_{\vec{k}}(t)} \bra{\cancel{0}_{\vec{k}}(t)}$ is written down explicitly in \App{subsec:tracingrhored}, where it is shown that the tracing-out procedure of \Eq{eq:traceout} gives rise to
\begin{align}
\label{eq:rhoredfinal}
\widehat{\rho}_{\vec{k},\mathrm{red}} = \sum\limits_{n= 0}^\infty \sum\limits_{n' = -\infty}^\infty ~  \sum\limits_{s,t = - \min{(n,n')}}^\infty \Xi_{k}(n,n',s,t) \ket{(n+s)^{(1)}_{\vec{k}},(n+t)^{(1)}_{-\vec{k}}}\bra{(n'+s)^{(1)}_{\vec{k}},(n'+t)^{(1)}_{-\vec{k}}}
\end{align}
with 
\begin{eqnarray}
\label{eq:rhoredampl}
	\begin{aligned}
		\Xi_{k}(n,n',s,t) = \sum\limits_{m = \max{(0,s,t)}}^\infty ~  \sum\limits_{m' = - m}^{n'} c_{k}(n,m,s,t) c^*_{k}(n'-m',m+m',s+m',t+m')\, .
	\end{aligned} 
\end{eqnarray}
This expression differs from the uncoupled case~\eqref{eq:trref} due to the presence of the $s$ and $t$ indices, related to the transfer of excitations from one sector to the other. 
One can check that the invariance of the $c_k$ coefficients under exchanging $s$ and $t$, which was noted below \Eq{eq:ampl} as a manifestation of the statistical isotropy of the state, guarantees that the $\Xi_k$ coefficients are also invariant under swapping $s$ and $t$, hence the reduced state is also statistically isotropic. 

Let us also recall that if one performs a perturbative expansion in the interaction Hamiltonian, that is, as argued in \Sec{sec:vacevol}, an expansion in $\tau_k$, then the corrections to $c_{k}(n,m,s,t)$ are of order $\mathcal{O}(\vert\tau_{k}\vert^{|s|+|t|+2q})$, where $q$ is a non-negative integer number that stands for the number of particles that change sectors and then change back. As a consequence, the corrections to $\Xi_{k}(n,n',s,t)$ are of order $\mathcal{O}[\vert\tau_{k}\vert^{2(|s|+|t|+Q)}]$, where $Q=m'+q+q'$, where $q$ and $q'$ are associated with the two $c_k$ coefficients appearing in \Eq{eq:rhoredampl}. This implies that, while we saw that the diagonal elements of the full quantum state received even corrections in $\vert \tau_k\vert$ only, see the discussion at the very end of \Sec{sec:vacevol}, we have now showed that all the entries of the reduced density matrix receive even corrections  in $\vert \tau_k\vert$ only. As a consequence, the leading correction to observables performed on the system is always quadratic in the coupling constants, in agreement with the results of \Refa{Martin:2018zbe}.

From \Eq{eq:rhoredfinal}, the purity~\eqref{eq:puritydef} can also be obtained, and in \App{subsec:tracingrhored} it is shown that
\begin{eqnarray}\label{eq:purity}
	\begin{aligned}
\gamma_{\vec{k}}(t) =&\sum\limits_{n=0}^\infty ~
\sum\limits_{n', u =- \infty}^\infty ~
\sum\limits_{s,t = - \min(n,n')}^\infty 
  \Xi_{k}(n,n',s,t) \Xi_{k}(n'-u,n-u,s+u,t+u).
	\end{aligned}
\end{eqnarray}
Let us stress that the above formulas are not perturbative and allow one to compute the reduced density matrix and its purity up to arbitrary order in the interaction terms.
\subsubsection*{Small-coupling limit}
In order to gain more insight, one can however derive the leading-order result in the interaction parameter $\vert\tau_k\vert$. After a lengthy though straightforward calculation, the reduced density matrix derived from the perturbed evolved vacuum state~\eqref{eq:perturbstate} is given by
\begin{align}
\label{eq:rhoredperturb}
		&\kern-1em \widehat{\rho}_{\vec{k},\mathrm{red}}(t) = \sum\limits_{n,n'= 0}^\infty c_{1,k}(n) c_{1,k}^*(n') \Ket{n^{(1)}_{\vec{k}},n^{(1)}_{-\vec{k}}}\Bra{n^{\prime(1)}_{\vec{k}},n^{\prime(1)}_{-\vec{k}}} 
+ \left|\tau_{k}\right|^2 \sum\limits_{n,n',m= 0}^\infty c_{1,k}(n) c_{1,k}^*(n')  \nonumber \\
		\bigg\{
&\vert c_{2,k}(m)\vert^2 \mathcal{F}_k(n,m+1)\mathcal{F}^*_k(n',m+1) 
\nonumber \\ & \qquad \times
\left[ \Ket{(n-1)^{(1)}_{\vec{k}},n^{(1)}_{-\vec{k}}}\Bra{(n'-1)^{(1)}_{\vec{k}},n^{\prime(1)}_{-\vec{k}}} + \Ket{n^{(1)}_{\vec{k}},(n-1)^{(1)}_{-\vec{k}}}\Bra{n^{\prime(1)}_{\vec{k}},(n'-1)^{(1)}_{-\vec{k}}} \right] \nonumber\\
		+ & \vert c_{2,k}(m)\vert^2 \mathcal{F}^*_k(n+1,m)\mathcal{F}_k(n'+1,m)	 
\nonumber \\ & \qquad \times
\left[ \Ket{(n+1)^{(1)}_{\vec{k}},n^{(1)}_{-\vec{k}}}\Bra{(n'+1)^{(1)}_{\vec{k}},n^{\prime(1)}_{-\vec{k}}} + \Ket{n^{(1)}_{\vec{k}},(n+1)^{(1)}_{-\vec{k}}}\Bra{n^{\prime(1)}_{\vec{k}},(n'+1)^{(1)}_{-\vec{k}}} \right] \nonumber\\
		- & c_{2,k}(m) c_{2,k}^*(m+1)\mathcal{F}_k(n,m+1) \mathcal{F}_k(n'+1,m+1) 
		\nonumber \\ & \qquad \times
		\left[ \Ket{(n-1)^{(1)}_{\vec{k}},n^{(1)}_{-\vec{k}}}\Bra{n^{\prime(1)}_{\vec{k}},(n'+1)^{(1)}_{-\vec{k}}} + \Ket{n^{(1)}_{\vec{k}},(n-1)^{(1)}_{-\vec{k}}}\Bra{(n'+1)^{(1)}_{\vec{k}},n^{\prime(1)}_{-\vec{k}}} \right] \nonumber\\
		- &c_{2,k}(m) c_{2,k}^*(m-1)  \mathcal{F}^*_k(n+1,m)  \mathcal{F}^*_k(n',m) 
\nonumber \\ & \qquad \times
		\left[ \Ket{(n+1)^{(1)}_{\vec{k}},n^{(1)}_{-\vec{k}}}\Bra{n^{\prime(1)}_{\vec{k}},(n'-1)^{(1)}_{-\vec{k}}} + \Ket{n^{(1)}_{\vec{k}},(n+1)^{(1)}_{-\vec{k}}}\Bra{(n'-1)^{(1)}_{\vec{k}},n^{\prime(1)}_{-\vec{k}}} \right] \nonumber\\
		+ &c_{2,k}(m) c_{2,k}^*(m-1) 
		\left[\mathcal{F}^*_k(n',m)\right]^2\Ket{n^{(1)}_{\vec{k}},n^{(1)}_{-\vec{k}}}\Bra{(n'-1)^{(1)}_{\vec{k}},(n'-1)^{(1)}_{-\vec{k}}} \nonumber\\
		+ & c_{2,k}(m) c_{2,k}^*(m+1)  \left[\mathcal{F}_k(n,m+1)\right]^2 \Ket{(n-1)^{(1)}_{\vec{k}},(n-1)^{(1)}_{-\vec{k}}}\Bra{n^{\prime(1)}_{\vec{k}},n^{\prime(1)}_{-\vec{k}}} \nonumber\\
		+ &c_{2,k}(m) c_{2,k}^*(m+1) \left[\mathcal{F}_k(n'+1,m+1)\right]^2 \Ket{n^{(1)}_{\vec{k}},n^{(1)}_{-\vec{k}}}\Bra{(n'+1)^{(1)}_{\vec{k}},(n'+1)^{(1)}_{-\vec{k}}} \nonumber\\
		+ & c_{2,k}(m) c_{2,k}^*(m-1)    \left[\mathcal{F}_k^*(n+1,m)\right]^2  \Ket{(n+1)^{(1)}_{\vec{k}},(n+1)^{(1)}_{-\vec{k}}}\Bra{n^{\prime(1)}_{\vec{k}},n^{\prime(1)}_{-\vec{k}}} \nonumber\\
		+& \vert c_{2,k}(m)\vert^2\left[\mathcal{G}_k^*(n',m) + \mathcal{G}_k(n,m) \right]\Ket{n^{(1)}_{\vec{k}},n^{(1)}_{-\vec{k}}}\Bra{n^{\prime(1)}_{\vec{k}},n^{\prime(1)}_{-\vec{k}}} \bigg\}.
\end{align}
One can see that the leading correction in $\vert\tau_k\vert$ is indeed of quadratic order [despite the full state~\eqref{eq:perturbstate} having linear contributions], in agreement with the above discussion. This expression allows one to evaluate the purity~\eqref{eq:puritydef} and one obtains
\begin{align}
\label{eq:puritypert}
		\gamma_{\vec{k}}(t) = &1 + 4 \left|\tau_{k}\right|^2 \sum\limits_{n, m = 0}^\infty \Rea  \Big\{	\vert c_{1,k}(n) \vert^2 \vert c_{2,k}(m) \vert^2 \mathcal{G}_k(n,m)
\nonumber \\ &
		+2	c_{1,k}(n+1)c_{1,k}^*(n)  c_{2,k}(m)c_{2,k}^*(m+1) \left[\mathcal{F}_k(n+1,m+1)\right]^2 \Big\}\, ,
\end{align}
where the relation $\sum_n |c_{i,k}(n)|^2=1$ has been used. Making use of \Eqs{eq:calF:def} and~\eqref{eq:calG:def}, the sums appearing in the expression can be carried out explicitly, and in terms of the squeezing and rotation parameters, one obtains
\begin{align}
\label{eq:puritypert:squeezing}
		\gamma_{\vec{k}}= &1 -4  \left|\tau_{k}\right|^2 \mathrm{sinc}^2\left(\theta_4^k\right)
\left[ \sinh^2\left(r_1^k-r_2^k\right)+\cos^2\left(\theta_4^k -\arg\tau_k \right)\sinh\left(2r_1^k\right)\sinh\left(2r_2^k\right)\right] 		\, .
\end{align}
One can check that, as $\vert \tau_k\vert$ increases away from $0$, $\gamma_{\vec{k}}$ decreases away from $1$, since the term inside the squared braces in \Eq{eq:puritypert:squeezing} is always positive. One can also see that larger squeezing amplitudes $r_1^k$ and $r_2^k$ lead to more efficient decoherence, as usually encountered~\cite{Kiefer:1998qe}. It is finally worth pointing out that in the specific configuration where $r_1^k=r_2^k$ and $\theta_4^k=\arg\tau_k\pm \pi$, the leading-order correction to the purity vanishes.
\subsection{Reduced Wigner function}
\label{subsec:Wred}
As argued in \Sec{subsec:wigner}, a complementary (and sometimes simpler from a computational standpoint) tool to analyse the four-mode squeezed state is provided by the Wigner function. Let us first establish how the Wigner function of the reduced system can be obtained from the one of the full two-field setup.

We consider again an operator of the form $\widehat{O}_{\vec{k}}=\widehat{O}_{1,\vec{k}}\otimes \widehat{\mathbb{I}}_2$. From \Eq{eq:Wigner-Weyl}, its Wigner-Weyl transform is simply given by $\widetilde{O}_{\vec{k}} = \widetilde{O}_{1,\vec{k}} / (2\pi)^{2} $, where $\widetilde{O}_{1,\vec{k}}$ is the Wigner-Weyl transform of $\widehat{O}_{1,\vec{k}}$ within the first-field sector,
\begin{align}
		\widetilde{\mathcal{O}}_{1,\vec{k}} (\boldsymbol{\mathfrak{q}}_{1,\vec{k}}) = \int_{\mathbb{R}^2} \frac{\dd x_1 \dd y_1}{(2\pi)^2} \ee^{- i \mathfrak{p}_{1, \vec{k}} x_1  - i \mathfrak{p}_{1, -\vec{k}} y_1} \left< \mathfrak{q}_{1, \vec{k}} + \frac{x_1}{2},  \mathfrak{q}_{1,- \vec{k}} + \frac{y_1}{2} \right|\widehat{O}_{1,\vec{k}} \left| \mathfrak{q}_{1, \vec{k}} - \frac{x_1}{2}, \mathfrak{q}_{1,- \vec{k}} - \frac{y_1}{2} \right> ,
\end{align}
with $\boldsymbol{\mathfrak{q}}_{1,\vec{k}} \equiv (\mathfrak{q}_{1, \vec{k}}, \mathfrak{q}_{1, -\vec{k}}, \mathfrak{p}_{1, \vec{k}}, \mathfrak{p}_{1, -\vec{k}})^{\mathrm{T}}$. Plugging this result into \Eq{eq:expectvalue}, the expectation value of $\widehat{O}_{\vec{k}}$ is given by
\begin{eqnarray}
\label{eq:expectvaluered}
	\bra{\cancel{0}_{\vec{k}}(t)}\widehat{\mathcal{O}}_{\vec{k}}\ket{\cancel{0}_{\vec{k}}(t)} = (2\pi)^2 \int \dd \boldsymbol{\mathfrak{q}}_{1, \vec{k}}~ \widetilde{\mathcal{O}}_{1,\vec{k}} (\boldsymbol{\mathfrak{q}}_{1,\vec{k}}) W_{\vec{k},\mathrm{red}}(\boldsymbol{\mathfrak{q}}_{1,\vec{k}}, t),
\end{eqnarray}
where we have defined the reduced Wigner function 
\begin{eqnarray}\label{eq:Wredmarg}
	W_{\vec{k},\mathrm{red}}(\boldsymbol{\mathfrak{q}}_{1,\vec{k}}, t) =  \int_{\mathbb{R}^4}\dd^4\boldsymbol{\mathfrak{q}}_{2,\vec{k}}~W_{\vec{k}}(\boldsymbol{\mathfrak{q}}_{1,\vec{k}}, \boldsymbol{\mathfrak{q}}_{2,\vec{k}}, t).
\end{eqnarray}
Comparing \Eq{eq:expectvaluered} with \Eq{eq:expectvalue}, one can see that $W_{\vec{k},\mathrm{red}}$ can be used to compute quantum expectation values of observables in the first-field space. Therefore, it corresponds to the Wigner function in the reduced phase space (where the different powers of $2\pi$ come from the different dimensions of the phase spaces). In other words, $W_{\vec{k},\mathrm{red}}$ is the Wigner-Weyl transform of $\widehat{\rho}_{\vec{k},\mathrm{red}}(t)$, which is shown explicitly in \App{sec:directproof}.  This is why partial trace in the Hilbert space is equivalent to partial integration in the phase space.

If the Wigner function is Gaussian, this partial integration can be easily done, and marginalisation over a phase-space variable simply corresponds to removing the associated lines and columns in the covariance matrix. From \Eq{eq:gaussianWigner}, this implies that
\begin{eqnarray}
\label{eq:Wred}
W_{\vec{k},\mathrm{red}}(\boldsymbol{\mathfrak{q}}_{1,\vec{k}}, t) = \frac{1}{(2\pi)^2 \sqrt{\det{\bs{\mathrm{Cov}}_{\vec{k},\mathrm{red}}}}}e^{-\frac{1}{2}\boldsymbol{\mathfrak{q}}_{1,\vec{k}}^{\mathrm{T}} \bs{\mathrm{Cov}}^{-1}_{\vec{k},\mathrm{red}} \boldsymbol{\mathfrak{q}}_{1,\vec{k}}}\, ,
\end{eqnarray}
where $\bs{\mathrm{Cov}}_{\vec{k},\mathrm{red}}$ is obtained from \Eq{eq:covmat} by removing the lines and columns related to the second sector (\ie the third, the fourth, the seventh and the eighth lines and columns).

The purity of the reduced system can then be computed from the reduced Wigner function, by using the property of the Wigner-Weyl transform \cite{doi:10.1119/1.2957889}
\begin{eqnarray}
\label{eq:propWeyl}
\tr\left(\widehat{A}\widehat{B}\right) = (2\pi)^2\int_{\mathbb{R}^4} \dd^4\boldsymbol{\mathfrak{q}}_{1,\vec{k}} \widetilde{A}\widetilde{B}\, ,
\end{eqnarray}
where $\widehat{A}$ and $\widehat{B}$ are two quantum operators acting on the first-field system and $\widetilde{A}$ and $\widetilde{B}$ are their Wigner-Weyl transforms. With $\widehat{A}=\widehat{B}=\widehat{\rho}_{\vec{k},\mathrm{red}}$, \Eq{eq:propWeyl} gives rise to the following expression for the purity~\eqref{eq:puritydef},
\begin{eqnarray}
\gamma_{\vec{k}}(t) = (2\pi)^2\int_{\mathbb{R}^4} \dd^4\boldsymbol{\mathfrak{q}}_{1,\vec{k}}  W^2_{\vec{k},\mathrm{red}}(\boldsymbol{\mathfrak{q}}_{1,\vec{k}}, t).
\end{eqnarray}
Plugging \Eq{eq:Wred} into that formula, one obtains 
\begin{eqnarray}
\label{eq:purityfinal}
	\gamma_{\vec{k}}(t) = \left(16 \det{\bs{\mathrm{Cov}}_{\vec{k},\mathrm{red}}}\right)^{-1/2} = \frac{1}{4} \left[\mathrm{Cov}^{(\phi \phi)}_{11, k} \mathrm{Cov}^{(\pi \pi)}_{11, k} - \left(\mathrm{Cov}^{(\phi \pi)}_{11, k}\right)^2\right]^{-1}.
\end{eqnarray}
As a consequence, for a Gaussian state, the purity of the system can be directly evaluated from the knowledge of the power spectra in the observable sector~\cite{Serafini:2003ke}. More precisely, the power spectra appear through a specific combination, \ie the determinant of the (reduced) covariance matrix, which makes the purity invariant under canonical transformations~\cite{Grain:2019vnq}. Let us indeed recall that in a two-dimensional (\ie single-field) system, there is a single symplectic invariant~\cite{Serafini:2003ke, 2012arXiv1209.2748D}, the so-called symplectic eigenvalue $\sigma_{\vec{k}}(t)$, which is such that the eigenvalues of $ \bs{\mathrm{Cov}}_{\vec{k},\mathrm{red}} \boldmathsymbol{\Omega} $ are given by $\pm i \sigma_{\vec{k}}(t)$. This leads to $\sigma_{\vec{k}}=1/(2 \sqrt{\gamma_{\vec{k}}})$, and the condition $\gamma_{\vec{k}}<1$ is equivalent to $\sigma_{\vec{k}}>1/2$. In passing, let us note that, for Gaussian states, the entanglement entropy introduced below \Eq{eq:puritydef} is related to the symplectic eigenvalue by~\cite{Eisert:2008ur} $S_{\vec{k},\mathrm{ent}} = (\sigma_{\vec{k}} + 1/2) \log_2 ( \sigma_{\vec{k}} +1/2) - (\sigma_{\vec{k}} - 1/2) \log_2 ( \sigma_{\vec{k}} - 1/2)$. This allows one to relate the linear and entanglement entropies, and check the above statement that the former provides a lower bound to the latter. 

If the two fields are uncoupled, the reduced system is in a pure state and one has~\cite{Grain:2019vnq} $\mathrm{Cov}^{(\phi \phi)}_{11, k} \mathrm{Cov}^{(\pi \pi)}_{11, k} - \left(\mathrm{Cov}^{(\phi \pi)}_{11, k}\right)^2 = 1/4$. This implies that $\gamma_{\vec{k}}(t) = 1$, $\sigma_{\vec{k}}(t) = 1/2$ and $S_{\vec{k},\mathrm{lin}}(t) =S_{\vec{k},\mathrm{ent}}(t)= 0$. Otherwise, $\gamma_{\vec{k}}<1$ signals the presence of decoherence. 
An obvious, yet crucial consequence of \Eq{eq:purityfinal} is that decoherence, \ie the reduction of $\gamma_{\vec{k}}$ away from unity, cannot be achieved without modifying the power spectra. In other words, for any system that undergoes decoherence, the observational predictions are necessarily altered, and an important question that will be addressed below is whether or not decoherence can proceed while keeping this alteration negligible~\cite{Martin:2018zbe,Martin:2018lin}. 

Let us finally stress that unlike the approach presented in \Sec{subsec:rhored}, which relies on a detailed analysis of the mathematical structure of $\mathrm{Sp}(4,\mathbb{R})$ and leads to a formula, \Eq{eq:purity}, that involves nine infinite sums; the Wigner function formalism only makes use of Gaussian integrals. It can therefore be straightforwardly generalised to higher-dimension systems (\ie containing more fields), while the approach of \Sec{subsec:rhored} would require further analyses of the groups $\mathrm{Sp}(2n,\mathbb{R})$. By plugging \Eqs{eq:P11}, \eqref{eq:P33} and \eqref{eq:P13} into \Eq{eq:purityfinal}, one can finally obtain a fully non-perturbative expression for the purity in terms of the squeezing parameters
\begin{align}
\label{eq:puritysqueeze}
    \gamma_{\vec{k}}(t) =&\Bigg\{\left|\widetilde{\tau}_k\right|^4 + \mathrm{sinc}^4(\theta^k)\left|\tau_k\right|^4 
	+ 2\mathrm{sinc}^2(\theta^k)\left|\tau_k\right|^2 \left|\widetilde{\tau}_k\right|^2
	\times \nonumber  \\ & \quad
	\bigg[ \cosh(2r^k_1)\cosh(2r_2^k)
	+\cos\left(2\mathrm{arg}\widetilde{\tau}_k-2\mathrm{arg}\tau_k\right) \sinh(2r^k_1)\sinh(2r^k_2) \bigg] \Bigg\}^{-1},
\end{align}
where we recall that $\widetilde{\tau}_k$ is defined in \Eq{eq:tau1}, and $\tau_k$ and $\theta^k$ in \Eq{eq:thetavarphi}.
\subsubsection*{Small-coupling limit}\label{sec:decosmall}
Similarly to what was done at the end of \Sec{subsec:rhored}, let us now further expand the purity in the limit where the system and the environment fields are weakly coupled. The power spectra in the observable sector are given by \Eqs{eq:P11}, \eqref{eq:P33} and \eqref{eq:P13}. Expanding these expressions up to second order in the interaction parameter $\vert \tau_k \vert$ [recalling that $(\theta^{k}_6 -i \theta^{k}_5) = \vert\tau_{k}\vert e^{i \mathrm{arg}\tau_k}$ and $\theta^{k} = \sqrt{(\theta^{k}_4)^2 + \vert\tau_{k}\vert^2}$ ], one obtains
\begin{align}
	\label{eq:P11O2} \mathrm{Cov}^{(\phi \phi)}_{11,k} =& \frac{1}{2k} \Bigg( \bigg[\cosh(2 r^{k}_1) + \cos(2 \theta^{k}_3+2 \theta^{k}_4) \sinh(2 r^{k}_1)\bigg] + \left|\tau_{k}\right|^2 \bigg\{- \frac{\sin^2 \theta^{k}_4}{\theta^{k}_4} \cosh(2 r^{k}_1) \nonumber \\
	&+ \frac{1}{2 (\theta^{k}_4)^2} \left[\cos(2 \theta^{k}_3) - \cos(2 \theta^{k}_3+2 \theta^{k}_4) - 2 \theta^{k}_4 \sin(2 \theta^{k}_3+2 \theta^{k}_4) \right] \sinh(2 r^{k}_1)\nonumber \\
	&+ \frac{\sin^2 \theta^{k}_4}{(\theta^{k}_4)^2}\left[\cosh(2 r^{k}_2) - \cos(2\theta^{k}_3 + 2\mathrm{arg}\tau_k) \sinh(2 r^{k}_2)\right]\bigg\}\Bigg) 
	\\
	\label{eq:P33O2} \mathrm{Cov}^{(\pi \pi)}_{11,k} =& \frac{k}{2} \Bigg( \bigg[\cosh(2 r^{k}_1) - \cos(2 \theta^{k}_3+2 \theta^{k}_4) \sinh(2 r^{k}_1)\bigg] + \left|\tau_{k}\right|^2 \bigg\{- \frac{\sin^2 \theta^{k}_4}{\theta^{k}_4} \cosh(2 r^{k}_1) \nonumber\\
	&- \frac{1}{2 (\theta^{k}_4)^2} \left[\cos(2 \theta^{k}_3) - \cos(2 \theta^{k}_3+2 \theta^{k}_4) - 2 \theta^{k}_4 \sin(2 \theta^{k}_3+2 \theta^{k}_4) \right] \sinh(2 r^{k}_1) \nonumber\\
	&+ \frac{\sin^2 \theta^{k}_4}{(\theta^{k}_4)^2}\left[\cosh(2 r^{k}_2) + \cos(2\theta^{k}_3 + 2\mathrm{arg}\tau_k) \sinh(2 r^{k}_2)\right]\bigg\}\Bigg)
	\\
	\label{eq:P13O2} \mathrm{Cov}^{(\phi \pi)}_{11,k} =& \frac{1}{2} \Bigg( -\sin(2 \theta^{k}_3+2 \theta^{k}_4)\sinh(2 r^{k}_1) + \left|\tau_{k}\right|^2 \bigg\{ \frac{\sin^2 \theta^{k}_4}{(\theta^{k}_4)^2} \sin(2\theta^{k}_3 + 2\mathrm{arg}\tau_k) \sinh(2 r^{k}_2) \nonumber\\
	&+ \frac{1}{2 (\theta^{k}_4)^2} \left[-\sin(2 \theta^{k}_3) + \sin(2 \theta^{k}_3+2 \theta^{k}_4) - 2 \theta^{k}_4 \cos(2 \theta^{k}_3+2 \theta^{k}_4) \right] \sinh(2 r^{k}_1) \bigg\}\Bigg).
\end{align}
In the limit where the two fields are uncoupled, $\tau_k=0$, one recovers the result obtained for two-mode squeezed states in \Refa{Grain:2019vnq}. One can also check that in agreement with the discussion of \Sec{subsec:rhored}, the leading correction to the power spectra is of quadratic order in $\vert\tau_{k}\vert$. Let us stress that those corrections involve parameters that describe the environment sector, such as $r_2^k$, hence observations carried out on the system alone can a priori lead to indirect information about the microphysical evolution of the traced over field(s). One should also note that, through the dynamical evolution, the presence of the interaction modifies all Bogolyubov coefficients, hence all squeezing and rotation parameters. This is why formally, in the above expressions, the leading terms (\ie the ones before $\vert\tau_k\vert^2$) need also be expanded. 

By plugging \Eqs{eq:P11O2}, \eqref{eq:P33O2} and \eqref{eq:P13O2} into \Eq{eq:purityfinal}, one can finally derive an expression for the purity expanded at quadratic order in $\vert\tau_k\vert^2$, and by doing so one exactly recovers \Eq{eq:puritypert:squeezing}. This is an important consistency check as the two methods employed to derive this result are completely independent (and, as already argued, the approach based on the Wigner function is computationally less heavy).

This calculation also makes explicit that as one increases the interaction strength, one decreases the purity and hence makes decoherence more efficient, but one also induces larger corrections to the observable power spectra. 
This allows one to answer the question asked above, namely whether or not decoherence can proceed without affecting too much the power spectra. As noticed below \Eq{eq:puritypert:squeezing}, decoherence becomes more efficient as quantum squeezing increases, and in the large-squeezing limit, the correction to $\gamma_{\vec{k}}=1$ is controlled by $\vert \tau_k\vert^2 \ee^{2 r}$. However, from \Eqs{eq:P11O2}, \eqref{eq:P33O2} and \eqref{eq:P13O2}, one can see that the relative correction to the power spectra is rather controlled by $\vert \tau_k\vert^2$ in that limit. As a consequence, if the interaction strength is such that 
\begin{align}
\label{eq:cond:decoh:without:ruining:PowerSpectra}
    \ee^{-r}\ll \vert \tau_k\vert \ll 1\, ,
\end{align}
decoherence takes place while keeping corrections to observable predictions tiny.
\section{Conclusion}
\label{sec:conclu}
In this work, we have performed a detailed study of the quantum dynamics of two scalar fields, quadratically coupled, and embedded in a homogeneous and isotropic background. Their dynamics is generated by a quadratic Hamiltonian with time-dependent coefficients. Evolution of such systems (either classical or quantum) is obtained by applying elements of the symplectic group $\mathrm{Sp}(4,\mathbb{R})$ to the initial configuration, so we have first investigated the mathematical structure of this group by presenting various descriptions of it. In particular, using the Bloch-Messiah decomposition that we further developed using the commutation relations of the Lie algebra, we have derived fully factorised expressions for the group elements of $\mathrm{Sp}(4,\mathbb{R})$. The ten parameters entering these expressions are dubbed the squeezing and rotation parameters, as per the three parameters entering the decomposition of $\mathrm{Sp}(2,\mathbb{R})$. Alternatively, this group is described by Bogolyubov coefficients, which we explicitly related to the squeezing and rotation parameters. 

We then provided the quantum representation of the Lie algebra of $\mathrm{Sp}(4,\mathbb{R})$, from which the quantum Hamiltonian can be easily expressed and interpreted. Couplings between the two fields manifest themselves through exchanges of quanta from one sector to the other (hence preserving the total number of quanta), and through direct productions of pairs in which each quantum belongs to a different sector. The latter provides a direct way to entangle the two sectors and those particles add up to the direct pair production occurring within each sector separately. The former also leads to entanglement but in an indirect way by transferring quanta which have been previously created in a given sector. Using this group-theoretic approach, we then showed that the evolution operator can be interpreted as the successive application of three blocks of quantum operations on the initial state. This sequence of operations schematically consists in first exchanging quanta between the two sectors, then creating pairs within each sector separately, and finally mixing again these newly-created quanta between the two sectors. 

Applying the evolution operator to the vacuum state allowed us to derive the most general expression for the four-mode squeezed states, see \Eqs{eq:vacevolfinal} and~\eqref{eq:ampl}, which, to our knowledge, has not been presented in the literature so far. It can be viewed as the copy of two two-mode squeezed states, one for each sector, which then exchange quanta according to their couplings. Its mathematical structure exhibits a power expansion in the coupling between the two fields, in which the power at each order gives the number of transfers between the two sectors. As an example, we provided explicit formulas for an expansion truncated at second order, \ie including up to two exchanges of particles between the two fields. We finally described the four-mode squeezed states in terms of their Wigner functions, which were shown to be Gaussian. Their covariance is built from all the cross-spectra and we expressed them using either the Bogolyubov coefficients or the ten squeezing and rotation parameters. 

In cases where one of the two sectors is unobserved (say the second sector, which we referred to as the ``environment''), entanglement between the two fields leads to quantum decoherence in the first sector (dubbed the ``system''), as well as modifications of its observable predictions. We studied this mechanism by first computing the reduced density matrix starting from the four-mode squeezed state. We showed that the environment induces corrections to observables that are necessarily of even power in the coupling strength. 
This is because, in order to preserve statistical isotropy, any particle transfer between the two sectors must be compensated by the inverse transfer of a particle with the same wavenumber, or by the transfer of a particle with opposite wavenumber, so the number of transfers is even.
The purity of the system, $\gamma_{\vec{k}}$, was also calculated from the reduced density matrix. We then investigated decoherence using the Wigner function and found that it substantially simplifies calculations. Indeed, we showed that tracing out the environment is readily obtained in the phase-space representation by marginalising the Wigner function  over the phase-space of the environment. Since the Wigner function of a four-mode squeezed state is Gaussian, this operation is trivial. This allowed us to obtain a non-perturbative expression of the purity in terms of the power spectra of the system. We have finally expanded the result at second-order in the coupling parameters in both approaches (where we have checked that the same result is obtained).

The fact that the purity can be expressed in terms of the power spectra of the system entails that decoherence, \ie the decrease of the purity, cannot proceed without affecting the observable power spectra. However, we have shown that in the large squeezing limit, there exists a regime, given by \Eq{eq:cond:decoh:without:ruining:PowerSpectra}, where the interaction strength is large enough to make the system decohere but small enough to keep the observables mostly unchanged, shedding some light on the results of \Refs{Martin:2018zbe,Martin:2018lin}. This also confirmed that squeezed states are more easily subject to decoherence~\cite{Kiefer:1998qe}.

Though limited to quadratic coupling, let us stress that our approach to decoherence does not rely on any approximation scheme, since the full quantum state of the joint system-plus-environment setup is first derived exactly, before tracing out the environment. It thus provides an ideal case study to frame the range of applicability of approximate approaches to decoherence, such as the Lindblad formalism (at least in the simple situation of quadratic couplings). One may indeed compare the exact results obtained in this work with the ones derived from those effective methods, and this is the topic of a future work.

As mentioned in \Sec{sec:intro}, our results are directly relevant for cosmology, in order to describe scalar fields in flat and non-flat Friedmann-Lema\^itre-Robertson-Walker geometries, quantum fields in curved spaces possibly with derivative couplings, and in the context of primordial cosmology, adiabatic and isocurvature perturbations in multiple-field inflation scenarios. But more generally, our results are relevant for any time-dependent, quadratic Hamiltonian that couples two degrees of freedom, regardless of the origin of these degrees of freedom. They thus offer a wide range of applications. In particular, the phase-space approach can be readily extended to cases where the environment is made of more than one scalar field. Suppose indeed that $N$ scalar fields, initially in their vacuum state, compose the environment. The Wigner function of the system-plus-environment setup is described by an $(N+1)$-dimensional Gaussian. Tracing out ({\textsl{i.e.}, in phase space, marginalising) over $N$ scalar fields is thus as trivial as tracing out over one single field. This may be used to understand how isocurvature modes can lead to the decoherence of the adiabatic sector, and should be compared with effective-field theory approaches for such systems~\cite{Pinol:2020kvw}. 

Finally, since squeezed states feature quantum entanglement, they are an interesting playground to discuss possible setups for Bell and Leggett-Garg inequality violations in continuous systems, see \Refs{Martin:2016tbd, Martin:2016nrr, Martin:2017zxs, Ando:2020kdz}. These analyses have been carried out for two-mode squeezed states, where only one type of entanglement can be harvested, namely the one between modes $\vec{k}$ and $-\vec{k}$ of the same field. Since the two subsystems $\vec{k}$ and $-\vec{k}$ are not locally distinct in real space, this necessarily restricts the analysis to those Bell inequalities where locality is not part of the assumptions being tested (such as, for instance, the Leggett-Garg or the temporal-Bell inequalities). The situation is however different for multiple-field systems, where one may chose to measure a field $\phi_1$ at position $\vec{x}_1$ and another field $\phi_2$ at position $\vec{x}_2$. This is because, on top of the entanglement between modes $\vec{k}$ and $-\vec{k}$ of the same field, one now has entanglement between quanta in different fields. This thus opens up the possibility to test for a wider class of Bell inequalities.
Let us also mention that the present work would additionally allow one to assess how decoherence affects the ability to test Bell inequalities with squeezed states. 


\appendix
\section{Commutation relations in the Lie algebra $\mathfrak{sp}(4,\mathbb{R})$}
\label{sec:commutrel}
Following the procedure exemplified at the end of \Sec{subsec:liealg}, which relies on combining the multiplication rule of the Kronecker product, \Eq{eq:Kronecker:Multiplication:Rule}, with the commutation relations between the Pauli matrices, \Eq{eq:Commutators:Pauli}, the commutators between the generators of $\mathfrak{sp}(4,\mathbb{R})$, as listed in Table~\ref{tab:gen}, are given by
\begin{eqnarray}
\label{eq:commutsqsq} \text{Sq./Sq.} &:& ~~~~\left[\bs{K}_1, \bs{K}_2\right] = 0;\\[10pt]
\label{eq:u1rot} \text{Rot./Rot.} &:& ~~~~\left[\bs{K}_3, \bs{K}_5\right] = 0,~~~~~~~\left[\bs{K}_3, \bs{K}_6\right] = 0, ~~~~~~~~\left[\bs{K}_3, \bs{K}_4\right] = 0,\\
&~&~~~~\left[\bs{K}_5, \bs{K}_6\right] = 2 \bs{K}_4,~~ \left[\bs{K}_6, \bs{K}_4\right] = 2 \bs{K}_5,~~\left[\bs{K}_4, \bs{K}_5\right] = 2 \bs{K}_6;\\[10pt]
\text{Boost/Boost} &:& ~~~~\left[\bs{K}_7, \bs{K}_8\right] = 0,~~~~~~~\left[\bs{K}_7, \bs{K}_{10}\right] = 0, ~~~~~~~~\left[\bs{K}_9, \bs{K}_8\right] = 0,\\
&~&~~~~\left[\bs{K}_9, \bs{K}_7\right] = 2 \bs{K}_6,~~~ \left[\bs{K}_8, \bs{K}_{10}\right] = 2 \bs{K}_5,~~~\left[\bs{K}_9, \bs{K}_{10}\right] = 2 \bs{K}_3;\\[10pt]
\text{Sq./Rot.} &:& ~~~~\left[\bs{K}_1, \bs{K}_6\right] = 0,~~~~~~\left[\bs{K}_2, \bs{K}_5\right] = 0,\\
&~&~~~~\left[\bs{K}_1, \bs{K}_3\right] = 2 \bs{K}_8,~~\left[\bs{K}_1, \bs{K}_4\right] = 2 \bs{K}_7,~~\left[\bs{K}_1, \bs{K}_5\right] = 2 \bs{K}_9,\\
&~&~~~~\left[\bs{K}_2, \bs{K}_3\right] = 2 \bs{K}_7,~~\left[\bs{K}_2, \bs{K}_4\right] = 2 \bs{K}_8,~~\left[\bs{K}_2, \bs{K}_6\right] = 2 \bs{K}_{10};\\[10pt]
\text{Sq./Boost} &:& ~~~~\left[\bs{K}_1, \bs{K}_{10}\right] = 0,~~~~~~\left[\bs{K}_2, \bs{K}_9\right] = 0,\\
&~&~~~~\left[\bs{K}_1, \bs{K}_7\right] = 2 \bs{K}_4,~~\left[\bs{K}_1, \bs{K}_8\right] = 2 \bs{K}_3,~~\left[\bs{K}_1, \bs{K}_9\right] = 2 \bs{K}_5,\\
&~&~~~~\left[\bs{K}_2, \bs{K}_7\right] = 2 \bs{K}_3,~~\left[\bs{K}_2, \bs{K}_8\right] = 2 \bs{K}_4,~~\left[\bs{K}_2, \bs{K}_{10}\right] = 2 \bs{K}_6;\\[10pt]
\text{Rot./Boost} &:& ~~~~\left[\bs{K}_4, \bs{K}_9\right] = 0,~~~~~~\left[\bs{K}_4, \bs{K}_{10}\right] = 0,\\
&~& ~~~~\left[\bs{K}_5, \bs{K}_7\right] = 0,~~~~~~\left[\bs{K}_6, \bs{K}_8\right] = 0,\\[10pt]
&~& ~~~~\left[\bs{K}_3, \bs{K}_7\right] = 2\bs{K}_2,~~~~~~\left[\bs{K}_3, \bs{K}_8\right] = 2\bs{K}_1,\\
&~& ~~~~\left[\bs{K}_4, \bs{K}_7\right] = 2\bs{K}_1,~~~~~~\left[\bs{K}_4, \bs{K}_8\right] = 2\bs{K}_2,\\
&~& ~~~~\left[\bs{K}_5, \bs{K}_9\right] = 2\bs{K}_1,~~~~~~\left[\bs{K}_6, \bs{K}_{10}\right] = 2\bs{K}_2,\\[10pt]
&~& ~~~~\left[\bs{K}_3, \bs{K}_{10}\right] = 2\bs{K}_9,~~~~~~\left[\bs{K}_9, \bs{K}_3\right] = 2\bs{K}_{10},\\
&~& ~~~~\left[\bs{K}_5, \bs{K}_{10}\right] = 2\bs{K}_8,~~~~~~\left[\bs{K}_8, \bs{K}_5\right] = 2\bs{K}_{10},\\
&~& ~~~~\left[\bs{K}_6, \bs{K}_7\right] = 2\bs{K}_9,~~~~~~\left[\bs{K}_9, \bs{K}_6\right] = 2\bs{K}_7.
\end{eqnarray}
One can identify various subalgebras, which is particularly useful when it comes to factorising down elements of the group. Since all subalgebras are three dimensional, we can use Bianchi classification to sort them. In the following equation, each line corresponds to a subalgebra, \ie a set of closed generators by the adjoint operation:
\begin{eqnarray}
\label{eq:su2} \text{Type IX} &:& ~~~~\left[\bs{K}_5, \bs{K}_6\right] = 2 \bs{K}_4,~~ \left[\bs{K}_6, \bs{K}_4\right] = 2 \bs{K}_5,~~\left[\bs{K}_4, \bs{K}_5\right] = 2 \bs{K}_6;\\[10pt]
\text{Type VIII} &:& ~~~~\left[\bs{K}_1, \bs{K}_3\right] = 2 \bs{K}_8,~~\left[\bs{K}_3, \bs{K}_8\right] = 2\bs{K}_1,~~\left[\bs{K}_8, \bs{K}_1\right] = - 2 \bs{K}_3\,; \\ 
&~& ~~~~\left[\bs{K}_1, \bs{K}_4\right] = 2 \bs{K}_7,~~\left[\bs{K}_4, \bs{K}_7\right] = 2\bs{K}_1,~~\left[\bs{K}_7, \bs{K}_1\right] = - 2 \bs{K}_4\,;\\
&~& ~~~~\left[\bs{K}_1, \bs{K}_5\right] = 2 \bs{K}_9,~~\left[\bs{K}_5, \bs{K}_9\right] = 2\bs{K}_1, ~~\left[\bs{K}_9, \bs{K}_1\right] = - 2 \bs{K}_5\,;\\[10pt]
&~& ~~~~\left[\bs{K}_2, \bs{K}_3\right] = 2 \bs{K}_7,~~\left[\bs{K}_3, \bs{K}_7\right] = 2\bs{K}_2, ~~\left[\bs{K}_7, \bs{K}_2\right] =- 2 \bs{K}_3\,; \\ 
&~& ~~~~\left[\bs{K}_2, \bs{K}_4\right] = 2 \bs{K}_8,~~\left[\bs{K}_4, \bs{K}_8\right] = 2\bs{K}_2, ~~\left[\bs{K}_8, \bs{K}_2\right] = - 2 \bs{K}_4\,;\\
&~& ~~~~\left[\bs{K}_2, \bs{K}_6\right] = 2 \bs{K}_{10},~~\left[\bs{K}_6, \bs{K}_{10}\right] = 2\bs{K}_2,~~\left[\bs{K}_{10}, \bs{K}_2\right] = -2 \bs{K}_6\,;\\[10pt]
&~&~~~~\left[\bs{K}_9, \bs{K}_6\right] = 2\bs{K}_7,~~\left[\bs{K}_6, \bs{K}_7\right] = 2 \bs{K}_9,~~\left[\bs{K}_7, \bs{K}_9\right] = -2 \bs{K}_6\,;\\
&~& ~~~~\left[\bs{K}_8, \bs{K}_5\right] = 2\bs{K}_{10},~~\left[\bs{K}_5, \bs{K}_{10}\right] = 2 \bs{K}_8,~~\left[\bs{K}_{10}, \bs{K}_8\right] = -2 \bs{K}_5\,;\\
&~&~~~~\left[\bs{K}_9, \bs{K}_3\right] = 2\bs{K}_{10},~~\left[\bs{K}_3, \bs{K}_{10}\right] = 2 \bs{K}_9,~~\left[\bs{K}_{10}, \bs{K}_9\right] = -2 \bs{K}_3\, .
\end{eqnarray}
The type IX Bianchi algebra is related to a $\mathfrak{su}(2)$ subalgebra that contains three rotation generators while the type VIII Bianchi algebra is isomorphic to $\mathfrak{sl}(2,\mathbb{R})$ subalgebras. The first six type-VIII subalgebras contain a rotation, a squeezing and a boost generator, so they can be related to $\mathfrak{sp}(2,\mathbb{R}) \cong \mathfrak{sl}(2,\mathbb{R})$, the last three type-VIII subalgebras are made of a rotation and two boost generators, making their interpretation less obvious.
\section{Evolved vacuum state}
\label{sec:vacappendix}
This appendix presents the computation that leads to the expression of the evolved vacuum state in the occupation-number representation given in \Eqs{eq:vacevolfinal} and~\eqref{eq:ampl}. Starting from the initial vacuum state~\eqref{eq:inivac4}, we follow the different operations displayed in the quantum circuits of \Sec{subsec:evolop}. Let us first consider $\widehat{\mathcal{R}}_{\vec{k}}(\boldsymbol{\varphi}_{k})$, which for convenience we reproduce here:
{\footnotesize
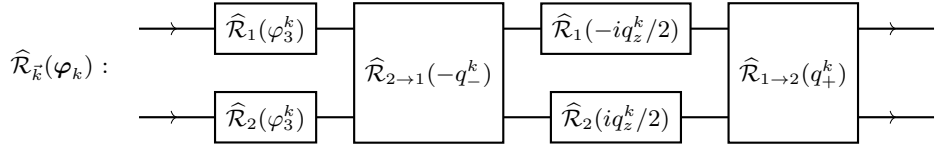
{
\begin{center}
		\begin{tikzpicture}
		\node[scale=1.0] {
		$\widehat{\mathcal{R}}_{\vec{k}}(\boldsymbol{\varphi}_{k})\ $:
			\begin{quantikz}
				\arrow[r] & \qw & \gate{\widehat{\mathcal{R}}_1(\varphi^{k}_3)} & \gate[wires=2]{\widehat{\mathcal{R}}_{2 \rightarrow 1}(-q^{k}_{-})} & \gate{\widehat{\mathcal{R}}_1(-i q^{k}_z /2)} & \gate[wires=2]{\widehat{\mathcal{R}}_{1 \rightarrow 2}(q^{k}_{+})} \arrow[r] & \qw & \qw \\
				\arrow[r] & \qw & \gate{\widehat{\mathcal{R}}_2(\varphi^{k}_3)} & & \gate{\widehat{\mathcal{R}}_2(i q^{k}_z /2)} & \arrow[r] & \qw & \qw
			\end{quantikz}
		};
	\end{tikzpicture}
\end{center}
}}
Since the operators contained in $\widehat{\mathcal{R}}_{\vec{k}}(\boldsymbol{\varphi}_{k})$ do not create particles, the components $\widehat{\mathcal{R}}_{2\to 1}$ and $\widehat{\mathcal{R}}_{1\to 2}$ leave the initial vacuum state unchanged. Moreover, the operator $\widehat{\mathcal{R}}_1(-i q^{k}_z /2)$ generates an overall factor $\ee^{- q^{k}_z/2}$, which is exactly compensated by the operator $\widehat{\mathcal{R}}_2(i q^{k}_z /2)$ that generates an overall factor $\ee^{ q^{k}_z/2}$. As a consequence, only the overall phase shift controlled by $\varphi^{k}_3$ remains, and the evolved vacuum state can be obtained as 
{\footnotesize
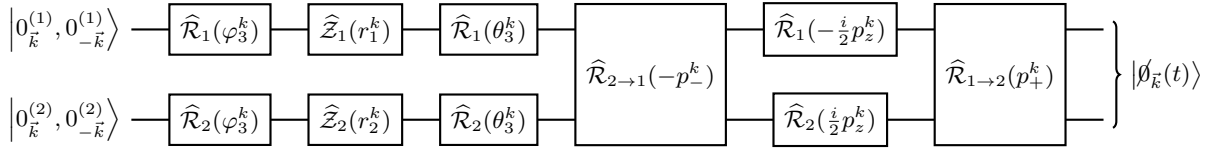
{
\begin{center}
	\begin{tikzpicture}
		\node[scale=1.0] {
			\begin{quantikz}
				\lstick{$\left|0^{(1)}_{\vec{k}},0^{(1)}_{-\vec{k}}\right>$} & \gate{\widehat{\mathcal{R}}_1(\varphi^{k}_3)} & \gate{\widehat{\mathcal{Z}}_1(r^{k}_1)} & \gate{\widehat{\mathcal{R}}_1(\theta^{k}_3)} & \gate[wires=2]{\widehat{\mathcal{R}}_{2 \rightarrow 1}(-p^{k}_{-})} & \gate{\widehat{\mathcal{R}}_1(-\frac{i}{2} p^{k}_z )} & \gate[wires=2]{\widehat{\mathcal{R}}_{1 \rightarrow 2}(p^{k}_{+})} & \qw\rstick[wires=2]{$\left|\cancel{0}_{\vec{k}}(t)\right>$}\\
				\lstick{$\left|0^{(2)}_{\vec{k}},0^{(2)}_{-\vec{k}}\right>$} & \gate{\widehat{\mathcal{R}}_2(\varphi^{k}_3)} & \gate{\widehat{\mathcal{Z}}_2(r^{k}_2)}& \gate{\widehat{\mathcal{R}}_2(\theta^{k}_3)} & & \gate{\widehat{\mathcal{R}}_2(\frac{i}{2} p^{k}_z )} & & \qw &
			\end{quantikz}
		};
	\end{tikzpicture}
\end{center}
}}
\noindent where we recall that $\widehat{\mathcal{R}}_i$, $\widehat{\mathcal{Z}}_i$ and $\widehat{\mathcal{R}}_{i \rightarrow j}$ with $i,j = 1,2$ are defined in \Eqs{eq:phase}, \eqref{eq:squeezing} and \eqref{eq:transfer}. Let us see how these operators act one after the other.

The operators $\widehat{\mathcal{R}}_1(\varphi^{k}_3)$ and $\widehat{\mathcal{R}}_2(\varphi^{k}_3)$ simply add a global phase factor $\ee^{-2i \varphi^{k}_3}$, so the initial vacuum state is first transformed according to
\bea 
\ee^{-2i\varphi^k_3}\ket{0^{(1)}_{\vec{k}},~0^{(1)}_{-\vec{k}},~ 0^{(2)}_{\vec{k}},~ 0^{(2)}_{-\vec{k}}}\, .
\eea

To derive the action of the squeezing operators $\widehat{\mathcal{Z}}_i$, we recall that the two squeezing generators commute, and that they act on each sector separately. As a consequence, we can use the result for two-mode squeezed states in $\mathrm{Sp}(2,\mathbb{R})$, as derived \eg in \Refa{Grain:2019vnq}, and write 
\bea
	\widehat{\mathcal{Z}}_1(r^{k}_1) \left[ \ket{0^{(1)}_{\vec{k}},~ 0^{(1)}_{-\vec{k}}} \otimes \chi^{(2)}_{\vec{k}}\right] &=&  \frac{1}{\cosh{r^{k}_1}} \sum\limits_{n=0}^\infty (-1)^n \tanh^n r^{k}_1 \ket{n^{(1)}_{\vec{k}},~n^{(1)}_{-\vec{k}}} \otimes \chi^{(2)}_{\vec{k}}\, ,\\
	\widehat{\mathcal{Z}}_2(r^{k}_2) \left[\chi^{(1)}_{\vec{k}} \otimes\ket{0^{(2)}_{\vec{k}},~ 0^{(2)}_{-\vec{k}}}\right] &=&  \chi^{(1)}_{\vec{k}} \otimes \frac{1}{\cosh{r^{k}_2}} \sum\limits_{m=0}^\infty (-1)^m \tanh^m r^{k}_2 \ket{m^{(2)}_{\vec{k}},~m^{(2)}_{-\vec{k}}},
\eea
where $\chi^{(i)}_{\vec{k}}$ is any vector belonging to $\mathcal{E}^{(i)}_{\vec{k}} \otimes \mathcal{E}^{(i)}_{-\vec{k}}$ for $i=1,2$ and the squeezing parameters $r^{k}_1$ and $r^{k}_2$ control the two-mode creation in each sector. At this stage the state is thus given by
\bea
\frac{e^{-2i\varphi^{k}_3}}{\cosh{r^{k}_1} \cosh{r^{k}_2}} \sum\limits_{n,m = 0}^\infty (-1)^{n+m} \tanh^n r^{k}_1 \tanh^m r^{k}_2 \ket{n^{(1)}_{\vec{k}},n^{(1)}_{-\vec{k}}, m^{(2)}_{\vec{k}}, m^{(2)}_{-\vec{k}}} \, .
\eea

Then comes the contribution from $\widehat{\mathcal{R}}_i(\theta_3^k)$, which simply involves number counting operators, see \Eq{eq:phase}, so the state becomes
\bea
\label{eq:state:interm:1}
\frac{e^{-2i(\varphi^{k}_3+\theta_3^k)}}{\cosh{r^{k}_1} \cosh{r^{k}_2}} \sum\limits_{n,m = 0}^\infty (-1)^{n+m}\ee^{-2i\theta_3^k(n+m)} \tanh^n r^{k}_1 \tanh^m r^{k}_2 \ket{n^{(1)}_{\vec{k}},n^{(1)}_{-\vec{k}}, m^{(2)}_{\vec{k}}, m^{(2)}_{-\vec{k}}}\, .
\eea

The action of $\widehat{\mathcal{R}}_{2 \rightarrow 1}$ is more involved since it transfers excitations from one sector to the other. We first note that, in $\widehat{\mathcal{R}}_{2 \rightarrow 1}$ and $\widehat{\mathcal{R}}_{1 \rightarrow 2}$, the domains $\vec{k}$ and $-\vec{k}$ can be factorised out since $\left[\widehat{a}^\dag_{1, \vec{k}}\widehat{a}_{2, \vec{k}}, \widehat{a}^\dag_{1, -\vec{k}} \widehat{a}_{2, -\vec{k}}\right] =\left[\widehat{a}^\dag_{2, \vec{k}}\widehat{a}_{1, \vec{k}}, \widehat{a}^\dag_{2, -\vec{k}} \widehat{a}_{1, -\vec{k}}\right] =0$. This implies that 
\bea
\widehat{\mathcal{R}}_{2 \rightarrow 1}(-p^{k}_{-}) &=& \exp\left[-i p^{k}_- \left(\widehat{a}^\dag_{1, \vec{k}}\widehat{a}_{2, \vec{k}}\right)\right]\exp\left[-i p^{k}_-\left( \widehat{a}^\dag_{1, -\vec{k}} \widehat{a}_{2, -\vec{k}}\right)\right]\\
&=&\left[\sum\limits_{i=0}^\infty \frac{(-i p^{k}_-)^i}{i!}\left(\widehat{a}^\dag_{1, \vec{k}}\widehat{a}_{2, \vec{k}}\right)^i\right]\left[\sum\limits_{j=0}^\infty \frac{(-i p^{k}_-)^j}{j!}\left( \widehat{a}^\dag_{1, -\vec{k}} \widehat{a}_{2, -\vec{k}}\right)^j\right],  \label{eq:opep1}
\eea
where the exponentials have been Taylor expanded, and a similar expression for $\widehat{\mathcal{R}}_{1 \rightarrow 2}(p^{k}_{+})$ can be written down for future use, namely
\bea
\widehat{\mathcal{R}}_{1 \rightarrow 2}(p^{k}_{+}) 
&=& \left[\sum\limits_{k=0}^\infty \frac{(i p^{k}_+)^k}{k!}\left(\widehat{a}^\dag_{2, \vec{k}}\widehat{a}_{1, \vec{k}}\right)^k\right]\left[\sum\limits_{\ell=0}^\infty \frac{(i p^{k}_+)^\ell}{\ell !}\left( \widehat{a}^\dag_{2, -\vec{k}} \widehat{a}_{1, -\vec{k}}\right)^\ell \right]. \label{eq:opep2}
\eea
The action of $\widehat{a}^\dag_{i,\pm\vec{k}}\widehat{a}_{j_{\pm\vec{k}}}$ is to transfer an excitation from sector $j$ to sector $i$, so upon applying \Eq{eq:opep1} onto \Eq{eq:state:interm:1}, with the schematic normalisation $\widehat{a}\ket{n} = \sqrt{n}\ket{n-1}$ and $\widehat{a}^\dag \ket{n} = \sqrt{n+1}\ket{n+1}$, the state becomes
\begin{align}
\label{eq:state:interm:2}
&\frac{e^{-2i(\varphi^{k}_3+\theta_3^k)}}{\cosh{r^{k}_1} \cosh{r^{k}_2}} \sum\limits_{n,m = 0}^\infty (-1)^{n+m}\ee^{-2i\theta_3^k(n+m)}
 \tanh^n r^{k}_1 \tanh^m r^{k}_2
 \sum\limits_{i=0}^m \frac{(-i p^{k}_-)^i}{i!} \sum\limits_{j=0}^m \frac{(-i p^{k}_-)^j}{j!}
  \nonumber \\
  &~~~~
\sqrt{\frac{m!}{(m-i)!}}\sqrt{\frac{m!}{(m-j)!}}\sqrt{\frac{(n+i)!}{n!}}\sqrt{\frac{(n+j)!}{n!}}
 \ket{(n+i)^{(1)}_{\vec{k}},(n+j)^{(1)}_{-\vec{k}}, (m-i)^{(2)}_{\vec{k}}, (m-j)^{(2)}_{-\vec{k}}}\, .
\end{align}

The next step is to apply $\widehat{\mathcal{R}}_1(-\frac{i}{2} p^{k}_z )$ and $\widehat{\mathcal{R}}_2(\frac{i}{2} p^{k}_z )$, which add imaginary phases (\ie exponential modulation) to each term that depend on their number of particles, and the state~\eqref{eq:state:interm:2} becomes
\bea
\label{eq:state:interm:3}
& &\frac{e^{-2i(\varphi^{k}_3+\theta_3^k)}}{\cosh{r^{k}_1} \cosh{r^{k}_2}} \sum\limits_{n,m = 0}^\infty (-1)^{n+m}\ee^{-2i\theta_3^k(n+m)}
 \tanh^n r^{k}_1 \tanh^m r^{k}_2
 \sum\limits_{i=0}^m \frac{(-i p^{k}_-)^i}{i!} \sum\limits_{j=0}^m \frac{(-i p^{k}_-)^j}{j!}
  \nonumber \\
 &  &~~~~
\sqrt{\frac{m!}{(m-i)!}}\sqrt{\frac{m!}{(m-j)!}}\sqrt{\frac{(n+i)!}{n!}}\sqrt{\frac{(n+j)!}{n!}}
\ee^{-\frac{p_z^k}{2}\left(2n+i+j\right)}
\ee^{\frac{p_z^k}{2}\left(2m-i-j\right)}
 \nonumber \\ &  &~~~~
 \ket{(n+i)^{(1)}_{\vec{k}},(n+j)^{(1)}_{-\vec{k}}, (m-i)^{(2)}_{\vec{k}}, (m-j)^{(2)}_{-\vec{k}}}\, .
 \nonumber\\
\eea
Finally, the application of $\widehat{\mathcal{R}}_{1 \rightarrow 2}(p^{k}_{+}) $ can be done using \Eq{eq:opep2}, and one obtains
\bea
\label{eq:state:interm:4}
& &\frac{e^{-2i(\varphi^{k}_3+\theta_3^k)}}{\cosh{r^{k}_1} \cosh{r^{k}_2}} \sum\limits_{n,m = 0}^\infty (-1)^{n+m}\ee^{-2i\theta_3^k(n+m)}
 \tanh^n r^{k}_1 \tanh^m r^{k}_2
 \sum\limits_{i=0}^m \frac{(-i p^{k}_-)^i}{i!} \sum\limits_{j=0}^m \frac{(-i p^{k}_-)^j}{j!}
  \nonumber \\
 &  &~~~~
\sqrt{\frac{m!}{(m-i)!}}\sqrt{\frac{m!}{(m-j)!}}\sqrt{\frac{(n+i)!}{n!}}\sqrt{\frac{(n+j)!}{n!}}
\ee^{-\frac{p_z^k}{2}\left(2n+i+j\right)}
\ee^{\frac{p_z^k}{2}\left(2m-i-j\right)}
 \nonumber \\ &  &~~~~
 \sum\limits_{k=0}^{n+i} \frac{(i p^{k}_+)^k}{k!}
 \sum\limits_{\ell=0}^{n+j} \frac{(i p^{k}_+)^\ell}{\ell !}
 \sqrt{\frac{(n+i)!}{(n+i-k)!}} \sqrt{\frac{(n+j)!}{(n+j-\ell)!}}
  \sqrt{\frac{(m-i+k)!}{(m-i)!}}  \sqrt{\frac{(m-j+\ell)!}{(m-j)!}}
   \nonumber \\ &  &~~~~
 \ket{(n+i-k)^{(1)}_{\vec{k}},(n+j-\ell)^{(1)}_{-\vec{k}}, (m-i+k)^{(2)}_{\vec{k}}, (m-j+\ell)^{(2)}_{-\vec{k}}}\, .
 \nonumber\\
\eea

This expression can be slightly simplified by replacing the sum over $k$ and $\ell$ with a sum over $s\equiv i-k$ and $t=j-\ell$, and one obtains
\bea
\label{eq:state:interm:5}
\ket{\cancel{0}_{\vec{k}}(t)}
& =&\frac{e^{-2i(\varphi^{k}_3+\theta_3^k)}}{\cosh{r^{k}_1} \cosh{r^{k}_2}} \sum\limits_{n,m = 0}^\infty (-1)^{n+m}\ee^{-2i\theta_3^k(n+m)}\ee^{p_z^k(m-n)}
 \tanh^n r^{k}_1 \tanh^m r^{k}_2
   \nonumber \\
 &  &
 \sum\limits_{i,j=0}^m \frac{(-i p^{k}_- \ee^{-p_z^k})^{i+j}}{i! j!}
\frac{(n+i)!(n+j)!m!}{(m-i)!(m-j)!n!}
 \sum\limits_{s=-n}^{i} 
 \sum\limits_{t=-n}^{j} \frac{(i p^{k}_+)^{i+j-s-t}}{ (i-s)! (j-t) !}
 \sqrt{\frac{(m-s)!(m-t)!}{(n+s)!(n+t)!}}
   \nonumber \\ &  &
 \ket{(n+s)^{(1)}_{\vec{k}},(n+t)^{(1)}_{-\vec{k}}, (m-s)^{(2)}_{\vec{k}}, (m-t)^{(2)}_{-\vec{k}}}\, .
 \nonumber\\
\eea
In order to first sum over the indices appearing in the number of particle eigenstates, one can flip the ordering of the sums over $i,j$ and $s,t$, using that $\sum_{i=0}^m \sum_{s=-n}^{i} = \sum_{s=-n}^m \sum_{i=\umax(s,0)}^{m}$ and $\sum_{j=0}^m \sum_{t=-n}^{j} = \sum_{t=-n}^m \sum_{j=\umax(t,0)}^{m}$, leading to
\bea
\label{eq:state:interm:6}
\ket{\cancel{0}_{\vec{k}}(t)}
& =&\frac{e^{-2i(\varphi^{k}_3+\theta_3^k)}}{\cosh{r^{k}_1} \cosh{r^{k}_2}} \sum\limits_{n,m = 0}^\infty (-1)^{n+m}\ee^{-2i\theta_3^k(n+m)}\ee^{p_z^k(m-n)}
 \tanh^n r^{k}_1 \tanh^m r^{k}_2
 \frac{m!}{n!}
   \nonumber \\
 &  &
  \sum\limits_{s,t=-n}^{m} 
  \sqrt{\frac{(m-s)!(m-t)!}{(n+s)!(n+t)!}} 
  (i p_+^k)^{-s-t}
 \sum\limits_{i=\umax(s,0)}^m 
 \sum\limits_{j=\umax(t,0)}^m 
 \frac{(p^{k}_- p^{k}_+ \ee^{-p_z^k})^{i+j}}{i! j! (i-s)! (j-t) !}
    \nonumber \\ &  &
\frac{(n+i)!(n+j)!}{(m-i)!(m-j)!}
 \ket{(n+s)^{(1)}_{\vec{k}},(n+t)^{(1)}_{-\vec{k}}, (m-s)^{(2)}_{\vec{k}}, (m-t)^{(2)}_{-\vec{k}}}\, .
 \nonumber\\
\eea
This is the result presented in \Eqs{eq:vacevolfinal} and~\eqref{eq:ampl} in the main text.
\section{Reduced density matrix}
\label{subsec:tracingrhored}
In this appendix, we explicitly trace out the environmental degrees of freedom in Fock space as discussed in \Sec{subsec:rhored}, leading to the expression for the reduced density matrix and the purity given in \Eqs{eq:rhoredfinal} and \eqref{eq:purity}. For the evolved vacuum state given in \Eq{eq:vacevolfinal}, the density matrix $\widehat{\rho}_{\vec{k}}(t)= \ket{\cancel{0}_{\vec{k}}(t)} \bra{\cancel{0}_{\vec{k}}(t)}$ reads
\begin{eqnarray}
\label{eq:rho}
	\begin{aligned}
		\widehat{\rho}_{\vec{k}} =&   \sum\limits_{n,m = 0}^\infty  ~ \sum\limits_{s,t = -n}^m ~ \sum\limits_{n',m' = 0}^\infty ~ \sum\limits_{s',t' = -n'}^{m'} c_{k}(n,m,s,t) c^*_{k}(n',m',s',t')\\ &\ket{(n+s)^{(1)}_{\vec{k}},(n+t)^{(1)}_{-\vec{k}},(m-s)^{(2)}_{\vec{k}},(m-t)^{(2)}_{-\vec{k}}}\bra{(n'+s')^{(1)}_{\vec{k}},(n'+t')^{(1)}_{-\vec{k}},(m'-s')^{(2)}_{\vec{k}},(m'-t')^{(2)}_{-\vec{k}}}.
	\end{aligned}
\end{eqnarray}
Following \Eq{eq:traceout}, one can trace out the environmental degrees of freedom in the Fock basis. When doing so, the only non-vanishing terms are such that $m-s=m'-s'$ and $m-t=m'-t'$. This allows one to fix $s'$ and $t'$, and since the conditions $-n'\leq s',t' \leq m'$ imposed by the sum boundaries in \Eq{eq:rho} imply that $m-m'-n'\leq s,t \leq m$, the reduced density matrix reads
\begin{eqnarray}
\begin{aligned}
\widehat{\rho}_{\vec{k}, \mathrm{red}}(t) = &  \sum\limits_{n,m,n',m' = 0}^\infty ~  \sum\limits_{s,t = \max(-n,m-m'-n')}^m  c_{k}(n,m,s,t) c^*_{k}(n',m',s+m'-m,t+m'-m)\\ 
&\ket{(n+s)^{(1)}_{\vec{k}},(n+t)^{(1)}_{-\vec{k}}}\bra{(n'+s+m'-m)^{(1)}_{\vec{k}},(n'+t+m'-m)^{(1)}_{-\vec{k}}}.
\end{aligned}
\end{eqnarray}
Let us now replace $n'$ and $m'$ by the new indices $N=n'+m'-m$ and $M=m'-m$. This gives rise to
\begin{eqnarray}
\begin{aligned}
\widehat{\rho}_{\vec{k}, \mathrm{red}}(t) = &  \sum\limits_{n,m= 0}^\infty ~  \sum\limits_{M= -m}^\infty  ~ \sum\limits_{N= M}^\infty ~ \sum\limits_{s,t = \max(-n,-N)}^m  c_{k}(n,m,s,t) c^*_{k}(N-M,m+M,s+M,t+M)\\ 
&\ket{(n+s)^{(1)}_{\vec{k}},(n+t)^{(1)}_{-\vec{k}}}\bra{(N+s)^{(1)}_{\vec{k}},(N+t)^{(1)}_{-\vec{k}}}.
\end{aligned}
\end{eqnarray}
One can note that the indices $m$ and $M$ do not appear explicitly in the elements of the Fock basis, which is the reason why we have performed the above change of indices. In order to rewrite the sums over $m$ and $M$ as internal sums, similarly to what was done between \Eqs{eq:state:interm:5} and~\eqref{eq:state:interm:6}, one can re-order the various indices and write
\begin{eqnarray}
\label{eq:rhoredpartial}
\begin{aligned}
\widehat{\rho}_{\vec{k}, \mathrm{red}}(t) = &  \sum\limits_{n=0}^\infty ~  
\sum\limits_{N=-\infty}^\infty ~
\sum\limits_{s,t=-\min(n,N)}^\infty ~
\ket{(n+s)^{(1)}_{\vec{k}},(n+t)^{(1)}_{-\vec{k}}}\bra{(N+s)^{(1)}_{\vec{k}},(N+t)^{(1)}_{-\vec{k}}}
\\ &
\underbrace{
\sum\limits_{m=\max(0,s,t)}^\infty ~
\sum\limits_{M=-m}^N ~
c_k(n,m,s,t) c^*_{k}(N-M,m+M,s+M,t+M)}_{\Xi_k(n,N,s,t)}\, ,
\end{aligned}
\end{eqnarray}
which defines the coefficients $\Xi_k$, and which matches \Eqs{eq:rhoredfinal} and \eqref{eq:rhoredampl} given in the main text (where the indices have been renamed for notational convenience). 

From this expression, the purity of the system can also be computed. Squaring \Eq{eq:rhoredpartial}, one obtains
\begin{eqnarray}
\begin{aligned}
\widehat{\rho}^2_{\vec{k},\mathrm{red}} = &\sum\limits_{n, n' = 0}^\infty ~
\sum\limits_{N, N' = -\infty}^\infty ~
\sum\limits_{s,t = - \min(n,N)}^\infty 
    \Xi_{k}(n,N,s,t) \Xi_{k}(n',N',s+N-n',t+N-n')\\
&\ket{(n+s)^{(1)}_{\vec{k}},(n+t)^{(1)}_{-\vec{k}}}\bra{(N'+N-n'+s)^{(1)}_{\vec{k}},(N'+N-n'+t)^{(1)}_{-\vec{k}}},
\end{aligned} 
\end{eqnarray}
from which the purity~\eqref{eq:puritydef} can be expressed as
\begin{align}
\gamma_{\vec{k}}(t) =&\sum\limits_{n=0}^\infty ~
\sum\limits_{N, N' = -\infty}^\infty ~
\sum\limits_{s,t = - \min(n,N)}^\infty 
  \Xi_{k}(n,N,s,t) \Xi_{k}(N+N'-n,N',s+n-N',t+n-N').
\end{align} 
This expression can be slightly simplified by replacing the sum over $N'$ by a sum over $u \equiv n-N'$, and one obtains
\begin{align}
\gamma_{\vec{k}}(t) =&\sum\limits_{n=0}^\infty ~
\sum\limits_{N, u =- \infty}^\infty ~
\sum\limits_{s,t = - \min(n,N)}^\infty 
  \Xi_{k}(n,N,s,t) \Xi_{k}(N-u,n-u,s+u,t+u).
\end{align} 
This matches \Eq{eq:purity} given in the main text, where the indices have been slightly renamed for notational convenience.
\section{Equivalence between tracing out in Hilbert space and marginalisation in the phase space}
\label{sec:directproof}
In \Sec{subsec:Wred}, we have shown that a partial trace of the density matrix in Hilbert space is associated to a partial integration over the corresponding degrees of freedom of the Wigner function in phase space. This proof was however indirect, and relied on the equivalence between different ways of computing the expectation value of quantum operators. In this appendix, for completeness, we provide a direct proof that the reduced Wigner function is nothing but the Wigner-Weyl transform of the reduced density matrix. 

For Hermitian operators, the Wigner-Weyl transform introduced in \Eq{eq:Wigner-Weyl} can be inverted, and the classical phase-space function $\widetilde{\mathcal{O}}_{\vec{k}} (\boldsymbol{\mathfrak{q}}_{\vec{k}})$ gives rise to the quantum operator $\widehat{O}_{\vec{k}}$ through
\begin{eqnarray}\label{eq:inverseWeyl}
	\widehat{O}_{\vec{k}}= \frac{1}{(2\pi)^4}\int_{\mathbb{R}^8}\dd^8 \boldsymbol{\zeta}_{\vec{k}}~\widehat{D}(\boldsymbol{\zeta}_{\vec{k}})\int_{\mathbb{R}^8}\dd^8\boldsymbol{\mathfrak{q}}_{\vec{k}}~\exp\left(i\boldsymbol{\zeta}_{\vec{k}}\cdot \boldsymbol{\mathfrak{q}}_{\vec{k}}\right)\widetilde{\mathcal{O}}_{\vec{k}}(\boldsymbol{\mathfrak{q}}_{\vec{k}}) \, .
\end{eqnarray}
It is obtained by integrating the four-mode displacement operator $\widehat{D}(\boldsymbol{\zeta}_{\vec{k}})$ against the Fourier transform of $\widetilde{\mathcal{O}}_{\vec{k}} (\boldsymbol{\mathfrak{q}}_{\vec{k}})$, where $\boldsymbol{\zeta}_{\vec{k}}$ is the conjugate vector to $\boldsymbol{\mathfrak{q}}_{\vec{k}}$, expanded as
\begin{eqnarray}\label{eq:defzeta}
	\boldsymbol{\zeta}_{\vec{k}} \equiv \left(\zeta_{1, \vec{k}}, \zeta_{1, -\vec{k}}, \zeta_{2, \vec{k}}, \zeta_{2, -\vec{k}}, \kappa_{1, \vec{k}}, \kappa_{1, -\vec{k}}, \kappa_{2, \vec{k}}, \kappa_{2, -\vec{k}}\right)^{\mathrm{T}}\, ,
\end{eqnarray}
and the displacement operator $\widehat{D}(\boldsymbol{\zeta}_{\vec{k}})$ is defined as
\begin{eqnarray}\label{eq:factordisplop}
	\widehat{D}(\boldsymbol{\zeta}_{\vec{k}}) = \widehat{D}_{1,\vec{k}}(\boldsymbol{\zeta}_{\vec{k}}) \widehat{D}_{1,-\vec{k}}(\boldsymbol{\zeta}_{\vec{k}}) \widehat{D}_{2,\vec{k}}(\boldsymbol{\zeta}_{\vec{k}}) \widehat{D}_{2,-\vec{k}}(\boldsymbol{\zeta}_{\vec{k}})\, , 
\end{eqnarray}
where $\widehat{D}_{i, \pm \vec{k}}(\boldsymbol{\zeta}_{\vec{k}})$ are the one-mode displacement operators
\begin{eqnarray}
\label{eq:Dk:OneMode}
	\widehat{D}_{i, \pm \vec{k}}(\boldsymbol{\zeta}_{\vec{k}})=\exp\left(\gamma_{i, \pm \vec{k}}\widehat{a}^\dag_{i, \pm \vec{k}}-\gamma^{*}_{i, \pm \vec{k}}\widehat{a}_{i, \pm \vec{k}}\right)
\end{eqnarray}
for $i = 1,2$ and
\begin{eqnarray}
\label{eq:defalpha}
	\gamma_{i, \pm \vec{k}}=-\frac{1}{\sqrt{2}}\left(\kappa_{i, \pm \vec{k}}+i\zeta_{i, \pm \vec{k}}\right).
\end{eqnarray}
The reason why $\widehat{D}(\boldsymbol{\zeta}_{\vec{k}})$ can be factorised in the form~\eqref{eq:factordisplop} is because the creation and annihilation operators commute across the four modes. 

Since the Wigner function $W_{\vec{k}}(\boldsymbol{\mathfrak{q}}_{\vec{k}}, t)$ is the Wigner-Weyl transform of the density matrix $\widehat{\rho}_{\vec{k}}(t)$, one can use \Eq{eq:inverseWeyl} to extract the density matrix from the Wigner function,
\begin{eqnarray}\label{eq:rhofromW}
	\widehat{\rho}_{\vec{k}}(t)=\frac{1}{(2\pi)^4}\int_{\mathbb{R}^8}\dd^8 \boldsymbol{\zeta}_{\vec{k}}~\widehat{D}(\boldsymbol{\zeta}_{\vec{k}})\int_{\mathbb{R}^8}\dd^8\boldsymbol{\mathfrak{q}}_{\vec{k}}~\exp\left(i\boldsymbol{\zeta}_{\vec{k}}\cdot \boldsymbol{\mathfrak{q}}_{\vec{k}}\right)W_{\vec{k}}(\boldsymbol{\mathfrak{q}}_{\vec{k}}, t)\, .
\end{eqnarray}
Similarly, the reduced density matrix must be connected to the reduced Wigner function (that we here aim at determining) through a relation of the form
\begin{eqnarray}
\label{eq:rhoredfrominvWeyl}
	\widehat{\rho}_{\vec{k},\mathrm{red}}(t)=\frac{1}{(2\pi)^2}\int_{\mathbb{R}^4}\dd^4\boldsymbol{\zeta}_{1,\vec{k}} ~\widehat{D}_{1}(\boldsymbol{\zeta}_{1,\vec{k}}) \int_{\mathbb{R}^4}\dd^4 \boldsymbol{\mathfrak{q}}_{1,\vec{k}}~\exp\left(i\boldsymbol{\mathfrak{q}}_{1,\vec{k}}.\boldsymbol{\zeta}_{1,\vec{k}}\right)W_{\vec{k},\mathrm{red}}(\boldsymbol{\mathfrak{q}}_{1,\vec{k}}, t),
\end{eqnarray}
where $\widehat{D}_{1}(\boldsymbol{\zeta}_{1,\vec{k}}) = \widehat{D}_{1,\vec{k}}(\boldsymbol{\zeta}_{1,\vec{k}}) \widehat{D}_{1,-\vec{k}}(\boldsymbol{\zeta}_{1,\vec{k}})$ is the first-sector displacement operator. 
Let us recall that $\widehat{\rho}_{\vec{k},\mathrm{red}}$ is defined in \Eq{eq:traceout} as the partial trace of the density matrix $\widehat{\rho}_{\vec{k}}$ over the environmental degrees of freedom. By plugging \Eq{eq:rhofromW} into \Eq{eq:traceout}, and expanding $\widehat{D}(\boldsymbol{\zeta}_{\vec{k}})$ as in \Eq{eq:factordisplop}, one finds
\begin{align}
\label{eq:fullrhored}
		\widehat{\rho}_{\vec{k},\mathrm{red}}(t) =& \frac{1}{(2\pi)^2}\int_{\mathbb{R}^4}\dd^4\boldsymbol{\zeta}_{1,\vec{k}} ~\widehat{D}_{1}(\boldsymbol{\zeta}_{1,\vec{k}}) \int_{\mathbb{R}^4}\dd^4\boldsymbol{\mathfrak{q}}_{1,\vec{k}}\exp\left(i\boldsymbol{\mathfrak{q}}_{1,\vec{k}}.\boldsymbol{\zeta}_{1,\vec{k}}\right) \nonumber \\
		& \Bigg[\frac{1}{(2\pi)^2}\int_{\mathbb{R}^4}\dd^4\boldsymbol{\zeta}_{2,\vec{k}} ~\sum\limits_{u,v = 0}^\infty \bra{u^{(2)}_{\vec{k}}, v^{(2)}_{-\vec{k}}(t)}\widehat{D}_{2}(\boldsymbol{\zeta}_{2,\vec{k}})\ket{u^{(2)}_{\vec{k}}, v^{(2)}_{-\vec{k}}(t)} \nonumber \\
		&\int_{\mathbb{R}^4}\dd^4\boldsymbol{\mathfrak{q}}_{2,\vec{k}}~\exp\left(i\boldsymbol{\mathfrak{q}}_{2,\vec{k}}.\boldsymbol{\zeta}_{2,\vec{k}}\right)W_{\vec{k}}(\boldsymbol{\mathfrak{q}}_{1,\vec{k}}, \boldsymbol{\mathfrak{q}}_{2,\vec{k}}, t)\Bigg]\, .
\end{align}
By comparing this expression with \Eq{eq:rhoredfrominvWeyl}, one can read off
\begin{align}
\label{eq:ReducedWigner:calN}
		W_{\vec{k},\mathrm{red}}(\boldsymbol{\mathfrak{q}}_{1,\vec{k}}, t) = \frac{1}{(2\pi)^2} \int_{\mathbb{R}^4}\dd^4\boldsymbol{\mathfrak{q}}_{2,\vec{k}}~W_{\vec{k}}(\boldsymbol{\mathfrak{q}}_{1,\vec{k}}, \boldsymbol{\mathfrak{q}}_{2,\vec{k}}, t)  \mathcal{N}(\boldsymbol{\mathfrak{q}}_{2,\vec{k}})\, ,
\end{align}
where 
\begin{eqnarray}
\label{eq:calN:def}
	\mathcal{N}(\boldsymbol{\mathfrak{q}}_{2,\vec{k}}) = \int_{\mathbb{R}^4}\dd^4\boldsymbol{\zeta}_{2,\vec{k}} ~\exp\left(i\boldsymbol{\mathfrak{q}}_{2,\vec{k}}.\boldsymbol{\zeta}_{2,\vec{k}}\right)\sum\limits_{u,v = 0}^\infty \bra{u^{(2)}_{\vec{k}},~ v^{(2)}_{-\vec{k}}(t)}\widehat{D}_{2}(\boldsymbol{\zeta}_{2,\vec{k}})\ket{u^{(2)}_{\vec{k}},~ v^{(2)}_{-\vec{k}}(t)}.
\end{eqnarray}
To evaluate $\mathcal{N}(\boldsymbol{\mathfrak{q}}_{2,\vec{k}})$, we first use the factorisation~\eqref{eq:factordisplop} and compute 
\begin{align}
\label{eq:sumdispl}
		&\sum\limits_{u,v = 0}^\infty \bra{u^{(2)}_{\vec{k}},~ v^{(2)}_{-\vec{k}}(t)}\widehat{D}_{2}(\boldsymbol{\zeta}_{2,\vec{k}})\ket{u^{(2)}_{\vec{k}},~ v^{(2)}_{-\vec{k}}(t)} = 
		\nonumber \\ & \qquad
		\left[\sum\limits_{u = 0}^\infty \bra{u^{(2)}_{\vec{k}}}\widehat{D}_{2,\vec{k}}(\boldsymbol{\zeta}_{2,\vec{k}})\ket{u^{(2)}_{\vec{k}}}\right]\cdot\left[\sum\limits_{v = 0}^\infty \bra{v^{(2)}_{-\vec{k}}}\widehat{D}_{2,-\vec{k}}(\boldsymbol{\zeta}_{2,\vec{k}})\ket{v^{(2)}_{-\vec{k}}}\right] .
\end{align}
Using the Baker–Campbell–Hausdorff formula, and recalling that $[\widehat{a}_{i, \pm \vec{k}},\widehat{a}^\dag_{i, \pm \vec{k}}]=1$, the one-mode displacement operator~\eqref{eq:Dk:OneMode} can be written as 
\begin{eqnarray}
	\widehat{D}_{i, \pm \vec{k}}(\boldsymbol{\zeta}_{i,\vec{k}})=\ee^{-\frac{1}{2}\left|\gamma_{i, \pm \vec{k}}\right|^2}\cdot\exp\left(\gamma_{i, \pm \vec{k}}\widehat{a}^\dag_{i, \pm \vec{k}}\right)\cdot\exp\left(-\gamma_{i, \pm \vec{k}}^*\widehat{a}_{i, \pm \vec{k}}\right),
\end{eqnarray}
where $\gamma_{i, \pm \vec{k}}$ is given in \Eq{eq:defalpha}. This means that, when evaluating $\bra{u^{(i)}_{\pm\vec{k}}}\widehat{D}_{i,\pm\vec{k}}(\boldsymbol{\zeta}_{i,\pm\vec{k}})\ket{u^{(i)}_{\pm\vec{k}}}$, one first has to compute
\begin{align}
	\exp\left(-\gamma_{i, \pm\vec{k}}^{*}\widehat{a}_{i,\pm \vec{k}}\right)\ket{u^{(i)}_{\pm\vec{k}}} = \ds\sum_{n=0}^u\frac{1}{n!}\left(-\gamma_{i, \pm\vec{k}}^{*}\right)^n\sqrt{u(u-1)\cdots(u-n+1)}\left|(u-n)^{(i)}_{\pm\vec{k}}\right>
\end{align}
where we have simply Taylor expanded the exponential function, and then
\begin{align}
	&\exp\left(\gamma_{i, \pm\vec{k}}\widehat{a}^\dagger_{i,\pm \vec{k}}\right)\ket{(u-n)^{(i)}_{\pm\vec{k}}} =
	\nonumber \\ & \qquad
	\sum_{m=0}^\infty\frac{1}{n!}\left(\gamma_{i, \pm\vec{k}}\right)^m\sqrt{(u-n+1)(u-n+2)\cdots(u-n+m)}\left|(u-n+m)^{(i)}_{\pm\vec{k}}\right>\, .
\end{align}
This leads to
\begin{align}
	\bra{u^{(i)}_{\pm\vec{k}}}\widehat{D}_{i,\pm\vec{k}}(\boldsymbol{\zeta}_{i,\pm\vec{k}})\ket{u^{(i)}_{\pm\vec{k}}} = \ee^{-\frac{1}{2}\left|\gamma_{i, \pm \vec{k}}\right|^2} \ds\sum_{n=0}^u\frac{1}{n!}\left(-\left|\gamma_{i, \pm\vec{k}}\right|^2\right)^n \frac{u(u-1)\cdots(u-n+1)}{n!}\, .
\end{align}
The remaining sum over $n$ can be performed by means of the Laguerre polynomials $L_u(z)$, see Eq.~(18.5.12) of \Refa{NIST:DLMF},
\begin{eqnarray}
	\bra{u^{(i)}_{\pm\vec{k}}}\widehat{D}_{i,\pm\vec{k}}(\boldsymbol{\zeta}_{i,\pm\vec{k}})\ket{u^{(i)}_{\pm\vec{k}}}= \ee^{-\frac{1}{2}\left|\gamma_{i, \pm \vec{k}}\right|^2} L_u\left(\left|\gamma_{i, \pm\vec{k}}\right|^2\right). 
\end{eqnarray}
Plugging this formula into \Eq{eq:sumdispl}, one obtains
\begin{align}
		&\sum\limits_{u,v = 0}^\infty \bra{u^{(2)}_{\vec{k}},~ v^{(2)}_{-\vec{k}}(t)}\widehat{D}_{2}(\boldsymbol{\zeta}_{2,\vec{k}})\ket{u^{(2)}_{\vec{k}},~ v^{(2)}_{-\vec{k}}(t)} =
		\nonumber \\ & \qquad \ee^{-\frac{1}{2}\left(\left|\gamma_{2,\vec{k}}\right|^2+\left|\gamma_{2,-\vec{k}}\right|^2\right)}
		\sum_{u=0}^\infty L_u\left(\left|\gamma_{2, \vec{k}}\right|^2\right)
			\sum_{v=0}^\infty L_v\left(\left|\gamma_{2, -\vec{k}}\right|^2\right)\, .
\end{align}

According to \Eq{eq:calN:def}, we now need to evaluate the Fourier transform of the above expression with respect to $\boldsymbol{\zeta}_{2,\vec{k}}$, in order to compute $\mathcal{N}(\boldsymbol{\mathfrak{q}}_{2,\vec{k}})$. To this end, we introduce the generating function of the Laguerre polynomials [see Eq.~(18.12.13) of \Refa{NIST:DLMF}],
\begin{eqnarray}
	G_z(t)=\ds\sum_{n=0}^\infty t^nL_n(z)=\frac{1}{1-t}\exp\left(-\frac{zt}{1-t}\right).
\end{eqnarray}
Evaluating this formula with $t=1-\varepsilon$ in the limit $\varepsilon\to 0$, one obtains
\begin{eqnarray}
	\sum\limits_{u = 0}^\infty L_u\left(\left|\gamma_{2, \pm\vec{k}}\right|^2\right) &=& \lim_{\varepsilon_{\pm\vec{k}} \to 0}\left[G_{\left|\gamma_{2, \pm\vec{k}}\right|^2}(1-\varepsilon_{ \pm\vec{k}})\right],
\end{eqnarray}
where we introduce $\varepsilon_{\pm\vec{k}}$ for each sector, $\vec{k}$ and $-\vec{k}$. 
This allows one to write $\mathcal{N}(\boldsymbol{\mathfrak{q}}_{2,\vec{k}})$ as the product of two limits of the Fourier transform of a Gaussian, \ie
\begin{align}
	\mathcal{N}(\boldsymbol{\mathfrak{q}}_{2,\vec{k}})=\prod_{s=+/-}\lim_{\varepsilon_{s\vec{k}} \to 0}\left[\frac{1}{\varepsilon_{s\vec{k}}} \int_{\mathbb{R}^2}\dd^2\boldsymbol{\zeta}_{2,s\vec{k}}~\exp\left(i\boldsymbol{\mathfrak{q}}_{2,s\vec{k}}.\boldsymbol{\zeta}_{2,s\vec{k}}\right) \exp\left(-\frac{2-\varepsilon_{s\vec{k}}}{4\varepsilon_{s\vec{k}}}\boldsymbol{\zeta}^2_{2,s\vec{k}}\right)\right],
\end{align}
where we have used the fact that $|\gamma_{2, \pm\vec{k}}|^2  = (\zeta^2_{2, \pm\vec{k}} + \kappa^2_{2, \pm\vec{k}} )/2 = \boldsymbol{\zeta}^2_{2,\pm\vec{k}}/2$.
The Fourier transform of a Gaussian is also a Gaussian and we have
\begin{eqnarray}
	\mathcal{N}(\boldsymbol{\mathfrak{q}}_{2,\vec{k}}) =\prod_{s=+/-}\lim_{\varepsilon_{s\vec{k}} \to 0}\left[\left(
	\frac{4\pi}{2-\varepsilon_{s\vec{k}}}\right)
	\exp\left(-\frac{\varepsilon_{s\vec{k}}}{2-\varepsilon_{s\vec{k}}}\boldsymbol{\mathfrak{q}}^2_{2,s\vec{k}}\right)\right]=(2\pi)^2.
\end{eqnarray}
We conclude that $\mathcal{N}(\boldsymbol{\mathfrak{q}}_{2,\vec{k}})$ simply corresponds to a global phase-space volume, with no dependence on the phase-space location. 

The reduced Wigner function is finally given by \Eq{eq:ReducedWigner:calN}, which leads to
\begin{eqnarray}\label{eq:tracingout}
	W_{\vec{k},\mathrm{red}}(\boldsymbol{\mathfrak{q}}_{1,\vec{k}}, t) =  \int_{\mathbb{R}^4}\dd^4\boldsymbol{\mathfrak{q}}_{2,\vec{k}}~W_{\vec{k}}(\boldsymbol{\mathfrak{q}}_{1,\vec{k}}, \boldsymbol{\mathfrak{q}}_{2,\vec{k}}, t),
\end{eqnarray}
\ie it is simply obtained by marginalising the full Wigner function over the environmental degrees of freedom in phase-space. This matches \Eq{eq:Wredmarg} given in the main text. Let us stress that the above result is fully generic and does not assume any specific shape for the Wigner function. It can be easily generalised to an arbitrary number of degrees of freedom both in the system and in the environment sectors, since the Wigner-Weyl transform is generated by a kernel that can be written in a fully factorisable form. Non-factorisability due to entanglement is entirely contained in the Wigner function. Therefore, the reduced Wigner function is always obtained from the full Wigner function by simply integrating over the phase-space variables describing the environmental sector. In a sense, marginalisation in phase space is the Wigner-Weyl representation of the partial trace. 

\bibliographystyle{JHEP}
\bibliography{biblio}

\end{document}